\documentclass[universe,article,accept,pdftex,moreauthors]{Definitions/mdpi} 
\firstpage{1} 
\pubvolume{12}
\issuenum{7}
\articlenumber{213}
\pubyear{2026}
\copyrightyear{2026}
\externaleditor{Stefano Vercellone} 
\datereceived{25 May 2026} 
\daterevised{3 July 2026} 
\dateaccepted{7 July 2026} 
\datepublished{16 July 2026} 

\usepackage{longtable}

\usepackage{booktabs}
\usepackage{tabularx}
\usepackage{array}
\usepackage{xcolor}

\Title{Far-Infrared Star Formation Rates of Quasar Host Galaxies from Multiwavelength Spectral Energy Distribution Decomposition}

\Author{{Xiaotong Feng} 
 $^{1,2,}$*\orcidA{}, {Xue-Bing Wu} 
 $^{1,2}$\orcidB{}, Yuming Fu $^{3,4}$\orcidC{}, Yuxuan Pang $^{5}$\orcidD{}, Rui Zhu $^{1,2}$\orcidE{} and Huimei Wang $^{1,2}$\orcidF{}}

\AuthorNames{Xiaotong Feng, Xue-Bing Wu, Yuming Fu, Yuxuan Pang, Rui Zhu and Huimei Wang}

\address{%
$^{1}$ \quad Department of Astronomy, School of Physics, Peking University, Beijing 100871, {China}\\

$^{2}$ \quad Kavli Institute for Astronomy and Astrophysics, Peking University, Beijing 100871, {China}\\
$^{3}$ \quad Leiden Observatory, Leiden University, Einsteinweg 55, 2333 CC Leiden, The Netherlands\\
$^{4}$ \quad Kapteyn Astronomical Institute, University of Groningen, P.O. Box 800, 9700 AV Groningen, The Netherlands\\
$^{5}$ \quad School of Astronomy and Space Science, University of Chinese Academy of Sciences (UCAS),\linebreak   Beijing 100049, {China}}

\corres{Correspondence: xtfeng@pku.edu.cn}

\abstract{
Reliable star formation rates (SFRs) are essential for studying the connection between black hole growth and quasar host galaxies. 
We study the far-infrared (FIR) SFRs and the host galaxy properties of 202 SDSS and PG quasars at $0.02<z\lesssim0.8$, spanning $\log({\rm SFR}_{\rm FIR}/M_\odot\,{\rm yr}^{-1})\simeq-0.45$--$2.76$, using multiwavelength spectral energy distribution (SED) decomposition. 
The photometry covers wavelengths from the optical to the FIR and is supplemented by JCMT/SCUBA-2 observations at 450 and 850~$\upmu$m.
We model the SEDs with CIGALE and AGNfitter and adopt multiple cold dust templates to quantify systematic uncertainties. 
The median model-dependent scatter among the five FIR SFR estimates is $0.14$ dex, and AGNfitter gives FIR SFRs lower than the mean CIGALE estimate by a median of $0.09$ dex. For the 58 quasars with SCUBA-2 coverage, including SCUBA-2 data changes the adopted FIR SFR by only $\sim$0.01 dex on average but can affect individual sources with limited {Herschel} 
coverage or radio-loud emission. Within our FIR-constrained sample, many quasar hosts lie on or above the star-forming main sequence, but the redshift-dependent FIR selection of the SDSS subsample limits conclusions about the full quasar-host population. We find no clear correlation between the main-sequence (MS) offset and the direct Eddington ratio, while the offset is positively related to the infrared-based $L_{\rm tor}/L_{\rm Edd}$ proxy. The minimum radiation field intensity in the dust model, $U_{\rm min}$, increases with bolometric luminosity and dust temperature. WISE W2 (4.6 $\upmu$m) and W3 (12 $\upmu$m) combined with Herschel bands can also provide useful empirical indicators of $f_{\rm AGN}$.
}


\begin{document}

\section{Introduction}\label{sec:introduction}

The growth of supermassive black holes (SMBHs) and the evolution of their host galaxies are closely linked.
Empirical correlations between the black hole mass and the properties of the host galaxy bulge (bulge luminosity, stellar mass, stellar velocity dispersion, etc.) suggest that the growth of SMBHs and galaxy formation are linked over a long time \citep{2000ApJ...539L..13G,2000ApJ...539L...9F,
2013ARA&A..51..511K}.
The luminous quasars provide an important phase for studying this connection, because rapid black hole accretion and the growth of their host galaxies can occur simultaneously. Therefore, the reliable measurement of the star formation rate (SFR) in a quasar host galaxy is essential to study whether black hole growth is related to star formation.

However, the measurement of SFRs of quasar host galaxies is not simple. Many SFR indicators can be contaminated by the strong emission of the central active galactic nucleus (AGN).
The accretion disk dominates optical and ultraviolet continuum, the narrow-line region (NLR) contributes to nebular emission lines, the dusty torus emission is strong at the near (NIR)- to-mid-infrared (MIR) band. 
Mid-infrared features, such as PAH emission and the [Ne II] line, can also trace star formation \citep{2016ApJ...818...60S,2007ApJ...658..314H}. However, in luminous quasars, strong AGN mid-infrared emission can make PAH features difficult to measure, while [Ne II] may include emission from the AGN narrow-line region \citep{2013ApJ...769...75S,2019ApJ...873..103Z}. 
X-ray emission in luminous quasars is also generally dominated by
accretion-powered AGN emission and therefore does not directly trace star formation \citep{2012MNRAS.419.2095M,2015A&ARv..23....1B}.
Radio emission and jets can also affect the radio continuum. So, most of the widely used SFR indicators of normal star-forming galaxies are unreliable (UV continuum, optical recombination lines, NIR emission, and radio continuum), unless the AGN contribution is carefully excluded \citep{1998ARA&A..36..189K,2012ARA&A..50..531K}.

Far-infrared (FIR) emission is one of the useful indicators of SFRs of AGN host galaxies. In star-forming galaxies, ultraviolet and optical photons from young stars are absorbed by dust and reradiated in the infrared (IR) band. So the FIR emission is generally less affected by the AGN torus than NIR and MIR emission and can be a relatively clean tracer of dust-obscured star formation. 
Many studies have shown that the FIR emission of normal AGNs is dominated by the host galaxy and can be used to probe the connection between star formation and AGN activity \citep{2012MNRAS.419...95M,2012A&A...545A..45R}. However, for luminous quasars, AGN torus may also contribute to FIR emission, and the radio emission from central AGN may affect the submillimeter-to-radio SED. Therefore, directly converting the FIR luminosity into an SFR may overestimate the SFR of quasar host galaxies if the AGN contribution is not properly~decomposed.

Therefore, a more robust method to measure the SFR is to do the multiwavelength spectral energy distribution (SED) fitting and decompose the stellar, dust, AGN, and radio components. 
The SED-fitting decomposition can separate the cold dust emission which is associated with star formation from the AGN hot dust torus. 
Then, the FIR luminosity of the cold dust component can be used to estimate the FIR SFR. 
However, the reliability of this method is related to the quality and coverage of the FIR photometric data.
Recent simulations have shown that the recovery of star-forming and AGN infrared luminosities depends on the wavelength coverage of the infrared SED. FIR data are especially important for constraining the cold dust component \citep{2026ApJ...997..150F}.
If there are additional submillimeter constraints, they can be important for reducing the uncertainties of the cold dust component, especially for the sources with no or limited Herschel detections.
In addition, different SED-fitting codes can give different SFRs even when they use the same photometric data, because they adopt different model assumptions \citep{2023ApJ...944..141P}.
The AGN contribution at FIR wavelengths is one source of this uncertainty. Different torus models use different dust geometries and grain properties and can therefore predict different FIR SED shapes. For example, the models of \citet{2006MNRAS.366..767F} and \citet{2015A&A...583A.120S} make different assumptions about the dust distribution and composition. As a result, the separation between AGN-heated dust and star-formation-heated dust can depend on the adopted model, especially when the FIR data are limited.
These issues motivate the use of multiple SED models and extended FIR/submillimeter coverage to obtain robust SFR estimates for quasar host galaxies.

The relation between star formation and AGN activity remains debated. 
Some studies found the correlation between the host galaxy SFR and AGN luminosity, black hole accretion rate, and Eddington ratio, while other studies also found weak or no correlations once redshift and stellar mass are controlled \citep{2013ApJ...773....3C,2014ApJ...782....9H,2015MNRAS.453..591S, 2018MNRAS.478.4238D,2020ApJ...901...66W,2022ApJ...934..130Z}. 
Part of this diversity may be caused by different timescales. The FIR emission traces average star formation over tens to hundreds of Myr, but the variability in AGN accretion has much shorter timescale. 
It is therefore important to derive a robust SFR of a quasar host galaxy. Then, we can compare them with the AGN properties while accounting for redshift and sample selection effects.

The main goal of this work is to derive robust FIR-based SFRs for low-redshift \mbox{type-I} quasars and to quantify the main uncertainties in these measurements. We study a sample of 202 quasars at $z\sim0.02\text{--}0.8$, including 115 SDSS and 87 PG quasars. We compile multiwavelength photometry from the optical to the FIR and supplement the archival data with JCMT/SCUBA-2 observations at 450 and 850~$\upmu$m.
We use two independent SED-fitting approaches, CIGALE and AGNfitter, together with multiple cold dust templates. We first examine how the adopted models and the inclusion of SCUBA-2 data affect the derived SFRs and dust properties. We then use the resulting host galaxy properties to study the locations of quasar hosts relative to the star-forming main sequence and their relations with black hole accretion and dust heating. As an additional application, we examine whether simple infrared colors can provide empirical estimates of the
infrared AGN fraction.  

This paper is organized as follows. Section \ref{sec:sampleandobs} describes the sample selection and observations. Section \ref{sec:sedfitting} presents the SED-fitting methods and adopted model parameters. Section~\ref{sec:results} gives the FIR SFRs and host galaxy properties, discusses the effect of the SCUBA-2 data, and examines the connection between star formation and AGN activity. We summarize our main results in Section \ref{sec:summaryconclusion}.

\section{Sample and Observations}\label{sec:sampleandobs}

\subsection{Sample Selection}\label{subsec:sample}

Our sample consists of 202 quasars. We started with the catalog of 105,783 quasars in the Sloan Digital Sky Survey Data Release 7 (SDSS DR7) \citep{2009ApJS..182..543A,2010AJ....139.2360S,2000AJ....120.1579Y} compiled by \citet{shen11}. This catalog provides a comprehensive set of physical measurements for SDSS DR7 quasars, including continuum measurements around the H$\alpha$, H$\beta$, {Mg {\sc ii}}
, and \mbox{{C {\sc iv}}} regions; measurements of broad and narrow emission lines; and black hole masses estimated from multiple broad emission lines. 
The parent SDSS DR7 catalog includes luminous quasars with $M_i<-22.0$ and, in general, at least one broad emission line with ${\rm FWHM}>1000~{\rm km~s^{-1}}$. A small number of spectroscopically confirmed quasars with complex absorption spectra were also retained. This exception does not refer to the inclusion of narrow-line AGNs.
Because this sample will also be used for a companion study of [{O {\sc ii}}] $\lambda3727$-based SFRs, and the SDSS DR7 spectral wavelength coverage is 3800--9200 \AA, we restrict the redshift range to $z \sim 0.02\text{--}0.8$, yielding 21,075 quasars.
To estimate far-infrared SFRs, we need far-infrared data. We therefore cross-match the 21,075 quasars with Herschel \citep{Pilbratt2010} catalogs provided by the NASA/IPAC Infrared Science Archive {(IRSA)}
\endnote{\url{https://irsa.ipac.caltech.edu/}, {accessed on 5 May 2021.} 
}.
The adopted far-infrared photometry includes the 70, 100, and 160 $\upmu$m bands from the Photoconductor Array Camera and Spectrometer (PACS) \citep{2010AA...518L...2P} and the 250, 350, and 500 $\upmu$m bands from the Spectral and Photometric Imaging Receiver (SPIRE) \citep{2010AA...518L...3G}. We require each source to be detected in at least two of these six Herschel bands to provide meaningful constraints on the far-infrared SED. After applying this criterion, we obtain a sample of 120 SDSS quasars.

\begin{figure}[H]
\vspace{+4pt}
\begin{minipage}{0.48\textwidth}
    \centering
    \includegraphics[width=\textwidth]{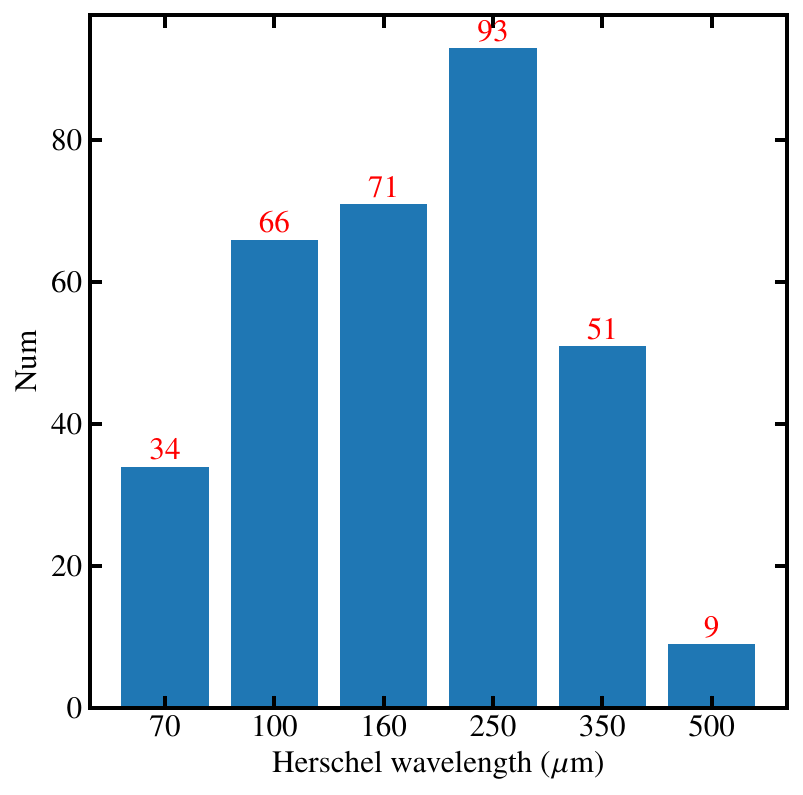}
  (\textbf{a}) SDSS quasars
\end{minipage}
\hfill
\begin{minipage}{0.48\textwidth}
    \centering
    \includegraphics[width=\textwidth]{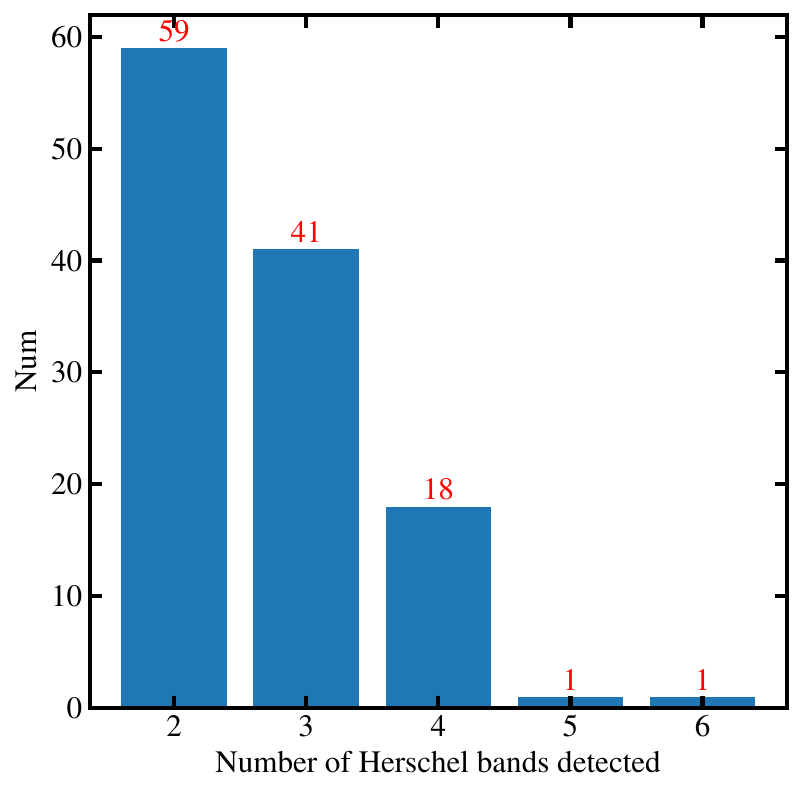}
    (\textbf{b}) SDSS quasars
\end{minipage}

\vspace{2mm}

\begin{minipage}{0.48\textwidth}
\centering
        \includegraphics[width=\textwidth]{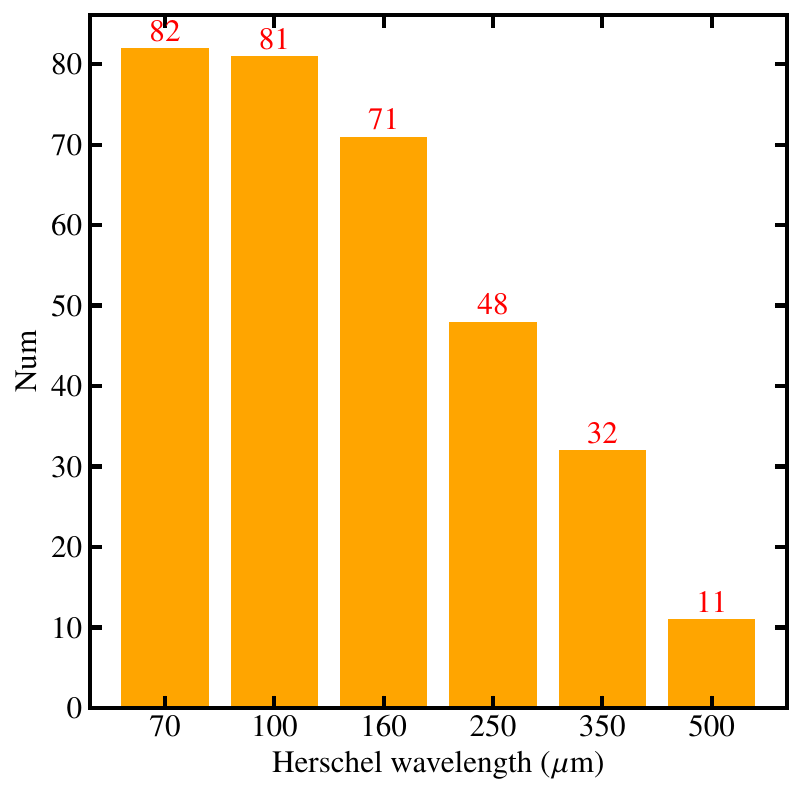}
   (\textbf{c}) PG quasars
\end{minipage}
\hfill
\begin{minipage}{0.48\textwidth}
    \centering
    \includegraphics[width=\textwidth]{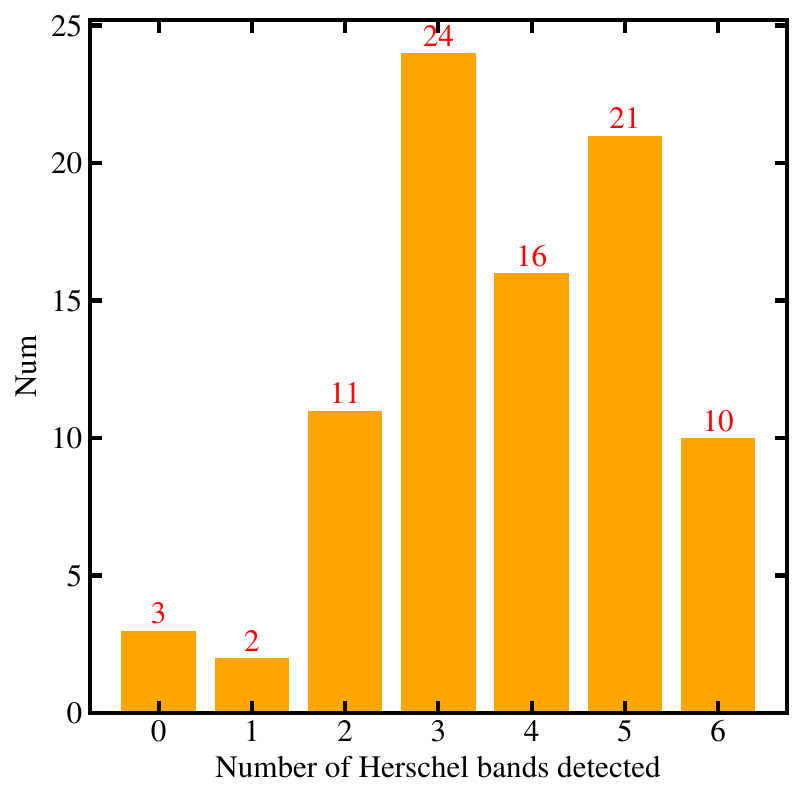}
(\textbf{d}) PG quasars
\end{minipage}

\caption{{Herschel} 
 detection statistics for the SDSS and PG quasars. 
{} 
 (\textbf{a},\textbf{b}) show the SDSS subsample: (\textbf{a}) distribution of the number of sources detected in six Herschel bands and (\textbf{b}) distribution of the number of detected Herschel bands for each source. 
{} (\textbf{c},\textbf{d}) show the same distributions for the PG~\mbox{quasars}.
}
\label{fig:hsbandhist}
\end{figure}

Figure \ref{fig:hsbandhist} shows the distribution of the number of sources detected in the six Herschel bands (70, 100, 160, 250, 350, and 500 $\upmu$m) and the distribution of the number of detected bands for each source. 
For our SDSS sample, the 250 $\upmu$m band has the highest detection number (93 of 120), while the number of detections decreases  at 350 $\upmu$m and 500 $\upmu$m.
Most of the sample (100 of 120) are detected in no more than three Herschel bands. As a result, the long-wavelength side of the Rayleigh--Jeans tail is poorly sampled for most sources, which may lead to relatively weak constraints on the far-infrared SED and affect the estimates of far-infrared luminosity, SFR, cold dust temperature, and other derived quantities.

To improve the constraints on the long-wavelength part of the far-infrared SED, we obtained additional submillimeter observations with the Submillimetre Common-User Bolometer Array 2 on the James Clerk Maxwell Telescope (JCMT/SCUBA-2) \citep{Holland2013}. The observations are described in Section \ref{subsec:scuba2}.

We also include 87 PG quasars with $z<0.5$ from the sample analyzed by \citet{shangguan2018} to enlarge the sample, because these objects have relatively extensive Herschel photometric coverage \citep{shangguan2018}. 
These objects originate from the Palomar--Green Bright Quasar Survey \citep{SG1983}, with optical spectroscopic properties compiled by \citet{BG1992}.
The PG quasars do not follow the same absolute-$i$-band magnitude selection as the SDSS subsample. While the SDSS parent catalog adopts $M_i<-22.0$, the PG survey was selected using bright optical magnitude and ultraviolet-excess criteria. Thus, the PG quasars should not be regarded as a homogeneous low-redshift extension of the SDSS sample, and possible differences in luminosity and color selection should be considered when interpreting the combined sample.
We adopt the Herschel photometry compiled by \mbox{\citet{shangguan2018}}, including the available PACS 70, 100, and 160~$\upmu$m and SPIRE 250, 350, and 500~$\upmu$m measurements, together with the reported uncertainties and upper limits. These data were not re-extracted in this work.
As shown in Figure \ref{fig:hsbandhist}, the PG sample has generally more complete PACS coverage than the SDSS subsample, whereas the SPIRE detection rate decreases toward longer wavelengths in both samples. 
Compared with the SDSS quasars, a larger fraction of PG quasars have three or more Herschel detections.
A summary of the selection criteria, redshift distributions, and luminosity ranges of the two subsamples is given in Table~\ref{tab:sample_comparison}.
A total of 5 of the 87 PG quasars are also included in the initially selected SDSS sample of 120 objects. To avoid double counting, these overlapping sources are assigned to the PG subsample, yielding a final sample of 202 unique quasars, comprising 115 SDSS quasars and 87 PG quasars.

\begin{table}[H]
\caption{{Comparison} 
 of the SDSS and PG subsamples. 
The listed numbers refer to unique quasars in the final sample. The initial SDSS selection contains 120 objects, five of which overlap with the PG sample. These duplicates are counted only once, resulting in 115 SDSS and 87 PG quasars.
For the numerical quantities, values are reported as the median with the 16th and 84th percentiles.}
\label{tab:sample_comparison}

\begin{adjustwidth}{-\extralength}{0cm}
\begin{tabularx}{\fulllength}{X>{\raggedright\arraybackslash}m{0.35\fulllength}>{\raggedright\arraybackslash}m{0.35\fulllength}}
\toprule
\textbf{Property} & \textbf{SDSS Subsample} & \textbf{PG Subsample} \\
\midrule
Number of sources & 115 & 87 \\ \midrule
Redshift range & $0.02 \leq z \leq 0.82$ & $z<0.5$ \\ \midrule
Median $z$ & $0.42_{-0.26}^{+0.31}$ & $0.14_{-0.08}^{+0.18}$ \\ \midrule
Median $\log(L_{\rm bol}/{\rm erg\,s^{-1}})$ & $45.53_{-0.32}^{+0.46}$ & $45.79_{-0.57}^{+0.78}$ \\ \midrule
Parent-sample selection & $M_i<-22$; broad-line SDSS DR7 quasars & Bright optical magnitude and UV excess PG/BQS selection \\ \midrule
Herschel data source & PACS/SPIRE photometry cross-matched from the IRSA Herschel point-source catalogs (HPPSC v1c and SPIRE PSC v2c) in this work & PACS/SPIRE fluxes, uncertainties, and upper limits adopted from \mbox{\citet{shangguan2018}} \\
\bottomrule
\end{tabularx}
\end{adjustwidth}
\end{table}

\subsection{SCUBA-2 Observations}\label{subsec:scuba2}

SCUBA-2 is the submillimeter continuum camera of the 15 m JCMT on Maunakea, Hawaii. It can observe simultaneously at 450 $\upmu$m and 850 $\upmu$m. The 450 $\upmu$m band provides a higher angular resolution and is therefore useful for resolving fine spatial structures, whereas the 850 $\upmu$m band has higher sensitivity and is more suitable for detecting faint~sources.

For our 120 selected SDSS quasars, we carried out several rounds of \mbox{JCMT/SCUBA-2} observations under the programs M18BP050, M19AP020, M19BP041, and M20AP044. Observations were obtained for 44 sources in total. 
The targets were not selected randomly from the parent sample. We first used the available Herschel photometry to perform preliminary cold dust SED fits and predict the expected SCUBA-2 flux densities. These predictions were used to estimate the integration time required to reach the planned sensitivity. In assigning observational priorities, we considered both the estimated observing time and the existing Herschel coverage, giving higher priority to sources with fewer Herschel detections in order to improve the constraints on the long-wavelength FIR SED. Owing to the limited telescope time, some sources with very low predicted submillimeter flux densities and correspondingly long estimated integration times were not observed. Thus, the SCUBA-2-observed subset was selected to balance observational feasibility with the need for improved long-wavelength constraints. It should not be regarded as a random or statistically complete subset of the 120 SDSS FIR-selected quasars.
The observations were performed in the ``CV Daisy'' (CV stands for constant velocity) observing mode, which is designed for small and compact sources of order 3 arcmin or less. The main beam size is $7.9^{\prime\prime}$ at 450 $\upmu$m and $13.0^{\prime\prime}$ at 850 $\upmu$m. 
The observations were obtained predominantly under moderate-to-good atmospheric conditions. The zenith atmospheric opacity at 225~GHz had a median value of $\tau_{225}=0.067$, with the 16th--84th percentile range of $\tau_{225}=0.056$--$0.085$. The full range of the start and end values recorded for the individual observations was $\tau_{225}=0.001$--$0.111$.
The effective on-source exposure time for each target ranges from $\sim$0.2 to $\sim$3.5 h. 

The SCUBA-2 data were reduced using the \texttt{ORAC-DR} pipeline \citep{2015A&C.....9...40J} within the {STARLINK 2018A} 
 software \citep{2014ASPC..485..391C}. Since most of our targets are faint, we adopted the configuration file \texttt{dimmconfig\_blank\_field.lis}, which is ideal for faint cosmological sources and blank fields. 
The flux calibration was performed using the standard SCUBA-2 flux conversion factors recommended by the JCMT/Starlink pipeline.
As the optical coordinates of all targets are well determined, source detection was performed around the known optical positions. We searched for submillimeter counterparts within one beam FWHM of the optical coordinates, corresponding to $7.9^{\prime\prime}$ at 450 $\upmu$m and $13.0^{\prime\prime}$ at 850 $\upmu$m.

A source is considered detected if there is a cluster of pixels with a signal-to-noise ratio $\geq$$3\sigma$ near the optical position and the emission morphology is consistent with the point spread function. Since the targets are expected to be unresolved point sources, the peak flux density was adopted as the flux density of the detected source. If no $\geq$$3\sigma$ signal is found near the optical position, we adopted the corresponding $3\sigma$ value as the upper limit on the flux density.
The observational results are listed in Table~\ref{tab:scuba2datasdss}. At 450~$\upmu$m\ and 850~$\upmu$m, two~and seven sources are detected at $\geq$$3\sigma$, respectively, while the remaining sources are assigned $3\sigma$ flux-density upper limits. 

\begin{table}[H]
\caption{SCUBA-2 observational data for the 44 SDSS quasars.}
\label{tab:scuba2datasdss}

\begin{adjustwidth}{-\extralength}{0cm}

\begin{tabularx}{\fulllength}{@{}p{4.5cm}CCC>{\centering\arraybackslash}p{2.5cm}>{\centering\arraybackslash}p{2.5cm}>{\centering\arraybackslash}p{1.3cm}@{}}
\toprule
\textbf{{Name} 
} & \textbf{R.A.} & \textbf{Dec.} & \textbf{\textit{z}} 
& \boldmath{$F_{450}$} & \boldmath{$F_{850}$} & \textbf{Obs.} \\
 & \textbf{({deg}
)} & \textbf{({deg})} &  
& \textbf{(mJy)} & \textbf{(mJy)} &  \\
\textbf{(1)} & \textbf{(2)} & \textbf{(3)} & \textbf{(4)} 
& \textbf{(5)} & \textbf{(6)} & \textbf{(7)} \\
\midrule
J011430.24+000420.94 & 18.6260 & 0.0725 & 0.455 & $<$136.90 & $<$6.92 & 20A \\
J014003.99{-}
002541.77 & 25.0166 & $-$0.4283 & 0.770 & $<$818.89 & $<$7.60 & 20A \\
J015243.15+002039.65 & 28.1798 & 0.3443 & 0.578 & $<$216.07 & 68.23 $\pm$ 2.64 & 20A \\
J015950.25+002340.82 & 29.9594 & 0.3947 & 0.163 & $<$237.88 & $<$7.23 & 19B \\
J021859.87+002855.80 & 34.7495 & 0.4822 & 0.351 & $<$317.81 & $<$8.24 & 20A \\
J090026.51+204158.70 & 135.1105 & 20.6996 & 0.706 & $<$86.87 & $<$4.51 & 19A \\
J090158.88+002313.87 & 135.4953 & 0.3872 & 0.196 & $<$285.36 & $<$6.20 & 19B \\
J092159.40+450912.38 & 140.4975 & 45.1534 & 0.235 & 126.15 $\pm$ 31.37 & $<$7.80 & 19A \\
J092635.12+072446.43 & 141.6463 & 7.4129 & 0.189 & $<$199.37 & $<$12.00 & 19A \\
J095017.06+215022.41 & 147.5711 & 21.8396 & 0.455 & $<$265.07 & $<$7.77 & 20A \\
J105959.93+574848.17 & 164.9997 & 57.8134 & 0.453 & $<$143.40 & 6.98 $\pm$ 1.79 & 20A \\
J110036.64+564134.89 & 165.1527 & 56.6930 & 0.834 & 148.23 $\pm$ 45.78 & $<$6.96 & 20A \\
J111830.29+402554.02 & 169.6262 & 40.4317 & 0.154 & $<$251.74 & $<$10.60 & 19A \\
J115253.68+000131.90 & 178.2237 & 0.0255 & 0.823 & $<$119.77 & $<$5.18 & 20A \\
J120226.76{-}012915.28 & 180.6115 & $-$1.4876 & 0.150 & $<$155.18 & 6.52 $\pm$ 2.04 & 19A \\
J120312.14+015321.30 & 180.8006 & 1.8892 & 0.296 & $<$83.32 & $<$4.77 & 20A \\
J121037.81+053805.88 & 182.6575 & 5.6350 & 0.436 & $<$87.25 & $<$3.74 & 19A \\
J121946.54+145259.37 & 184.9439 & 14.8832 & 0.401 & $<$278.92 & $<$7.74 & 19A \\
J122011.88+020342.21 & 185.0495 & 2.0617 & 0.240 & $<$236.71 & 131.57 $\pm$ 4.02 & 19A \\
J122102.95{-}000733.74 & 185.2623 & $-$0.1260 & 0.366 & $<$229.49 & $<$6.12 & 19A \\
J122317.80+092306.94 & 185.8242 & 9.3853 & 0.682 & $<$110.44 & $<$4.63 & 19A \\
J122526.21+141332.24 & 186.3592 & 14.2256 & 0.760 & $<$695.52 & $<$10.10 & 19A \\
J122822.10+114606.83 & 187.0921 & 11.7686 & 0.365 & $<$355.78 & $<$8.13 & 19A \\
J122839.20+035749.29 & 187.1633 & 3.9637 & 0.608 & $<$445.35 & $<$8.94 & 19A \\
J123436.54+123918.66 & 188.6522 & 12.6552 & 0.777 & $<$103.89 & 4.29 $\pm$ 1.32 & 19A \\
J124511.26+335610.12 & 191.2969 & 33.9361 & 0.711 & $<$92.09 & $<$4.90 & 20A \\
J125317.57+310550.64 & 193.3232 & 31.0974 & 0.782 & $<$99.35 & $<$6.32 & 20A \\
\bottomrule
\end{tabularx}
\end{adjustwidth}
\end{table}

\begin{table}[H]\ContinuedFloat
\caption{\textit{Cont.}}
\label{tab:scuba2datasdss}
\begin{adjustwidth}{-\extralength}{0cm}
\begin{tabularx}{\fulllength}{@{}p{4.5cm}CCC>{\centering\arraybackslash}p{2.5cm}>{\centering\arraybackslash}p{2.5cm}>{\centering\arraybackslash}p{1.3cm}@{}}
\toprule
\textbf{Name} & \textbf{R.A.} & \textbf{Dec.} & \textbf{\textit{z}} 
& \boldmath{$F_{450}$} & \boldmath{$F_{850}$} & \textbf{Obs.} \\
 & \textbf{({deg})} & \textbf{({deg})} &  
& \textbf{(mJy)} & \textbf{(mJy)} &  \\
\textbf{(1)} & \textbf{(2)} & \textbf{(3)} & \textbf{(4)} 
& \textbf{(5)} & \textbf{(6)} & \textbf{(7)} \\
\midrule

J125703.80+250457.56 & 194.2658 & 25.0827 & 0.821 & $<$133.65 & $<$4.49 & 19A \\
J125711.97+274216.45 & 194.2999 & 27.7046 & 0.793 & $<$85.01 & $<$4.19 & 20A \\
J130622.96+225752.95 & 196.5957 & 22.9647 & 0.758 & $<$80.81 & $<$6.64 & 20A \\
J131247.97+250756.71 & 198.1999 & 25.1324 & 0.425 & $<$89.84 & $<$4.87 & 20A \\
J132919.84+250626.45 & 202.3327 & 25.1073 & 0.781 & $<$133.21 & $<$6.69 & 20A \\
J140621.89+222346.54 & 211.5912 & 22.3963 & 0.098 & $<$116.46 & $<$4.95 & 19A \\
J140655.66+015712.88 & 211.7319 & 1.9536 & 0.427 & $<$108.39 & $<$4.87 & 20A \\
J140700.40+282714.65 & 211.7517 & 28.4541 & 0.077 & $<$124.33 & $<$6.70 & 20A \\
J141700.83+445606.39 & 214.2535 & 44.9351 & 0.113 & $<$167.56 & $<$5.70 & 19A \\
J145001.69+022006.75 & 222.5070 & 2.3352 & 0.520 & $<$113.76 & $<$4.31 & 19A \\
J145108.76+270926.92 & 222.7865 & 27.1575 & 0.064 & $<$105.18 & $<$4.91 & 19A \\
J145538.73+002238.06 & 223.9114 & 0.3772 & 0.434 & $<$81.94 & $<$4.67 & 20A \\
J152114.26+222743.87 & 230.3094 & 22.4622 & 0.136 & $<$102.20 & $<$4.76 & 19A \\
J161413.20+260416.21 & 243.5550 & 26.0712 & 0.131 & $<$149.57 & $<$5.52 & 19A \\
J163352.34+402115.66 & 248.4681 & 40.3544 & 0.782 & $<$94.68 & 4.07 $\pm$ 1.30 & 19A \\
J163915.81+412833.70 & 249.8159 & 41.4760 & 0.690 & $<$358.96 & 77.81 $\pm$ 6.01 & 19A \\
J171352.43+584201.25 & 258.4685 & 58.7003 & 0.521 & $<$212.11 & $<$5.67 & 19A \\
\bottomrule
\end{tabularx}
\end{adjustwidth}
\noindent\footnotesize{{Note}
: Column (1): SDSS name. Column (2): right ascension (J2000). Column (3): declination (J2000). Column (4): redshift. Columns (5) and (6): flux densities at 450 $\upmu$m and 850 $\upmu$m, respectively; upper limits are given at the $3\sigma$ level. Column (7): observation semester.}
\end{table}

For the PG quasars, we cross-matched our sample with the public JCMT/SCUBA-2 archival data available from the Canadian Astronomy Data Centre (CADC)\endnote{\url{https://www.cadc-ccda.hia-iha.nrc-cnrc.gc.ca/en/search/?collection=JCMT&Observation.instrument.name=SCUBA-2&Plane.calibrationLevel=2&noexec=true}, {accessed on 20 June 2020.} 
}.
Then, we reduced the data using STARLINK. This procedure yielded SCUBA-2 data for 14 PG quasars. 
For the 450~$\upmu$m\ band, some measurements are unavailable because of the missed raw observations or unreliable reduction results. 
So, only the 850~$\upmu$m\ measurements are available for part of our PG sample. 
The final usable SCUBA-2 data for these nine PG quasars include three $3\sigma$ flux-density upper limits at 450~$\upmu$m and two  $\geq$$3\sigma$ detections plus seven $3\sigma$ flux-density upper limits 
at 850~$\upmu$m.

\section{SED Fitting} \label{sec:sedfitting}

Since quasars emission is very strong across a broad range of the electromagnetic spectrum, SED fitting can give more reliable FIR SFRs and host galaxy properties for our quasar sample. 
By decomposing the observed SED into different individual components, we can account for the AGN-heated hot dust emission in the MIR band and possible contamination in the radio band. Then, we can isolate the FIR cold dust emission associated with star formation.

\subsection{Multiwavelength Data}
\label{subsec:multiwavelengthdata}

\textls[-15]{We compiled multiwavelength data from optical to FIR for each quasar. For the radio-loud quasars, we also collected available radio measurements. 
The complete photometric data used as input for both CIGALE and AGNfitter are provided in Supplementary Table~S1 in machine-readable format.
The catalog products, versions, and corresponding references are summarized in Table~\ref{tab:photometry_sources}.
The NASA/IPAC Infrared Science Archive (IRSA) and the NASA/IPAC Extragalactic Database (NED)\endnote{\url{https://ned.ipac.caltech.edu/}, {accessed on 16 June 2023.} 
} were used for data access and cross-identification.}

\begin{table}[H]
\caption{{Photometric} 
 data products adopted in this work. Source-by-source references for radio and millimeter measurements are provided in Supplementary Table~S1 in machine-readable format.}
\label{tab:photometry_sources}
\small
\begin{adjustwidth}{-\extralength}{0cm}
\begin{tabularx}{\fulllength}{@{}
>{\raggedright\arraybackslash}p{0.17\fulllength}
>{\raggedright\arraybackslash}p{0.23\fulllength}
>{\raggedright\arraybackslash}X
>{\raggedright\arraybackslash}p{0.20\fulllength}@{}}
\toprule
\textbf{Facility} & \textbf{Bands} & \textbf{Catalog or Data Product} & \textbf{References} \\
\midrule
\multicolumn{4}{@{}l}{{Optical to mid-infrared}}
 \\ \midrule

SDSS
& $ugriz$
& SDSS DR7 imaging photometry
& \citep{2009ApJS..182..543A,shen11} \\ \midrule
2MASS
& $J$, $H$, $K_s$
& 2MASS All-Sky Point Source Catalog, compiled in SDSS DR7 quasar-property~catalog
& \citep{2006AJ....131.1163S,2003AJ....126.1090C,shen11} \\ \midrule
WISE
& W1, W2, W3, W4
& WISE Preliminary Data Release,\linebreak   {14 April 2011} 
& \citep{2010AJ....140.1868W,shen11} \\ \midrule
{Spitzer}
& IRAC 3.6--8.0~$\upmu$m; MIPS 24~$\upmu$m
& SEIP Source List, Cryogenic Release~3.0
& \citep{2004ApJS..154...10F,2004ApJS..154...25R} \\ \midrule

\multicolumn{4}{@{}l}{{Far-infrared and submillimeter}} \\ \midrule

Herschel/PACS
& 70, 100, 160~$\upmu$m
& Herschel/PACS Point Source Catalogue, HPPSC v1c
& \citep{2010AA...518L...2P} \\ \midrule
Herschel/SPIRE
& 250, 350, 500~$\upmu$m
& Standard SPIRE Point Source Catalog v2c
& \citep{2010AA...518L...3G} \\ \midrule
PG infrared data
& NIR to FIR
& Published photometric compilation
& \citep{shangguan2018} \\ \midrule
SCUBA-2
& 450, 850~$\upmu$m
& JCMT observations and archival data
& This work \\ \midrule

\multicolumn{4}{@{}l}{{Radio and millimeter}} \\ \midrule

FIRST
& 1.4~GHz
& VLA FIRST Survey Catalog,\linebreak   14 December 2017 version
& \citep{1995ApJ...450..559B,1997ApJ...475..479W,2015ApJ...801...26H} \\ \midrule
Other radio/mm data
& 150~MHz--150~GHz; 1.2~mm
& Individual published measurements
& Supplementary Table~S1\\
\bottomrule
\end{tabularx}
\end{adjustwidth}
\footnotesize
{Note}. Source-by-source radio and millimeter measurements are provided in Supplementary Table~S1, together with their original literature sources \citep{1998AJ....115.1693C,1971AuJPA..19....1W,1995ApJS...97..347G,1978AJ.....83..685O,1990PKS...C......0W,1991ApJS...75....1B,1988MNRAS.234..919H,1981AAS...45..367K,1991ApJS...75.1011G,2011MNRAS.411...85A,1995AAS..113..409S,1994MNRAS.267..167G,2015MNRAS.454.3864B,2016MNRAS.462.1910H,2011ApJS..194...29R,2009ApJS..180..283W,2008AA...489...49V,2014MNRAS.438.3058R,2004MNRAS.352..673A,1992ApJS...79..331W,1966ApJS...13...65P,1991MNRAS.251...46H,1980MNRAS.190..903L,2004MNRAS.351..845G,2016AJ....152...82T,2011ApJ...732...45S,2010AA...520A..62A,1979AuJPA..46....1B,1994ApJS...90..179G,2003AA...402...87H,1996MNRAS.282..779W,2018ApJS..238....9K,2007MNRAS.379.1442W,2001AA...368..414V,1969ApJ...157....1K,1976AJ.....81.1084G}.
\end{table}
\vspace{-12pt}
\subsection{SED Fitting Using CIGALE}
\label{subsec:cigale}

\begin{figure}[H]
\begin{minipage}[t]{0.47\textwidth}
\centering
\includegraphics[width=\linewidth]{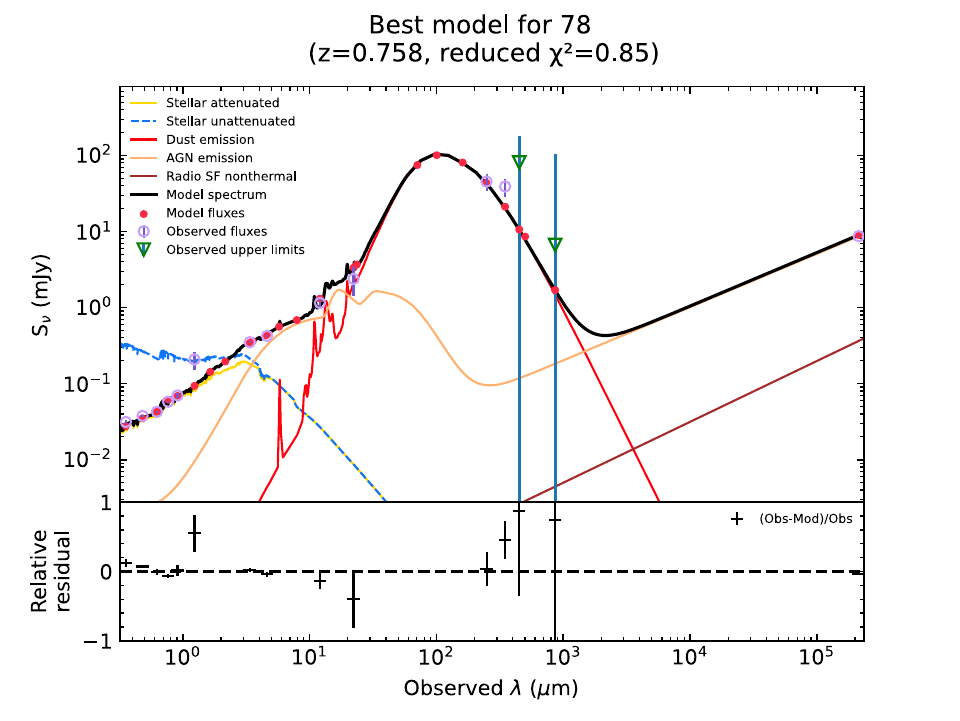}
(\textbf{a}) CIGALE: \texttt{dale2014}
\end{minipage}
\hfill
\begin{minipage}[t]{0.47\textwidth}
\centering
\includegraphics[width=\linewidth]{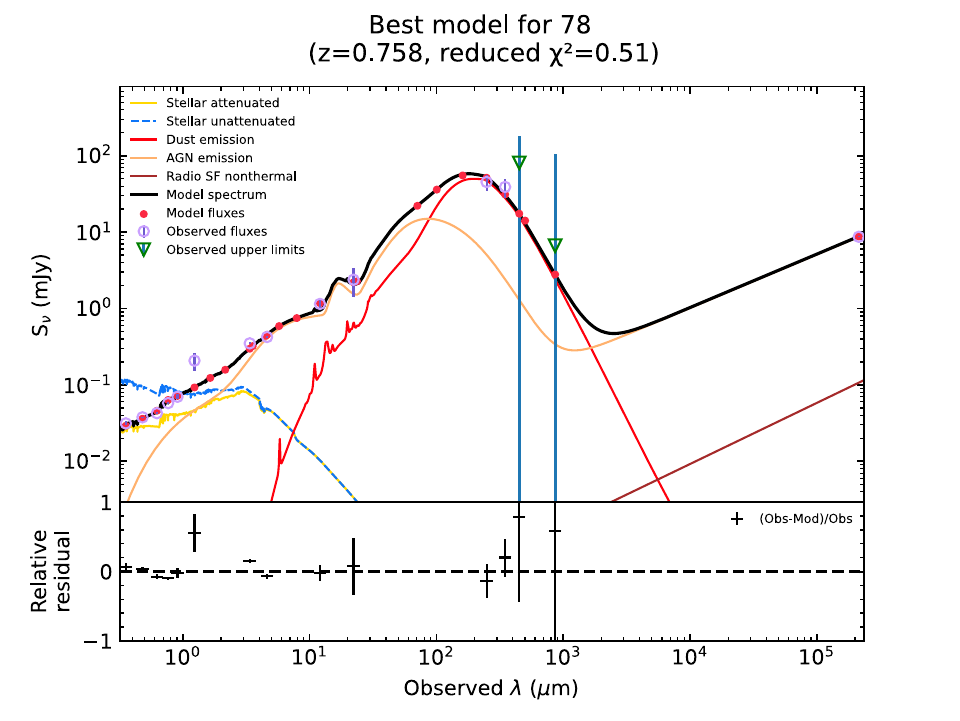}
(\textbf{b}) CIGALE: \texttt{dl2014}
\end{minipage}

%
\vspace{1.5mm}

\begin{minipage}[t]{0.47\textwidth}
\centering
\includegraphics[width=\linewidth]{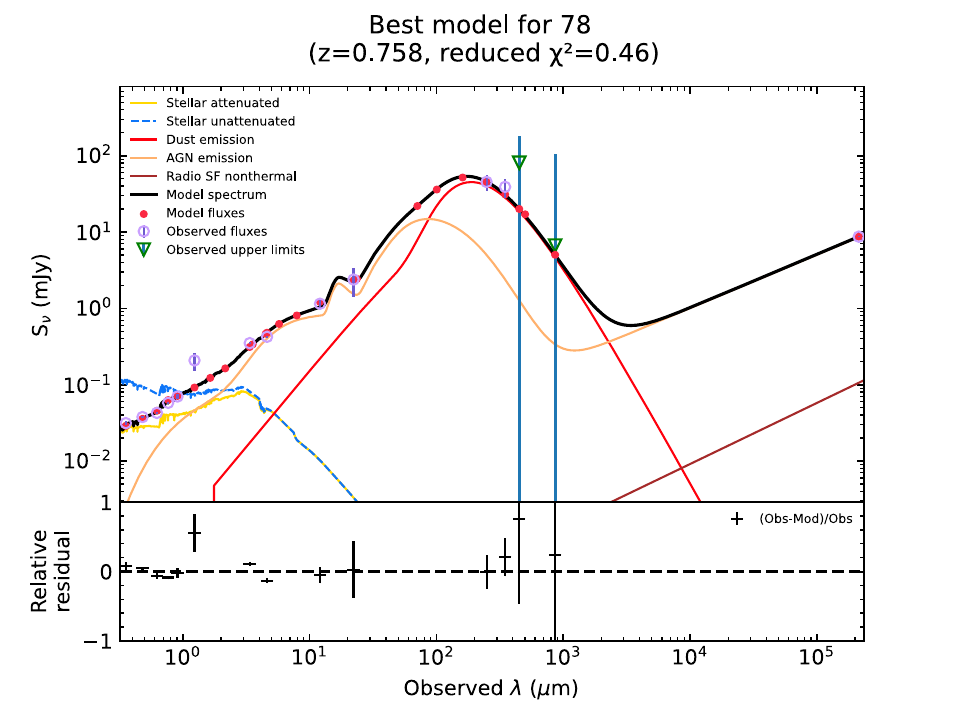}
(\textbf{c}) CIGALE: \texttt{casey2012}
\end{minipage}
\hfill
\begin{minipage}[t]{0.47\textwidth}
\centering
\includegraphics[width=\linewidth]{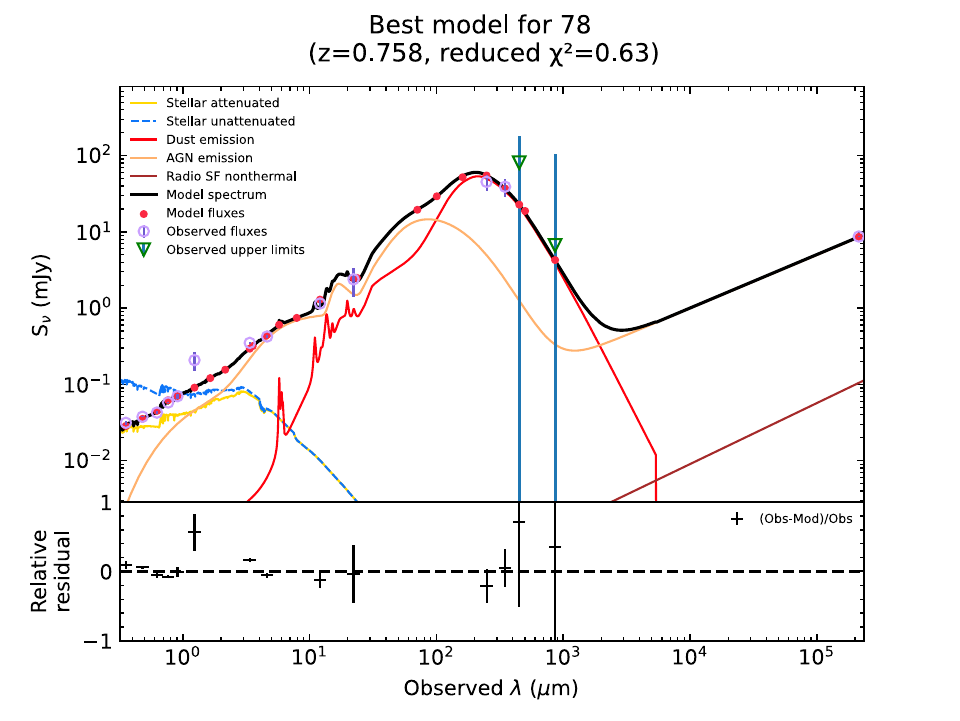}
(\textbf{d}) CIGALE: \texttt{THEMIS}
\end{minipage}

\vspace{2mm}
\makebox[\textwidth]{
\begin{minipage}[t]{0.76\textwidth}
\centering
\includegraphics[width=\linewidth]{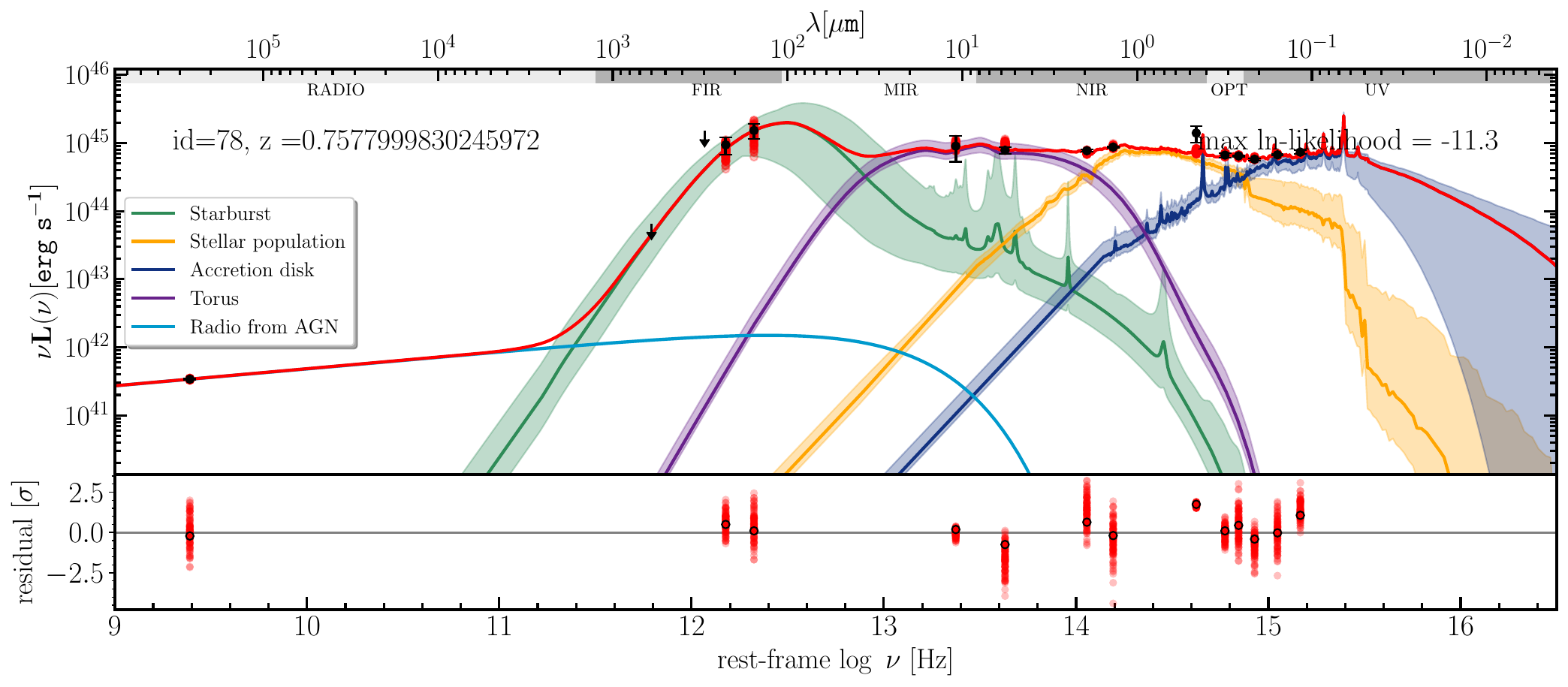}
(\textbf{e}) AGNfitter
\end{minipage}}

\caption{{Comparison} 
 of the SED decompositions of J130622.96+225752.95 obtained with CIGALE and AGNfitter. {} 
 (\textbf{a}--\textbf{d}) show the CIGALE results obtained with the four adopted cold dust emission models: \texttt{dale2014}, \texttt{dl2014}, \texttt{casey2012}, and \texttt{THEMIS}. {}(\textbf{e}) shows the AGNfitter result for the same quasar. All panels use the same observed photometry and upper limits. The lower panels show the relative residuals. The component colors follow the native output conventions of the two fitting codes; their physical meanings are indicated in the individual panels.}
\label{fig:sedcomparison78}
\end{figure}

\textls[-15]{CIGALE\endnote{\url{https://cigale.lam.fr/}, {accessed on 16 April 2022.} 
} (Code Investigating GALaxy Emission) \citep{2005MNRAS.360.1413B,2009A&A...507.1793N,2019A&A...622A.103B}
is a widely used SED-fitting code. 
It models multiwavelength photometric data from FUV to radio and estimates physical properties, such as SFR, stellar mass, dust attenuation, dust emission, and AGN contribution.
CIGALE relies on a predefined model grid, which is computed once and then applied to all sources. 
It is very efficient for large samples.
Reliable results require sufficiently dense sampling of the relevant parameter space.
However, the memory requirement increases as the number of models, photometric bands, and different physical parameters increases. 
Therefore, to keep the memory requirements manageable, we adopt a coarser sampling at first.
Then, we further refine the grid around the preliminary best-fit values for a few key parameters.
In CIGALE, the SED is constructed through a sequence of independent modules, each describing a specific physical component. 
These modules typically include star formation history (SFH), simple stellar population (SSP), nebular emission, dust attenuation, dust emission, AGN, X-ray, radio, rest-frame parameters, and intergalactic medium (IGM). 
We use the following models and parameters for each module: (1) SFH: the delayed SFH with optional exponential burst. (2) SSP: BC03 \citep{2003MNRAS.344.1000B}, with the \citet{2003PASP..115..763C} initial mass function (IMF) and solar metallicity, $Z=0.02$. 
Since the BC03 module in CIGALE does not include the \citet{2001MNRAS.322..231K} IMF which is adopted elsewhere in this work, the relevant CIGALE-derived quantities are converted to that IMF following \citet{2014ARA&A..52..415M}. (3) Dust attenuation: modified \citet{2000ApJ...533..682C} attenuation law. (4) Dust emission: we adopt four models, dale2014~\citep{2014ApJ...784...83D}, dl2014~\citep{2014ApJ...780..172D}, casey2012~\citep{2012MNRAS.425.3094C}, and THEMIS~\citep{2017AA...602A..46J}. (5) AGN emission: skirtor2016 \citep{2012MNRAS.420.2756S,2016MNRAS.458.2288S}. (6)~Radio: synchrotron emission based on \citet{1985ApJ...298L...7H}. 
The parameters of these models are listed in Table \ref{tab:cigalepar}. 
Since the FIR emission is the primary focus of this work, we adopt four dust emission models for the dust module: dale2014, dl2014, casey2012, and THEMIS.
As an example, Figure~\ref{fig:sedcomparison78} presents the CIGALE SED-fitting results of J130622.96+225752.95 obtained with the four adopted cold dust emission models. The AGNfitter decomposition of the same quasar is shown in panel~(e), enabling a direct comparison between the two fitting approaches.}

\begin{table}[H]
\caption{CIGALE model parameters.}
\label{tab:cigalepar}
\small
\begin{adjustwidth}{-\extralength}{0cm}
\begin{tabularx}{\fulllength}{@{}p{11cm} c p{5cm}@{}}
\toprule
\textbf{Parameter} & \textbf{Symbol} & \textbf{Input Values} \\
\midrule

\multicolumn{3}{@{}l}{{SFH: delayed star formation history with an additional burst} 
} \\
\midrule
E-folding time of the main stellar population model (Myr) 
& $\tau_{\rm main}$ 
& 100, 500, 1000, 5000, {10,000} 
 \\

E-folding time of the late starburst population model (Myr) 
& $\tau_{\rm burst}$ 
& {20,000} \\

Mass fraction of the late burst population 
& $f_{\rm burst}$ 
& 0.0, 0.01, 0.1, 0.3 \\

Age of the main stellar population in the galaxy (Myr) 
& $t_{\rm main}$ 
& 100, 500, 1000, 5000, {10,000} \\

Age of the late burst (Myr) 
& $t_{\rm burst}$ 
& 20 \\

\midrule
\multicolumn{3}{@{}l}{{SSP: Bruzual \& Charlot \citep{2003MNRAS.344.1000B}}} \\
\midrule
Initial mass function 
& -- 
& Chabrier \citep{2003PASP..115..763C} \\

Metallicity 
& $Z$ 
& 0.02 \\

Age of the separation between the young and the old stellar population (Myr) 
& $t_{\rm sep}$ 
& 10 \\

\midrule
\multicolumn{3}{@{}l}{{Dust attenuation: modified Calzetti et al. \citep{2000ApJ...533..682C} attenuation law}} \\
\midrule
Color excess of the nebular line emission 
& $E(B-V)_{\rm lines}$ 
& 0.05, 0.1, 0.2, 0.3, 0.4, 0.5, 0.7, 0.9 \\

Reduction factor applied to $E(B-V)_{\rm lines}$ to compute the stellar continuum attenuation {$E(B-V)_{\star}$} 
& $f_{E(B-V)}$ 
& 0.44 \\

\midrule
\multicolumn{3}{@{}l}{{Dust emission: dale2014 \citep{2014ApJ...784...83D}}} \\
\midrule
AGN fraction 
& $f_{\rm AGN}$ 
& 0 \\

Alpha slope 
& $\alpha$ 
& 1.0 to 3.0 \\

\midrule
\multicolumn{3}{@{}l}{{Dust emission: dl2014 \citep{2014ApJ...780..172D}}} \\
\midrule
Mass fraction of PAH 
& $q_{\rm PAH}$ 
& 0.47 to 5.95 \\

Minimum radiation field 
& $U_{\rm min}$ 
& 1.0 to 30.0 \\

Power-law slope $dU/dM \propto U^{\alpha}$ 
& $\alpha$ 
& 2.0 \\

Fraction illuminated from $U_{\rm min}$ to $U_{\rm max}$ 
& $\gamma$ 
& 0.02 to 0.20 \\

\midrule
\multicolumn{3}{@{}l}{{Dust emission: casey2012 \citep{2012MNRAS.425.3094C}}} \\
\midrule
Temperature of the dust 
& $T_{\rm dust}$ 
& 20 to 90 K \\

Emissivity index of the dust 
& $\beta$ 
& 1.0 to 3.0 \\

Mid-infrared power-law slope 
& $\alpha$ 
& 2.0 \\

\midrule
\multicolumn{3}{@{}l}{{Dust emission: THEMIS \citep{2017AA...602A..46J}}} \\
\midrule
Mass fraction of hydrocarbon solids, i.e., {a--C(:H)} 
 smaller than 1.5 nm, also known as HAC 
& $q_{\rm HAC}$ 
& 0.02 to 0.32 \\

Minimum radiation field 
& $U_{\rm min}$ 
& 1.0 to 30.0 \\

Power-law slope $dU/dM \propto U^{\alpha}$ 
& $\alpha$ 
& 2.0 \\

Fraction illuminated from $U_{\rm min}$ to $U_{\rm max}$ 
& $\gamma$ 
& 0.02 to 0.2 \\

\midrule
\multicolumn{3}{@{}l}{{AGN emission: skirtor2016 \citep{2012MNRAS.420.2756S,2016MNRAS.458.2288S}}} \\
\midrule
Average edge-on optical depth at 9.7 $\upmu$m 
& $\tau$ 
& 3 to 11 \\

Power-law exponent that sets radial gradient of dust density 
& $p$ 
& 1.0 \\

Index that sets dust density gradient with polar angle 
& $q$ 
& 1.0 \\

Angle measured between the equatorial plane and edge of the torus 
& $\theta$ 
& 40 \\

Ratio of outer to inner radius, $R_{\rm out}/R_{\rm in}$ 
& $R$ 
& 20 \\

Fraction of total dust mass inside clumps 
& $M_{\rm cl}$ 
& 0.97 \\

Inclination, i.e., viewing angle 
& $i$ 
& 0 to 50 \\

AGN fraction 
& $f_{\rm AGN}$ 
& 0.01 to 0.99 \\

Polar-dust color excess 
& $E(B-V)_{\rm PD}$ 
& 0, 0.2, 0.4 \\

\midrule
\multicolumn{3}{@{}l}{{Radio}} \\
\midrule
The value of the FIR/radio correlation coefficient for star formation 
& $q_{\rm IR,SF}$ 
& 2.58 \\

The slope of the power-law synchrotron emission related to star formation 
& $\alpha_{\rm SF}$ 
& 0.8 \\

The radio-loudness parameter for AGN 
& $R_{\rm AGN}$ 
& 0.01 to 5000 \\

The slope of the power-law AGN radio emission, assumed isotropic 
& $\alpha_{\rm AGN}$ 
& 0.7 or 0.6, 0.7, 0.8, 0.9, 1.0 \\
\bottomrule
\end{tabularx}
\end{adjustwidth}
\end{table}
\vspace{-15pt}

\subsection{SED Fitting Using AGNfitter}
\label{subsec:agnfitter}

AGNfitter\endnote{\url{https://github.com/GabrielaCR/AGNfitter/}, {accessed on 19 February 2024.} 
}
\citep{2016ApJ...833...98C,2024A&A...688A..46M} is an SED-fitting code specifically designed for AGNs. 
It can model multiwavelength photometric data from the radio, submillimeter, IR, optical, and UV bands to the X-ray.
It uses a Bayesian MCMC approach to decompose the observed SED into the physical components associated with the AGN and its host galaxy. 
This Bayesian framework estimates model parameters and their uncertainties from posterior probability distribution functions (PDFs).
This provides more information than a single best-fit solution. 
Owing to its computational efficiency, AGNfitter is very suitable for statistical studies of large samples of AGNs and their host galaxies.

\textls[-15]{The SED components of AGNfitter include stellar population, cold dust emission from star-forming regions, AGN dusty torus, accretion disk, and radio emission. 
Since no X-ray data are used in this work, we do not include the X-ray component. 
AGNfitter allows for theoretical, empirical, and semi-empirical templates for different components. 
The models and parameter ranges we used for each component are summarized in Table \ref{tab:agnfitterpar}. 
These components include (1)~stellar emission: BC03 stellar population synthesis models~\citep{2003MNRAS.344.1000B}; (2) cold dust emission: S17 model \citep{2018A&A...609A..30S}; (3) AGN dusty torus: SKIRTOR model
\citep{2016MNRAS.458.2288S}; (4)~accretion disk: THB21 model~\citep{2021MNRAS.508..737T}; and (5) radio emission: a simple power-law (SPL) model. 
As an example, AGNfitter fitting result for the quasar J130622.96+225752.95 shows in} {Figure~\ref{fig:sedcomparison78}e.} 

\begin{table}[H]
\caption{Models and parameters adopted in AGNfitter.}
\label{tab:agnfitterpar}

\begin{adjustwidth}{-\extralength}{0cm}

\begin{tabularx}{\fulllength}{@{}p{3cm} p{2cm} p{2.5cm} p{7cm} p{0.14\textwidth}@{}}
\toprule
\textbf{Component} & \textbf{Model} & \textbf{Parameter} & \textbf{Description} & \textbf{Range} \\
\midrule

Stellar emission 
& BC03 
& $\tau$ 
& Time scale of the exponential SFH [log year] 
& [0.05, age($z$)] \\

&  
& age 
& Galaxy age [log year] 
& [7, age($z$)] \\

&  
& $E(B-V)_{\rm gal}$ 
& Galaxy reddening 
& [0, 0.6] \\

&  
& GA 
& Galaxy normalization 
& [$-$10, 10] \\

\midrule

Cold dust emission 
& S17 
& $T_{\rm dust}$ 
& Dust temperature [K] 
& [14.24, 42] \\

&  
& fracPAH 
& PAH fraction 
& [0, 0.05] \\

&  
& SB 
& Starburst normalization 
& [$-$10, 10] \\

\midrule

Torus emission 
& SKIRTOR 
& incl 
& Torus inclination angle [$^\circ$] 
& [0, 90] \\

&  
& oa 
& Opening angle [$^\circ$] 
& [10, 80] \\

&  
& $\tau_\nu$ 
& Optical depth at 9.7 $\upmu$m 
& [3, 11] \\

&  
& TO 
& Torus normalization 
& [$-$10, 10] \\

\midrule

Accretion disk 
& THB21 
& $E(B-V)_{\rm bbb}$ 
& BBB reddening 
& [0, 1] \\

&  
& BB 
& BBB normalization 
& [$-$10, 10] \\

\midrule

Radio 
& SPL 
& $\alpha$ 
& Slope of synchrotron emission [log $\nu$] 
& [$-$2, 1] \\

&  
& RAD 
& Radio normalization 
& [$-$10, 10] \\
\bottomrule
\end{tabularx}
\end{adjustwidth}
\end{table}

\section{Results and Discussion} \label{sec:results}

\subsection{FIR SFR Estimates and Model Comparison} \label{subsec:firsfrandcompare}

We estimate the FIR SFRs of the quasar host galaxies using the infrared luminosity--SFR calibration of \citet{1998ARA&A..36..189K}, with the coefficient converted to the \citet{2001MNRAS.322..231K} IMF following
\citet{2014ARA&A..52..415M}:
\begin{equation}
\label{eq:firsfr}
{\rm SFR}~(M_\odot~{\rm yr}^{-1})
= 3.02 \times 10^{-44} L_{\rm FIR}~({\rm erg~s}^{-1}).
\end{equation}

Here, $L_{\rm FIR}$ is measured by integrating the cold dust component obtained from the SED fitting over the infrared wavelength range of 8--1000~$\upmu$m.
For each quasar, we obtain five FIR SFR estimates: four from CIGALE using different cold dust templates, and one from~AGNfitter.

\begin{figure}[H]
\includegraphics[width=0.6\textwidth]{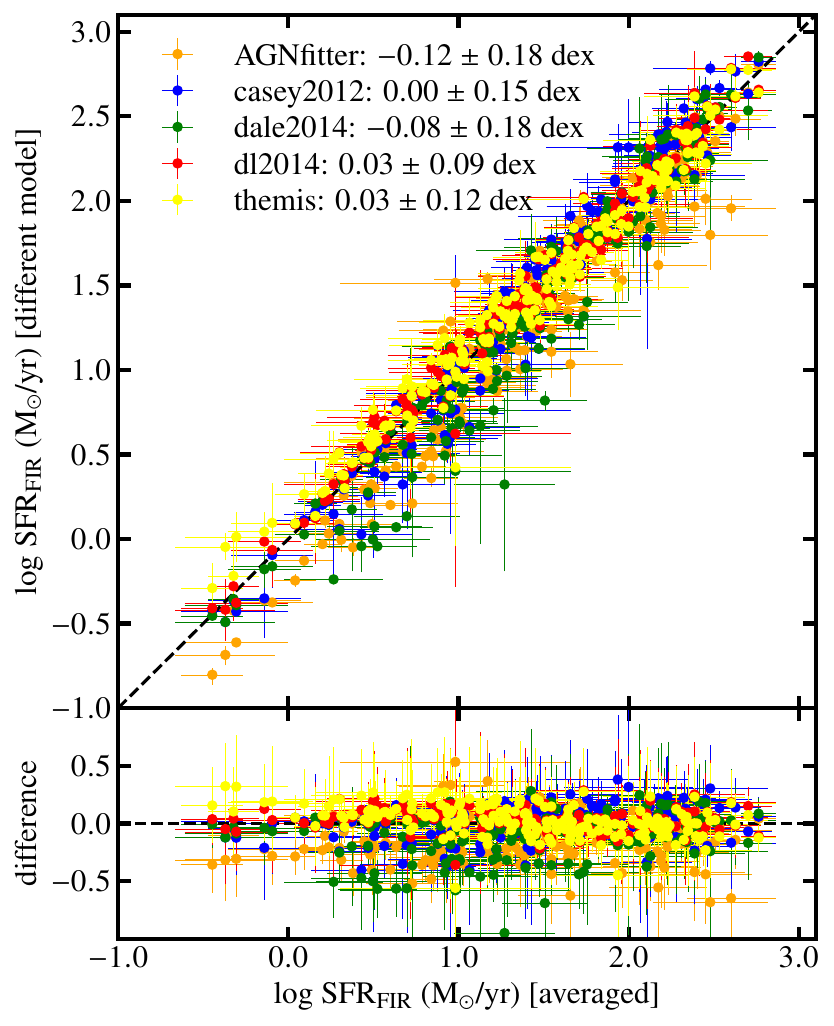}
\caption{
{Comparison} 
 of the five FIR SFR estimates. The abscissa shows the mean SFR, while the ordinate shows the FIR SFR derived from each model. The blue, green, red, and yellow points represent the CIGALE results obtained with the casey2012, dale2014, dl2014, and THEMIS cold dust templates, respectively. The orange points represent the AGNfitter results. The lower panel shows the residuals relative to the mean SFR.
}
\label{fig:compare5firsfr}
\end{figure}

Figure~\ref{fig:compare5firsfr} compares the FIR SFRs of five different models of each source. The mean SFR of the five estimations is shown on the abscissa, and each FIR SFR of the five different models are shown on the ordinate.

To quantify the model-dependent uncertainty of the FIR SFRs, we calculated, for each quasar, the standard deviation of the logarithmic SFR estimates obtained from the four CIGALE cold dust templates and AGNfitter. Among the four CIGALE templates, the median source-by-source scatter is 0.10 dex, corresponding to a multiplicative factor of 1.26 or an approximately 26\% difference in SFR. The 16th--84th percentile range is \mbox{0.02--0.23 dex}. Relative to the mean CIGALE estimate, AGNfitter gives a median offset of $-0.09$ dex, corresponding to an SFR approximately 19\% lower. When all five SFR estimates are considered together, the median model-dependent scatter is 0.14 dex, corresponding to a factor of 1.38 or approximately 38\%. The 16th--84th percentile range is 0.06--0.24 dex. Therefore, for quasars with photometric coverage and SED assumptions comparable to those adopted here, a systematic uncertainty of approximately 0.14 dex, or about 38\%, provides a practical estimate of the uncertainty associated with the choice of SED-fitting code and cold dust template. This uncertainty reflects model dependence and should be considered separately from the formal fitting uncertainty within an individual model.

The selected SED-fitting results are summarized in Table~\ref{tab:sedresultpart}. 
Only a portion of the table is shown here; the full table for all the 202 quasars is provided in Appendix~\ref{app:sedresults}. 
When the same parameter is produced by multiple templates, we adopt the average value as the final~estimate.

\subsection{Effect of SCUBA-2 Data on the SED-Fitting Results}
\label{subsec:resultwithorwithoutscubadata}

To examine the impact of the SCUBA-2 observations on the SED-fitting results, we performed two fits for each source with each template, one including and the other excluding the SCUBA-2 data. In total, five models were considered: the four CIGALE cold dust templates and AGNfitter. Since the influence of the SCUBA-2 data may depend on the available far-infrared constraints and radio properties, we divided the full sample into subsamples according to the number of detected Herschel bands and the radio classification.

\begin{figure}[H]
\includegraphics[width=0.6\textwidth]{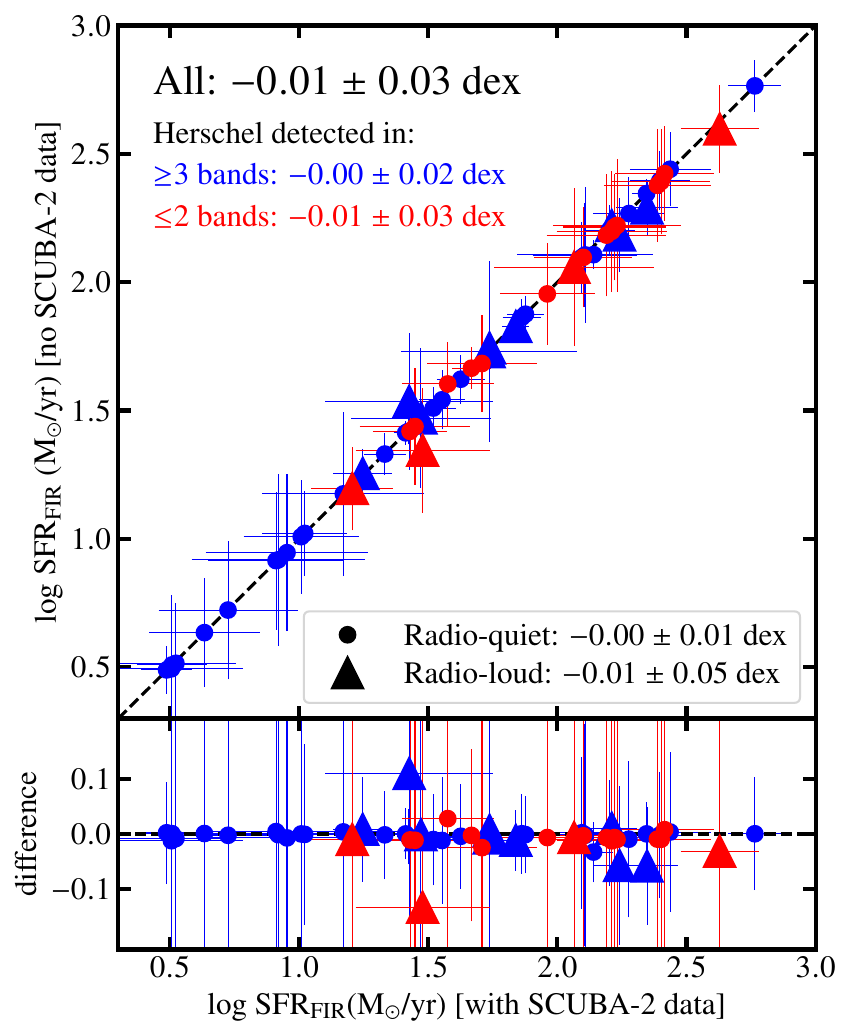}
\caption{
{Comparison} 
 of the FIR SFRs derived from SED fitting with and without the SCUBA-2 data. The abscissa shows the FIR SFR obtained when the SCUBA-2 data are included, while the ordinate shows the FIR SFR obtained when the SCUBA-2 data are excluded. Red symbols indicate sources detected in no more than two of the six Herschel bands, and blue symbols indicate sources detected in three or more Herschel bands. Triangles and circles represent radio-loud and radio-quiet sources, respectively. The mean offsets for the full sample and different subsamples are labeled in the upper left and lower right corners.
}
\label{fig:firsfrwithornotscuba2}
\end{figure}

We first compare the FIR SFRs derived with and without the SCUBA-2 data. For each source, the FIR SFR is taken as the average value from the five models. The result is shown in Figure \ref{fig:firsfrwithornotscuba2}. The abscissa shows the FIR SFR obtained from the SED fitting including the SCUBA-2 data, while the ordinate shows the FIR SFR obtained without the SCUBA-2 data, together with the corresponding residuals. Red symbols indicate sources detected in no more than two of the six Herschel bands, while blue symbols indicate sources detected in three or more Herschel bands. Triangles denote radio-loud sources, and circles denote radio-quiet sources. For the full sample and for each subsample divided by the number of Herschel detections and radio classification, we calculate the mean difference between the results obtained without and with the SCUBA-2 data. These mean offsets are labeled in the upper left and lower right corners of the figure. The same color and symbol scheme is used in the following comparisons of the other five model parameters.

\begin{table}[H]
\caption{{Selected} 
 SED-fitting results for the quasar sample. Only a portion of the table is shown here; the full table for all 202 quasars is provided in Appendix~\ref{app:sedresults}.}
\label{tab:sedresultpart}
\small
\begin{adjustwidth}{-\extralength}{0cm}
\begin{tabularx}{\fulllength}{@{}lc>{\centering\arraybackslash}p{2cm}>{\centering\arraybackslash}p{2cm}ccccc@{}}
\toprule
\textbf{Name} &
\boldmath{$z$} &
\boldmath{$\log M_\star$} &
\boldmath{$\log {\rm SFR}_{\rm FIR}$} &
\boldmath{$T_{\rm dust}$} &
\boldmath{$U_{\rm min}$} &
\boldmath{$f_{\rm AGN}$} &
\boldmath{$\log M_{\rm BH}$} &
\boldmath{$\log\lambda_{\rm Edd}$} \\
&
&
\textbf{(\boldmath{$M_{\odot}$})} &
\textbf{(\boldmath{$M_{\odot}~{\rm yr}^{-1}$})} &
\textbf{(K)} &
&
&
\textbf{(\boldmath{$M_{\odot}$})} &
\\
\textbf{(1)} &
\textbf{(2)} &
\textbf{(3)} &
\textbf{(4)} &
\textbf{(5)} &
\textbf{(6)} &
\textbf{(7)} &
\textbf{(8)} &
\textbf{(9)} \\
\midrule
J001115.69+011459.23&0.580&10.86 $\pm$ 0.53&2.34 $\pm$ 0.15&43.52 $\pm$ 2.99&14.53 $\pm$ 8.68&0.03 $\pm$ 0.01&8.37&$-$0.95\\
J002421.84+002508.49&0.599&10.22 $\pm$ 0.12&2.12 $\pm$ 0.15&30.89 $\pm$ 3.56&2.73 $\pm$ 2.37&0.29 $\pm$ 0.05&7.36&$-$0.16\\
J005905.51+000651.67&0.719&11.56 $\pm$ 0.31&2.00 $\pm$ 0.38&45.86 $\pm$ 9.04&11.14 $\pm$ 8.83&0.70 $\pm$ 0.07&8.94&$-$0.64\\
J011430.24+000420.94&0.455&11.70 $\pm$ 0.08&1.43 $\pm$ 0.14&21.02 $\pm$ 3.12&1.18 $\pm$ 1.02&0.20 $\pm$ 0.07&8.42&$-$1.32\\
J011536.92{-}
000011.01&0.532&11.05 $\pm$ 0.18&1.72 $\pm$ 0.14&21.92 $\pm$ 2.72&1.06 $\pm$ 0.40&0.34 $\pm$ 0.09&8.24&$-$1.15\\
\bottomrule
\end{tabularx}
\end{adjustwidth}
\noindent\footnotesize{{Note}: Column (1): source name. Column (2): redshift. 
Column (3): stellar mass. Column (4): FIR SFR derived from the cold dust component 
integrated over $8$--$1000~\upmu\mathrm{m}$. Column (5): dust temperature. 
Column (6): minimum radiation field intensity. Column (7): fractional AGN contribution 
to the total infrared luminosity, measured over $8$--$1000~\upmu\mathrm{m}$;
when a parameter is returned by multiple SED-fitting templates, we report the average value.
Columns (8) and (9) give the black hole mass and Eddington ratio, respectively. For the SDSS quasars, $M_{\rm BH}$ and $L_{\rm bol}$ are adopted from \citet{shen11}; for the PG quasars, they are adopted from \citet{shangguan2018}. The Eddington ratios are calculated as $\lambda_{\rm Edd}=L_{\rm bol}/L_{\rm Edd}$
}
\end{table}

Unless otherwise specified, the comparison with SCUBA-2 data includes all 58 quasars with available SCUBA-2 observations, including both direct detections and sources with SCUBA-2 upper limits. We define a SCUBA-2 detection as an ${\rm S/N}\geq3$ in either the 450 or 850~$\upmu$m band. The remaining sources with SCUBA-2 coverage are treated as upper-limit-only sources.

The FIR SFRs derived without the SCUBA-2 data are, on average, lower by only 0.01~dex than those obtained when the SCUBA-2 data are included. 
Sources with fewer Herschel detections show a larger scatter than those with better Herschel coverage. 
Similarly, radio-loud sources exhibit a larger scatter than radio-quiet sources. 
In particular, for some radio-loud sources, the difference between the two FIR SFR estimates can reach $\sim$0.12~dex. 
Overall, including the SCUBA-2 data has only a minor effect on the mean FIR SFR of the full sample. 
However, its impact can be appreciable for individual sources with limited Herschel detections or for radio-loud quasars.

To assess whether directly measured submillimeter fluxes provide stronger constraints than upper limits alone, we repeated the comparison separately for the SCUBA-2-detected and upper-limit-only subsets. The results are summarized in Table~\ref{tab:scuba_detection_test}. The mean change in the adopted FIR SFR is close to zero for both subsets. 
The SCUBA-2-detected sources show a substantially larger source-to-source scatter, $0.059$ dex, than the upper-limit-only sources, $0.011$ dex. Their individual changes range from $-0.110$ to $+0.133$ dex, whereas the upper-limit-only sources show a much narrower range of $-0.011$ to $+0.058$ dex. Thus, direct SCUBA-2 detections do not introduce a significant systematic shift in the mean FIR SFR but can lead to larger adjustments in the fitted FIR SFRs for individual sources.

In addition to the FIR SFR, we compare the dust-related parameters derived with and without the SCUBA-2 data for the five SED-fitting models, including the four CIGALE cold dust templates and AGNfitter. The detailed comparisons are presented in Appendix~\ref{app:sedparcompare}.

\vspace{-6pt}
\begin{table}[H]
\caption{Effect of SCUBA-2 constraints on the adopted FIR SFR. We define
$\Delta\log {\rm SFR}_{\rm FIR}=\log {\rm SFR}_{\rm with\ SCUBA}-\log {\rm SFR}_{\rm no\ SCUBA}$.
A SCUBA-2 detection is defined as an ${\rm S/N}\geq3$ in either the 450 or 850~$\upmu$m band. The quoted uncertainties are the standard deviations among sources.}
\label{tab:scuba_detection_test}
\begin{tabularx}{\textwidth}{p{3cm}>{\centering\arraybackslash}p{0.5cm}C>{\centering\arraybackslash}p{2cm}C}
\toprule
\textbf{Subsample} & \boldmath{$N$} & \textbf{Mean} \boldmath{$\Delta\log {\rm SFR}_{\rm FIR}$} & \textbf{Median} & \textbf{Range} \\
& & \textbf{({dex}
)} & \textbf{({dex})} & \textbf{({dex})} \\
\midrule
All observed
& 58 & $0.006 \pm 0.026$ & 0.003 & $-0.110$ to $+0.133$ \\
Detected
& 11 & $0.008 \pm 0.059$ & 0.001 & $-0.110$ to $+0.133$ \\
Upper-limit only
& 47 & $0.006 \pm 0.011$ & 0.003 & $-0.011$ to $+0.058$ \\
\bottomrule
\end{tabularx}
\end{table}

\textls[-15]{We can see that excluding the SCUBA-2 data has only a minor effect on the mean dust parameter estimates for the full sample. 
However, the impact becomes more noticeable for sources with limited Herschel detections and for radio-loud quasars. 
For example, when the SCUBA-2 data are excluded, the casey2012 dust temperature can differ by up to $\sim$18~K for individual sources, and the difference in $\log U_{\rm min}$ for the dl2014 template can reach $\sim$0.7~dex for some radio-loud quasars. 
Similar behavior is found for the THEMIS and AGNfitter results. 
These results indicate that the SCUBA-2 data do not significantly change the average properties of the full sample, but they provide useful additional constraints for individual sources, especially for those with sparse far-infrared coverage or radio-loud~sources.}

Although the comparison above quantifies the effect of including the SCUBA-2 measurements in the SED fitting, it does not address a separate observational uncertainty associated with the relatively large beams at FIR and submillimeter wavelengths. The Herschel/SPIRE and SCUBA-2 fluxes may include contributions from nearby dusty galaxies that are physically associated with the quasar or are unrelated projected sources. Such source confusion is expected to be more relevant at the higher-redshift end of the sample, where the physical area covered by a beam is larger. High-resolution ALMA observations of FIR-bright SDSS quasars at $1<z<4$ have shown that multiple submillimeter counterparts can occur within a single SPIRE beam \citep{2025A&A...702A.183H}.

We visually inspected the SCUBA-2 maps and did not identify obvious cases with multiple resolved submillimeter peaks near the target positions. Nevertheless, the angular resolutions of SPIRE and SCUBA-2 are insufficient to exclude unresolved companions or projected dusty galaxies. In particular, the SCUBA-2 beam FWHM values are $7.9''$ at 450~$\upmu$m and $13.0''$ at 850~$\upmu$m, so optically faint or obscured FIR contributors may remain unresolved. We therefore do not apply a uniform correction to the SPIRE or SCUBA-2 fluxes.
The quantitative results of \citet{2025A&A...702A.183H} cannot be directly transferred to our sample because their quasars are at substantially higher redshifts and were selected to be FIR bright, whereas our sample spans $0.02<z\lesssim0.8$. Source confusion should nevertheless be regarded as an additional source-level systematic uncertainty, particularly for the highest-redshift and most FIR-luminous quasars. It may contribute to the scatter in the inferred FIR luminosities, SFRs, and dust parameters, but spatially resolved submillimeter observations are required to quantify its effect for individual sources.

\subsection{Main-Sequence Offset and Black Hole Accretion} \label{subsec:msandbhar}

\subsubsection{Main-Sequence Location of Quasar Host Galaxies}
\label{subsubsec:ms_location}

\textls[-15]{Using SED fitting, we obtain far-infrared star formation rates and stellar masses for the 202 quasars in our sample.
These measurements allow us to investigate the connection between black hole accretion and galaxy evolution by examining the locations of quasar host galaxies relative to the star-forming main sequence (MS).
The possible model-related scatter in stellar mass and its impact on the following analysis are further discussed in Section~\ref{subsec:stellarmassdiscussion}.}

\textls[-15]{The main sequence describes a tight correlation between star formation rate and stellar mass for star-forming galaxies. 
It has been observed over a wide range of redshifts \citep{2004MNRAS.351.1151B,2011A&A...533A.119E,2014ApJS..214...15S, 2015ApJ...801L..29R,2019MNRAS.483.3213P,2019MNRAS.490.5285P}.
Previous studies have found different results regarding the locations of quasar host galaxies relative to the MS. 
Some studies found quasar hosts to lie below the MS \citep{2015MNRAS.452.1841S,2020ApJ...888...78S} and some found them to be consistent with the MS \citep{2014MNRAS.443..755H,2017ApJ...839..120W, 2022A&A...659A.125S}, while others found them to lie above the MS
\citep{2020MNRAS.498.1560J,2020ApJ...899..112S,2021ApJ...910..124X, 2022A&A...658A..35K}. 
These discrepancies may be caused by many things: differences in sample selection, uncertainties and methodological differences in SFR and stellar mass measurements, the adopted shape and normalization of the MS relation.}

\begin{figure}[H]
\includegraphics[width=0.75\textwidth]{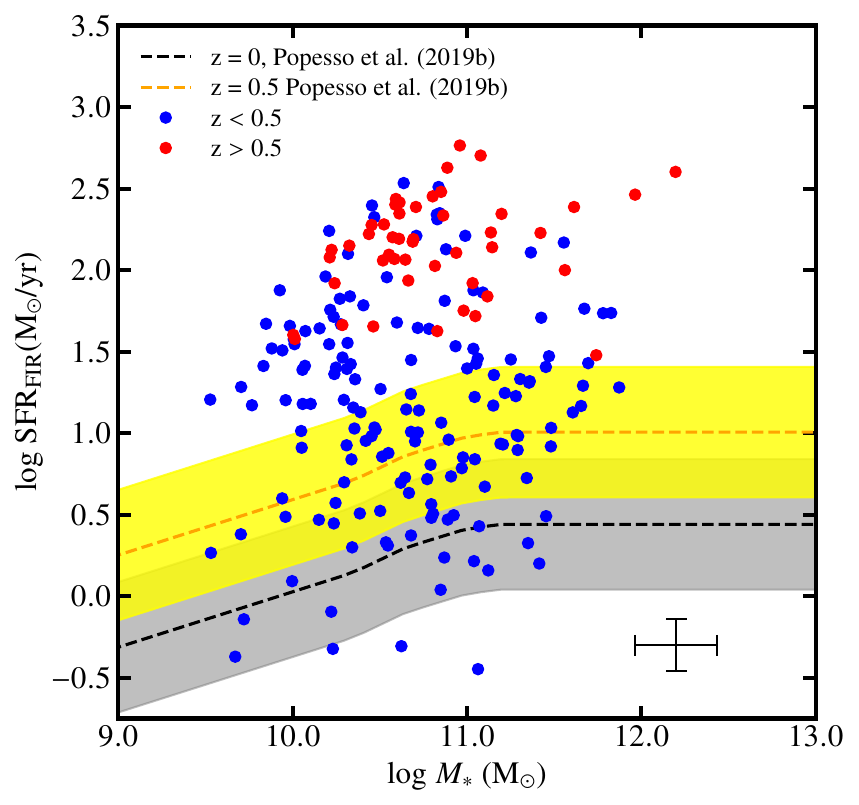}
\caption{
{Relation} 
 between star formation rate and stellar mass for the host galaxies of the 202 quasars. The black and orange dashed lines show the star-forming main-sequence relations at $z=0$ and $z=0.5$ from \citet{2019MNRAS.490.5285P}, respectively. The gray and yellow shaded regions indicate the corresponding main-sequence widths. The quasars are divided into two redshift bins: $z<0.5$ sources are shown as blue points, and $z>0.5$ sources are shown as red points.
}
\label{fig:mainsequence}
\end{figure}

The locations of our quasar host galaxies on the MS are shown in Figure~\ref{fig:mainsequence}. The black dashed line and gray shaded region represent the MS relation at $z=0$ and its width ($\pm 0.4$~dex) from \citet{2019MNRAS.490.5285P}, while the orange dashed line and yellow shaded region show the corresponding relation and width at $z=0.5$. Blue points denote quasars at $z<0.5$, and red points denote quasars at $z>0.5$.

For the $z<0.5$ subsample, most quasar hosts lie on or above the MS. The $z>0.5$ subsample is located at systematically higher SFRs, with most sources lying above the MS. 
However, the apparent high-redshift excess should be interpreted with caution because the SDSS subsample is selected to have detections in at least two Herschel bands. We quantify the resulting redshift-dependent FIR selection effect in Section~\ref{subsubsec:fir_selection}.

Since this redshift division is relatively coarse, and \citet{2019MNRAS.490.5285P} provide an MS relation that explicitly evolves with redshift, we further quantify the offset of each quasar from the MS at its own redshift. 
This reduces the effect of redshift evolution. 
We define the MS offset as
\begin{equation}
\Delta_{\rm MS} = \log {\rm SFR} - \log {\rm SFR}_{\rm MS},
\end{equation}
where $\log {\rm SFR}_{\rm MS}$ is the SFR expected from the MS relation at the
redshift and stellar mass of each quasar.

\subsubsection{Star Formation and Black Hole Accretion}
\label{subsubsec:sfr_bhar}

\begin{figure}[H]
\begin{minipage}{0.48\textwidth}
    \centering
    \includegraphics[width=\textwidth]{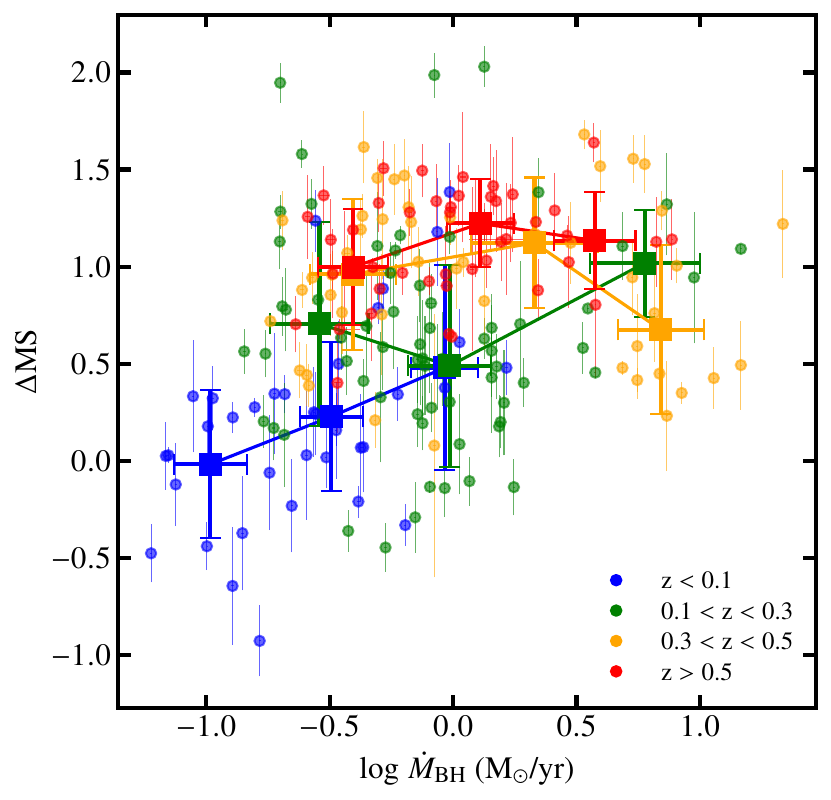}
    (\textbf{a})
\end{minipage}
\hfill
\begin{minipage}{0.48\textwidth}
    \centering
    \includegraphics[width=\textwidth]{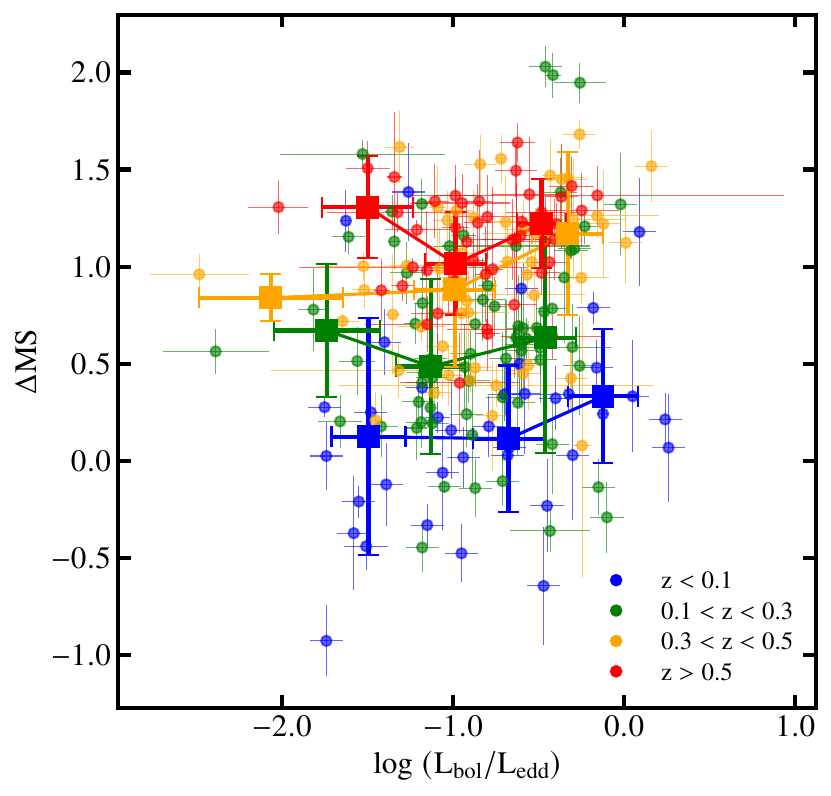}
    (\textbf{b})
\end{minipage}

\caption{
{Relation} 
 between main-sequence offset, $\Delta_{\rm MS}$, and (\textbf{a}) black hole accretion rate and (\textbf{b})~Eddington ratio. The sample is divided into four redshift bins: $z<0.1$ (blue), $0.1<z<0.3$ (green), $0.3<z<0.5$ (orange), and $z>0.5$ (red). In each redshift bin, the sources are further divided into three bins along the abscissa, and the median trends are shown.
}
\label{fig:bhardeltamsleddratiodeltams}
\end{figure}

Several studies of AGNs with different luminosities have reported correlations between host galaxy star formation and AGN luminosity, black hole accretion rate, or Eddington ratio \citep{2013ApJ...773....3C,2018MNRAS.478.4238D,2020ApJ...901...66W, 2022ApJ...934..130Z}. 
Therefore, we compare $\Delta_{\rm MS}$ with the black hole accretion rate and Eddington ratio, as shown in Figure \ref{fig:bhardeltamsleddratiodeltams}. 
The sample is divided into four redshift bins, $z<0.1$, $0.1<z<0.3$, $0.3<z<0.5$, and $z>0.5$, shown in blue, green, orange, and red, respectively. 
In each redshift bin, the sources are further divided into three bins along the abscissa, and the median trends are shown.
The left panel of Figure~\ref{fig:bhardeltamsleddratiodeltams} shows that 
a weak positive correlation is seen only in the lowest-redshift bin, $z<0.1$, whereas the higher-redshift bins show little or no correlation. 
Therefore, the full-sample trend should be interpreted with caution.
The right panel shows that $\Delta_{\rm MS}$ is largely uncorrelated with the Eddington ratio, both for the full sample and within individual redshift bins.

The FIR SFR traces star formation averaged over relatively long timescales, whereas $L_{\rm bol}$, BHAR, and the Eddington ratio trace more instantaneous black hole activity. 
The weak or nearly flat relation between AGN luminosity and the SFR of host galaxies does not necessarily mean there is no physical connection between black hole growth and star formation. Instead, it is more likely due to the strong variability in the AGN accretion rate on short timescales, whereas the SFR reflects the average star formation activity over longer timescales.
(e.g., \citep{2014ApJ...782....9H,2015MNRAS.453..591S}). 
In addition, although we account for the redshift evolution of the main sequence using the relation of \citet{2019MNRAS.490.5285P}, the higher-redshift quasars in our sample still tend to have larger $\Delta_{\rm MS}$. This may be affected by the residual redshift dependence or sample-selection effects.

\begin{figure}[H]
\begin{minipage}{0.48\textwidth}
    \centering
    \includegraphics[width=\textwidth]{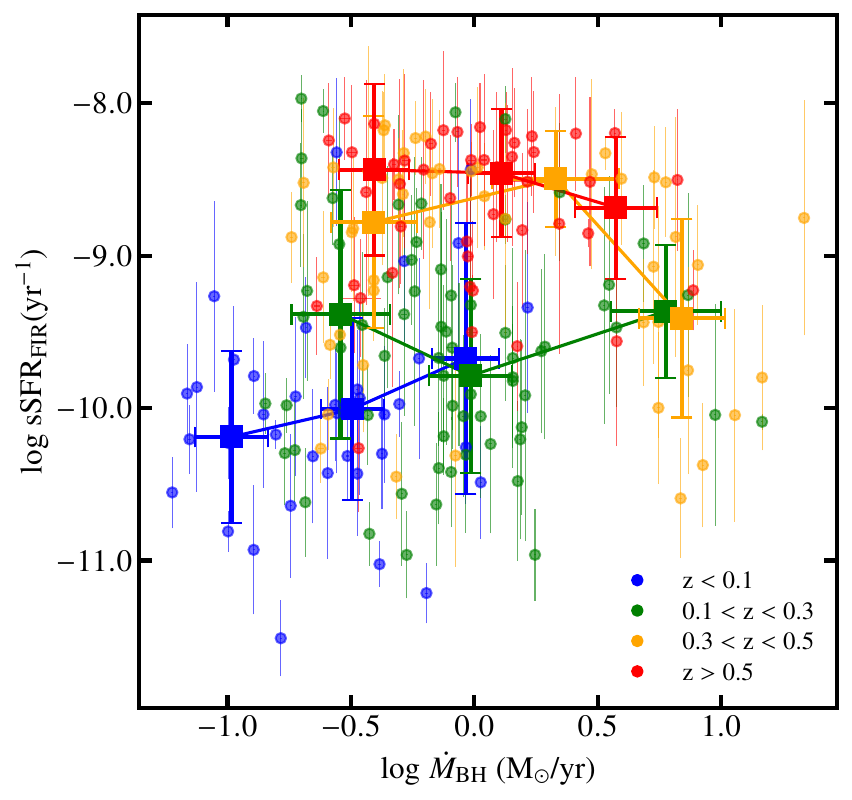}
    (\textbf{a})
\end{minipage}
\hfill
\begin{minipage}{0.48\textwidth}
    \centering
    \includegraphics[width=\textwidth]{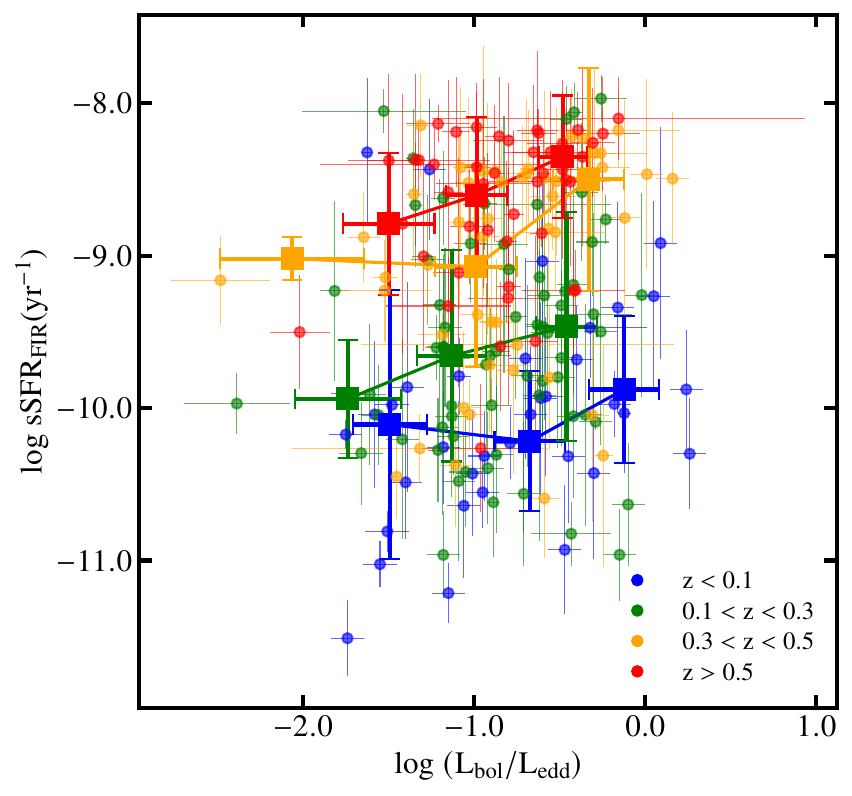}
    (\textbf{b})
\end{minipage}

\caption{
{Relation} 
 between specific star formation rate and (\textbf{a}) black hole accretion rate and (\textbf{b})~Eddington ratio. The sample is divided into four redshift bins: $z<0.1$ (blue), $0.1<z<0.3$ (green), $0.3<z<0.5$ (orange), and $z>0.5$ (red). In each redshift bin, the sources are further divided into three bins along the abscissa, and the median trends are shown.
}
\label{fig:bhar_sSFR_Leddratio_sSFR}
\end{figure}

\textls[-15]{We also compare the sSFR with the black hole accretion rate and Eddington ratio in Figure~\ref{fig:bhar_sSFR_Leddratio_sSFR}, using the same redshift bins as in Figure~\ref{fig:bhardeltamsleddratiodeltams}. The sSFR--BHAR relation is affected by redshift, as higher-redshift quasars tend to have both a higher sSFR and higher BHAR. However, the trends within individual redshift bins are weak or do not show. The sSFR--Eddington ratio relation also does not show a clear trend for the full sample or within individual redshift bins.}

\vspace{-6pt}

Therefore, for the nearby quasar subsample at $z<0.1$, there is a weak positive trend between the star formation of host galaxies and the black hole accretion. 
But for quasars at $z>0.1$, our data do not show clear evidence for a direct, instantaneous connection between global star formation and current black hole accretion activity.
However, these results do not rule out a long-term coevolution between quasars and their host galaxies. 
Instead, connection between star formation and black hole growth is likely affected by redshift evolution, sample selection, and the different timescales between FIR SFR and AGN accretion indicators. 
For samples spanning a broad redshift range, these effects should be carefully controlled when we study the relation between star formation and central black hole activity.

\begin{figure}[H]
\begin{minipage}{0.48\textwidth}
    \centering
    \includegraphics[width=\textwidth]{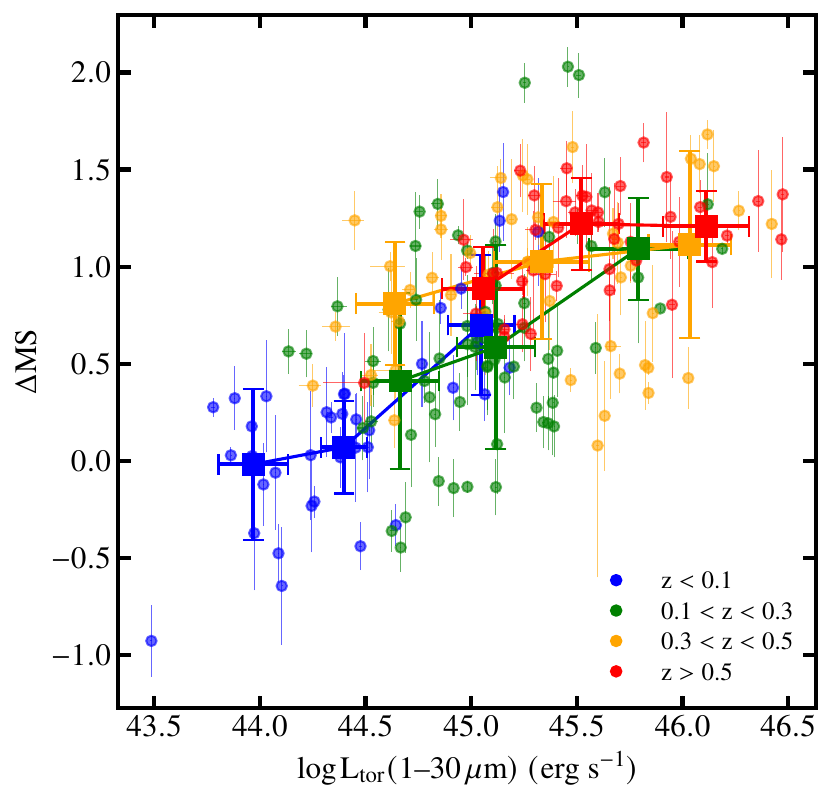}
    (\textbf{a})
\end{minipage}
\hfill
\begin{minipage}{0.48\textwidth}
    \centering
    \includegraphics[width=\textwidth]{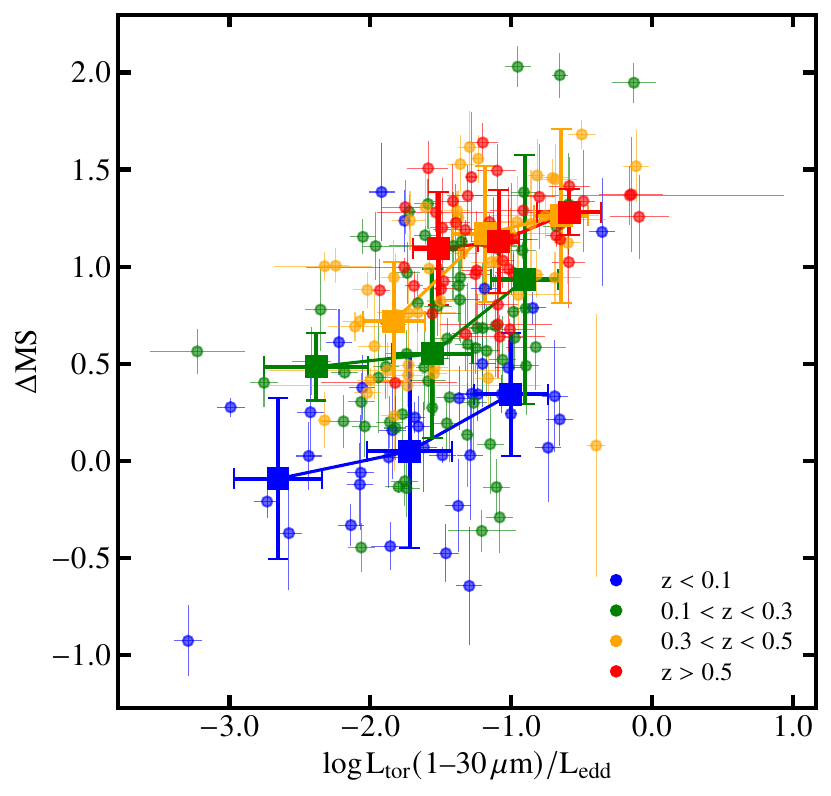}
    (\textbf{b})
\end{minipage}

\caption{
{Relation} 
 between the main-sequence offset, $\Delta_{\rm MS}$, and infrared-based AGN indicators derived from AGNfitter: (\textbf{a}) the torus luminosity integrated over $1$--$30~\upmu{\rm m}$, $L_{\rm tor}$, and (\textbf{b}) the torus-to-Eddington luminosity ratio, $L_{\rm tor}/L_{\rm Edd}$. 
The color scheme and redshift bins are the same as those in Figure~\ref{fig:bhardeltamsleddratiodeltams}.
In each redshift bin, the sources are further divided into three bins along the abscissa, and the median trends are shown.
The Kendall--$\tau$ correlation coefficients and the corresponding
two-sided $p$-values are listed in Table~\ref{tab:torus_kendall}.
}
\label{fig:torus_deltams}
\end{figure}

As an additional test, we examined the relation between $\Delta_{\rm MS}$ and two infrared-based AGN indicators derived from AGNfitter: the torus luminosity integrated over $1$--$30~\upmu{\rm m}$, $L_{\rm tor}$, and the ratio $L_{\rm tor}/L_{\rm Edd}$. These quantities are expected to be less affected by short-timescale optical variability than optical-continuum-based AGN luminosities. The results are shown in Figure~\ref{fig:torus_deltams}, and the Kendall--$\tau$ test results are listed in Table~\ref{tab:torus_kendall}.

For $L_{\rm tor}$, $\Delta_{\rm MS}$ shows positive correlations in the $z<0.1$, $0.1<z<0.3$, and $z>0.5$ bins, while no significant trend is found at $0.3<z<0.5$. The result is clearer when the torus luminosity is normalized by the Eddington luminosity. A positive correlation between $\Delta_{\rm MS}$ and $L_{\rm tor}/L_{\rm Edd}$ is found in all four redshift bins. The relation is strongest at $0.3<z<0.5$, while it is weaker at $z>0.5$.

We also tested whether these results could be caused by residual redshift trends within each redshift bin. The quantity $\log(L_{\rm tor}/L_{\rm Edd})$ does not show a significant correlation with redshift in any individual bin. Therefore, the positive $\Delta_{\rm MS}$--$L_{\rm tor}/L_{\rm Edd}$ relation cannot be explained simply by redshift mixing within the bins. In addition, we recalculated $\Delta_{\rm MS}$ using only the CIGALE-derived FIR SFRs and stellar masses. The correlations with the AGNfitter torus quantities remain qualitatively similar, suggesting that they are not mainly caused by covariance within one SED-fitting method.

We note that $L_{\rm tor}/L_{\rm Edd}$ is not a direct measurement of the Eddington ratio. It may also depend on torus geometry, covering factor, and uncertainties in the SED decomposition. In addition, the redshift-dependent FIR selection discussed in Section~\ref{subsubsec:fir_selection} should be considered when interpreting these relations.

We note that the Popesso et al. MS relation is based on a Chabrier IMF, whereas our FIR SFR calibration is converted to a Kroupa IMF. The difference between these two IMFs is small, at the level of $\lesssim$0.05 dex \citep{2014ARA&A..52..415M}. Moreover, the IMF conversion would shift both the SFR and stellar mass in nearly the same direction, so the relative position with respect to the MS and the derived $\Delta_{\rm MS}$ values are only weakly affected. This difference is much smaller than the typical uncertainties in our SED-based SFRs and stellar masses. Therefore, the IMF difference is expected to have only a minor effect on the relative location of our quasars with respect to the MS and on the derived $\Delta_{\rm MS}$ values.

More importantly, the MS comparison should be interpreted in the context of our FIR-constrained sample. For the SDSS subsample, we require detections in at least two Herschel bands, which naturally favors quasars with stronger FIR emission and higher FIR-based SFRs. Quasars without sufficient Herschel detections, which may have lower cold dust luminosities and lower SFRs, are not included in this comparison. 
Therefore, our result should not be interpreted as a general statement that all quasars lie on or above the MS. Instead, it shows that many FIR-constrained quasars in our sample are located on or above the MS. Although the PG quasars generally have better FIR coverage, the effects of FIR non-detections and upper limits should still be considered when drawing conclusions about the overall quasar population.

\subsubsection{Redshift-Dependent FIR Selection Effects}
\label{subsubsec:fir_selection}

Because the SDSS subsample requires detections in at least two Herschel bands, its effective FIR sensitivity is expected to vary with the redshift. To quantify this effect, we estimated an approximate two-band Herschel sensitivity for each source using its fitted SED and the available Herschel photometric uncertainties. We converted the corresponding FIR luminosity limit into an SFR limit and evaluated the associated MS offset at the median redshift and stellar mass in each redshift bin. The results are summarized in Table~\ref{tab:ms_sensitivity}.

\begin{table}[H]
\caption{Approximate effective two-band Herschel sensitivity of the SDSS subsample. For each source, we estimate the FIR luminosity required to reach the adopted detection threshold in at least two Herschel bands using its fitted SED and photometric uncertainties. The corresponding limiting SFR is evaluated at the median redshift and stellar mass in each redshift bin. We define $\Delta_{\rm MS,lim}=\log{\rm SFR}_{\rm lim}-\log{\rm SFR}_{\rm MS}(M_{\star,\rm med},z_{\rm med})$. These values are approximate effective selection thresholds rather than formal completeness limits.}
\label{tab:ms_sensitivity}
\begin{tabularx}{\textwidth}{p{2.5cm}CC>{\centering\arraybackslash}p{2cm}>{\centering\arraybackslash}p{2cm}C}
\toprule
\textbf{Redshift Bin} & \boldmath{$N$} & \boldmath{$z_{\rm med}$} & \boldmath{$\log M_{\star,\rm med}$} &
\boldmath{$\log {\rm SFR}_{\rm lim}$} & \boldmath{$\Delta_{\rm MS,lim}$} \\
& & & \boldmath{$(M_\odot)$} & \boldmath{$(M_\odot,{\rm yr}^{-1})$} & \textbf{({dex}
)} \\
\midrule
$z < 0.1$         & 5  & 0.09 & 10.24 & $-$0.20 & $-$0.42 \\
$0.1 < z < 0.3$   & 30 & 0.18 & 10.60 &  0.70 &  0.19 \\
$0.3 < z < 0.5$   & 33 & 0.37 & 10.30 &  1.43 &  0.86 \\
$z > 0.5$         & 47 & 0.70 & 10.69 &  1.93 &  0.87 \\
\bottomrule
\end{tabularx}
\end{table}

The estimated selection threshold is well below the MS at $z<0.1$, with $\Delta_{\rm MS,lim}\simeq-0.42$ dex. It increases to $\Delta_{\rm MS,lim}\simeq0.19$ dex at $0.1<z<0.3$ and reaches approximately $0.86$--$0.87$ dex at $z>0.3$. These values are approximate effective thresholds rather than formal completeness limits, because the Herschel band coverage and depths vary among sources. Nevertheless, they indicate that the SDSS selection becomes progressively less sensitive to MS and below-MS hosts with increasing redshift.

Therefore, the elevated $\Delta_{\rm MS}$ distribution at high redshift in Figure~\ref{fig:mainsequence} should not be interpreted as evidence that the full quasar-host population has intrinsically enhanced SFRs. This effect should also be considered when interpreting the $\Delta_{\rm MS}$--BHAR and $\Delta_{\rm MS}$--$\lambda_{\rm Edd}$ relations in Figure~\ref{fig:bhardeltamsleddratiodeltams}, as well as the sSFR trends in Figure~\ref{fig:bhar_sSFR_Leddratio_sSFR}. In particular, the higher-redshift bins mainly represent the FIR-selected, high-SFR portion of the quasar-host population, and may not be representative of quasars without Herschel detections and with lower SFRs.
This selection effect is also relevant to FIR-selected quasar samples at higher redshift, which preferentially include FIR-luminous hosts with high SFRs \citep{2016MNRAS.462.4067P}.

\subsection{Radiation Field Intensity and Dust Heating}
\label{subsec:umindiscussion}

\begin{figure}[H]
\begin{minipage}{0.48\textwidth}
    \centering
    \includegraphics[width=\textwidth]{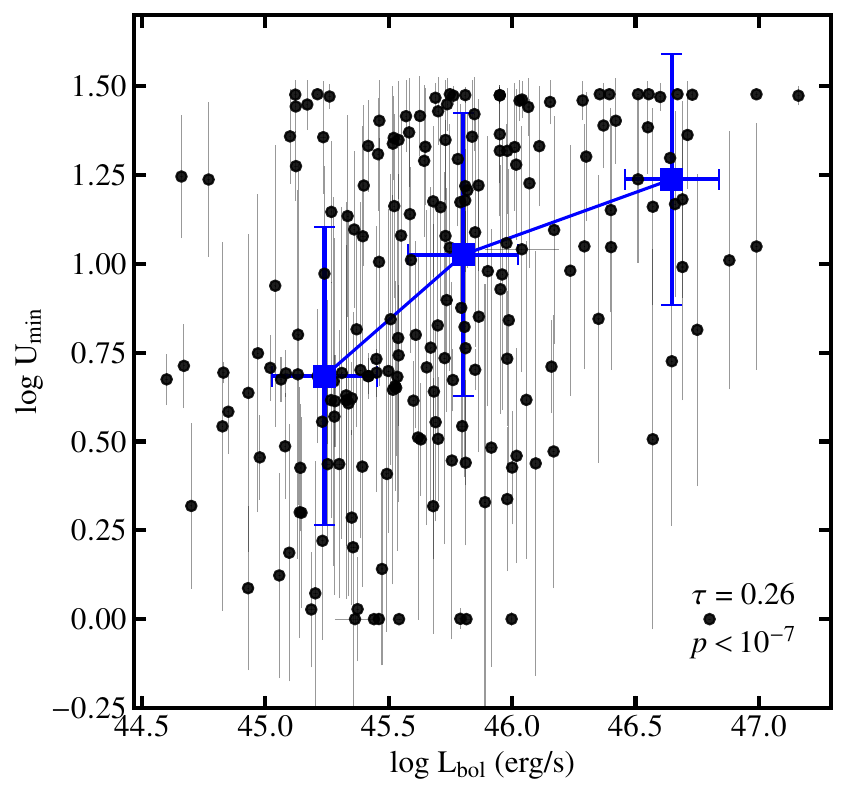}
    (\textbf{a})
\end{minipage}
\hfill
\begin{minipage}{0.48\textwidth}
    \centering
    \includegraphics[width=\textwidth]{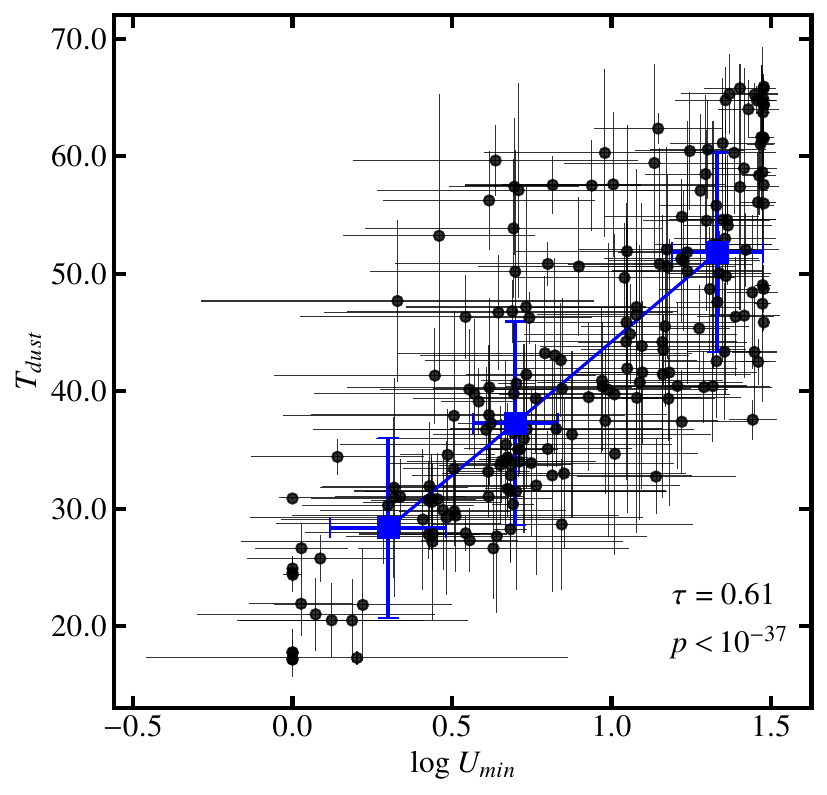}
    (\textbf{b})
\end{minipage}

\caption{
{Comparison} 
 of $U_{\rm min}$ with (\textbf{a}) bolometric luminosity, $L_{\rm bol}$, and (\textbf{b}) cold dust temperature, $T_{\rm dust}$. The Kendall--$\tau$ coefficient and the corresponding $p$ value are shown in the lower-right corner of each panel. Blue points show the median values in three bins of the full sample.
}
\label{fig:LbolUminaddbinmedianTUminaddbin}
\end{figure}

Based on the radiation field intensity, $U_{\rm min}$, derived from the SED fitting, we compare $U_{\rm min}$ with the quasar bolometric luminosity, $L_{\rm bol}$, and the cold dust temperature, $T_{\rm dust}$, in Figure \ref{fig:LbolUminaddbinmedianTUminaddbin}. The Kendall--$\tau$ coefficient and the corresponding $p$ value are shown in the lower right corner of each panel. To better illustrate the overall trends, we divide the full sample into three bins and show the median values as blue points.

The left panel shows that $U_{\rm min}$ tends to increase with increasing $L_{\rm bol}$, while the right panel shows a positive correlation between $U_{\rm min}$ and $T_{\rm dust}$. Since $U_{\rm min}$ traces the intensity of the radiation field heating the diffuse dust, and $T_{\rm dust}$ is also sensitive to dust heating, the positive correlation between these two quantities is expected. More importantly, the correlation between $U_{\rm min}$ and $L_{\rm bol}$ suggests that more luminous quasars may expose the cold dust in their host galaxies to a stronger radiation field.

This trend is consistent with the scenario in which AGN radiation contributes to heating dust on host galaxy scales. 
\citet{shangguan2018} found that quasar host galaxies tend to have higher $U_{\rm min}$ values than normal galaxies. It suggests that the enhanced radiation field may be caused by AGN heating of the interstellar medium. 
They also found that the inferred cold dust temperature may be biased high if the FIR SED contains a contribution from warmer dust.
Several recent studies further support the possibility that AGN radiation can heat dust on scales larger than the classical torus. \citet{2015A&A...579A..60S} used radiative transfer modeling of the $z\sim6$ quasar SDSS J1148+5251 and found that AGN radiation may power a substantial fraction of the observed FIR emission. \mbox{\citet{2021ApJ...921...55M}} showed that in dust-enshrouded AGN, host galaxy-scale diffuse dust heated by the AGN can increase the FIR flux at $\lambda \gtrsim 100~\upmu$m. \citet{2023MNRAS.518.3667D} further showed that AGN radiation can heat ISM dust in simulated $z\sim6$ quasar hosts and lead to overestimated TIR-based SFRs if this contribution is ignored.

Our results are broadly consistent with these results, but the correlation should be interpreted carefully. 
Our SED fitting does not provide spatial information, so we cannot determine where the dust is heated. If a large fraction of the cold dust is located far from the nucleus in star-forming regions, it may be heated mainly by local star formation and only weakly affected by AGN radiation. The enhanced $U_{\rm min}$ may instead be related to dust closer to the central region, more compact star formation, or both. In addition, the studies mentioned above mostly focus on very luminous, high-redshift, or dust-enshrouded systems, while our sample consists mainly of lower-redshift type-I quasars.
In addition, other factors may also contribute to this trend, such as redshift evolution, sample selection, enhanced star formation, and degeneracies in the SED fitting. 
Therefore, this relation provides suggestive evidence for a possible AGN effect, but it does not prove that AGN heating is the only origin of the enhanced radiation field.

\subsection{Possible Effects of Stellar Mass} \label{subsec:stellarmassdiscussion}

Stellar mass may also affect the relations discussed above, because it is connected with black hole mass and may therefore indirectly affect the correlations involving AGN luminosity or black hole accretion rate. We make additional tests about stellar mass. 
Before examining stellar mass-dependent trends, we first estimate the model related scatter in the stellar masses. We compared the stellar masses obtained from different cold dust templates. The median scatter among these stellar masses is about 0.24 dex. This shows that the stellar masses of type-I quasars are also model dependent. It is  likely because the bright AGN emission can contaminate the optical and NIR continuum, making it difficult to accurately separate the host galaxy stellar component in the SED fitting.
This scatter may introduce additional uncertainty in the sSFR and $\Delta{\rm MS}$, so the related trends should be interpreted statistically.
We divide the sample into three stellar mass bins ($\log(M_\ast/M_\odot)<10.5$, $10.5<\log(M_\ast/M_\odot)<11.0$, $\log(M_\ast/M_\odot)>11.0$) and repeat related analysis in these~bins.

\begin{figure}[H]
\begin{minipage}{0.48\textwidth}
    \centering
    \includegraphics[width=\textwidth]{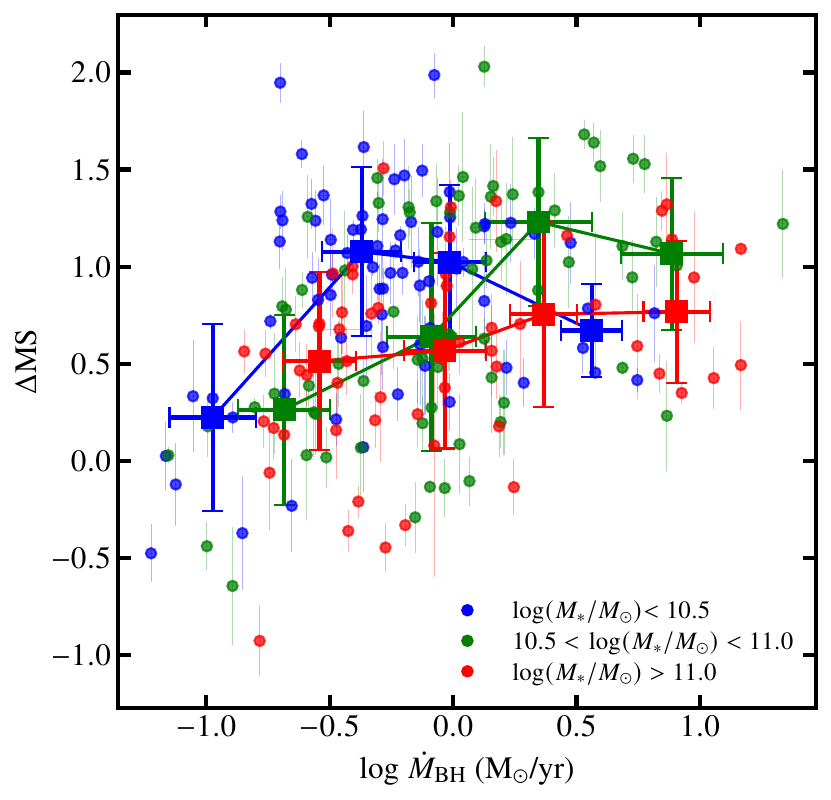}
    (\textbf{a})
\end{minipage}
\hfill
\begin{minipage}{0.48\textwidth}
    \centering
    \includegraphics[width=\textwidth]{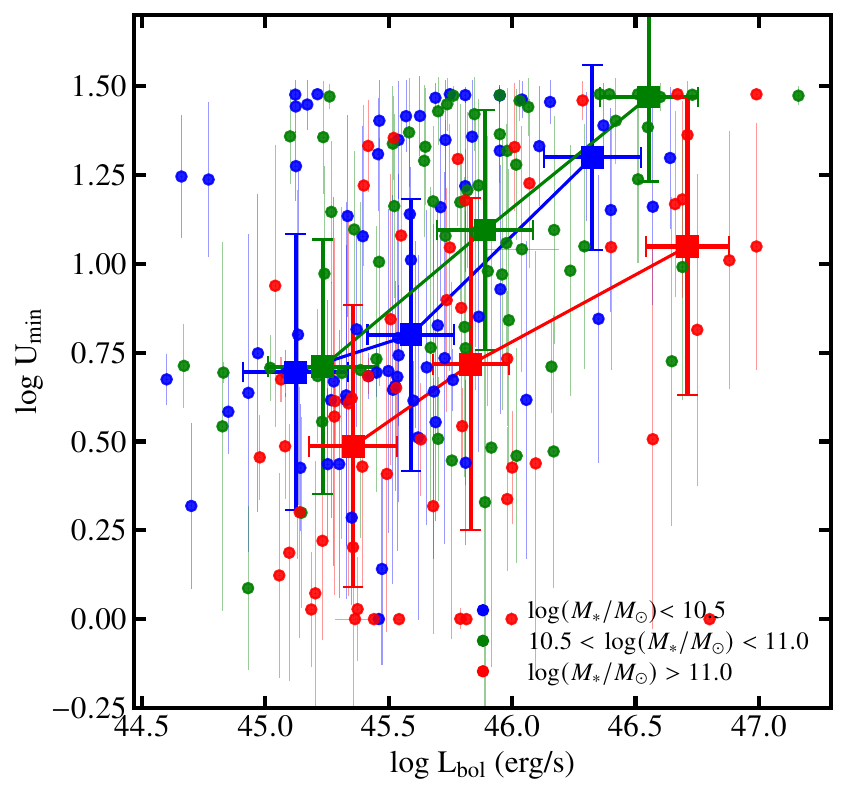}
    (\textbf{b})
\end{minipage}
\caption{
(\textbf{a}) {Relation} 
 between the main-sequence offset, $\Delta_{\rm MS}$, and black hole accretion rate in three stellar mass bins.
(\textbf{b}) Relation between the minimum radiation field intensity, $U_{\min}$, and AGN bolometric luminosity in the same stellar mass bins.
Blue, green, and red symbols represent sources with $\log(M_\ast/M_\odot)<10.5$, $10.5<\log(M_\ast/M_\odot)<11.0$, and $\log(M_\ast/M_\odot)>11.0$, respectively.
Large squares and connecting lines show the median values in bins along the abscissa, and the error bars indicate the corresponding scatter.}

\label{fig:bhardeltamsLbolUmin_addstellarmassbin}
\end{figure}

Figure~\ref{fig:bhardeltamsLbolUmin_addstellarmassbin}a shows the relation between $\Delta_{\rm MS}$ and the BHAR in three stellar mass bins. The median trends are not same among the different stellar mass bins, and the scatter is large. The low and intermediate mass subsamples show an increase in $\Delta_{\rm MS}$ from low-to-moderate BHAR, while the high mass subsample shows a weaker trend and a generally lower $\Delta_{\rm MS}$ at a given BHAR.

Figure~\ref{fig:bhardeltamsLbolUmin_addstellarmassbin}b shows the relation between $U_{\min}$ and $L_{\rm bol}$ in the same stellar mass bins. The increasing trend of $U_{\min}$ with $L_{\rm bol}$ is still visible within different stellar mass bins, suggesting that this relation is unlikely to be solely driven by stellar mass. We also directly checked the relations between stellar mass and dust-heating parameters. Stellar mass does not show a positive correlation with either $U_{\min}$ or $T_{\rm dust}$. Therefore, although stellar radiation is certainly an important heating source for cold dust, the observed $U_{\min}$--$L_{\rm bol}$ relation may also include a contribution from AGN radiation.

Additional stellar mass binned tests are shown in Appendix~\ref{app:mstar_tests}. The relation between $\Delta_{\rm MS}$ and the Eddington ratio also does not show no same trend in different stellar mass bins. In addition, the positive relation between $T_{\rm dust}$ and $U_{\min}$ remains visible within different stellar mass bins, indicating that this dust parameter relation is not mainly driven by stellar~mass.

\subsection{Effects of FIR Data Quality}
\label{subsec:spiretest}

The SDSS subsample was originally selected to have detections in at least two Herschel bands. However, the long-wavelength FIR coverage still varies substantially among the sources. To test whether the main results are strongly affected by sources with limited SPIRE constraints, we define two higher-quality SPIRE subsamples within the 115 unique SDSS quasars. The first requires an $S/N_{250}\geq5$ at 250 $\upmu$m, while the second requires detections with an $S/N\geq3$ in at least two SPIRE bands. Each subsample contains 51~quasars, and they overlap in 35 objects.

Table~\ref{tab:spire_subsamples} compares the full SDSS FIR-constrained sample with the two SPIRE-quality subsamples. Both SPIRE-selected subsamples have higher median redshifts, FIR SFRs, and main-sequence offsets than the full SDSS sample. In particular, their median $\log {\rm SFR}_{\rm FIR}$ values are about $0.4$ dex higher than that of the full SDSS sample, and all selected sources lie above the adopted main sequence. Therefore, these subsets preferentially select FIR-brighter quasars and should not be regarded as unbiased representations of the full SDSS quasar population.

\begin{table}[H]
\caption{Summary of the full SDSS FIR-constrained sample and the two SPIRE-quality subsamples. Values are reported as the median with the 16th and 84th percentiles. The two SPIRE-quality subsamples overlap in 35 sources. Kendall--$\tau$ statistics are calculated from individual sources.}
\label{tab:spire_subsamples}

\small
\begin{tabularx}{\textwidth}{p{4.5cm}ccc}
\toprule
\textbf{Quantity} & \textbf{Full SDSS Sample} & \boldmath{$S/N_{250}\geq5$} & \boldmath{$\geq$}\textbf{2 SPIRE Detections} \\
\midrule
Number of sources & 115 & 51 & 51 \\
Median $z$ & $0.42_{-0.26}^{+0.31}$ & $0.50_{-0.22}^{+0.26}$ & $0.60_{-0.27}^{+0.17}$ \\
Median $\log(M_\ast/M_\odot)$ & $10.65_{-0.48}^{+0.50}$ & $10.71_{-0.49}^{+0.49}$ & $10.85_{-0.58}^{+0.51}$ \\
Median $\log({\rm SFR}_{\rm FIR}/M_\odot\,{\rm yr}^{-1})$ & $1.71_{-0.54}^{+0.60}$ & $2.08_{-0.63}^{+0.31}$ & $2.11_{-0.63}^{+0.30}$ \\
Median $\Delta_{\rm MS}$ & $1.00_{-0.48}^{+0.34}$ & $1.23_{-0.51}^{+0.22}$ & $1.19_{-0.49}^{+0.19}$ \\
Fraction with $\Delta_{\rm MS}\geq0$ & $98\%$ & $100\%$ & $100\%$ \\
Kendall--$\tau$ for $\log U_{\min}$--$\log L_{\rm bol}$ & $0.24$ & $0.16$ & $0.28$ \\
$p$ value & <$10^{-3}$ & $0.103$ & <$10^{-2}$ \\
\bottomrule
\end{tabularx}
\end{table}

The corresponding comparisons are shown in Appendix~\ref{app:spiretests} (Figures~\ref{fig:mainsequence_2subsample}--\ref{fig:Lbol_Umin_2subsample}). The SPIRE-quality quasars remain predominantly above the star-forming main sequence, indicating that the elevated SFRs of the FIR-constrained SDSS sample are not driven solely by sources with weak long-wavelength constraints. However, because both tests are applied within the original Herschel-selected sample and preferentially retain FIR-brighter objects, they do not remove the underlying FIR selection bias. The main-sequence result should therefore still be interpreted only for quasars with adequate FIR constraints.

For the relations between $\Delta_{\rm MS}$ and the BHAR or Eddington ratio, the two SPIRE-quality subsamples occupy broadly similar regions of parameter space to the full SDSS sample. Because of the reduced sample sizes and the remaining redshift dependence, we do not repeat the redshift-binned median-trend analysis for these subsets. The redshift-resolved results in Figures~\ref{fig:bhardeltamsleddratiodeltams} and \ref{fig:bhar_sSFR_Leddratio_sSFR} therefore remain the basis for interpreting the relations between star formation and black hole accretion.

The $U_{\min}$--$L_{\rm bol}$ relation remains positive in both SPIRE-quality subsamples, but its statistical significance depends on the adopted FIR-quality criterion. The $S/N_{250}\geq5$ subsample gives $\tau=0.16$ with $p=0.103$, whereas the subsample with at least two SPIRE detections gives $\tau=0.28$ with $p<10^{-2}$. These results are qualitatively consistent with the positive trend found for the full sample but show that its statistical strength is sensitive to the FIR data selection. Thus, the observed $U_{\min}$--$L_{\rm bol}$ relation should be interpreted cautiously and may also be affected by sample selection, host galaxy properties, and SED-fitting degeneracies.

\subsection{Infrared Colors as Tracers of the AGN Contribution} \label{subsec:fracagndiscussion}

\begin{figure}[H]
\includegraphics[width=0.6\textwidth]{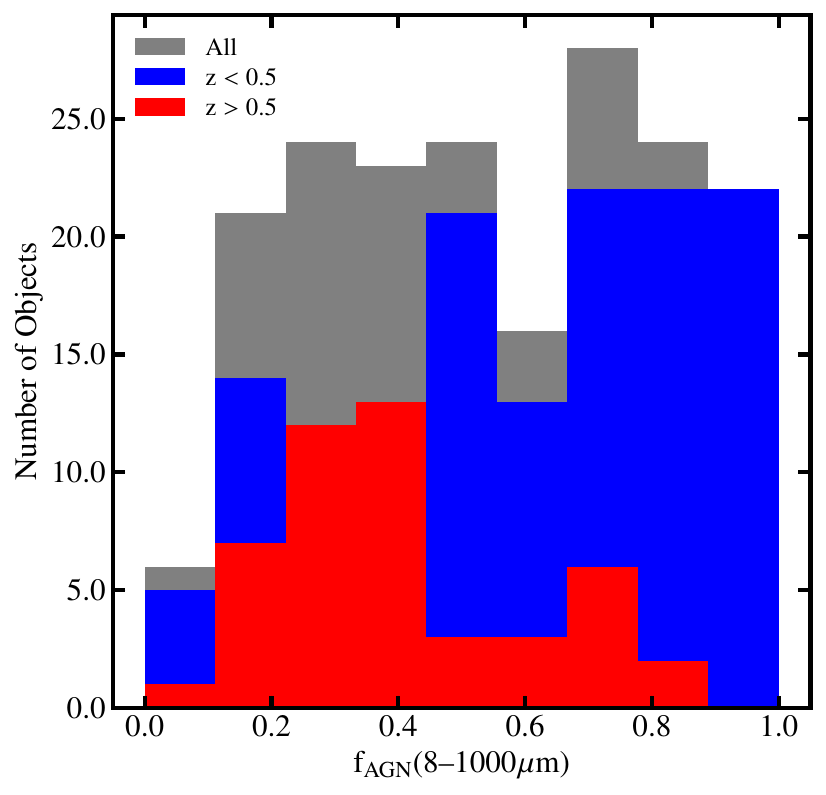}
\caption{
{Distribution} 
 of $f_{\rm AGN}(8$--$1000~\upmu\mathrm{m})$. The gray histogram shows the full sample, while the blue and red histograms show the $z<0.5$ and $z>0.5$ subsamples, respectively.
}
\label{fig:fracagn_hist}
\end{figure}

The SED-fitting results provide the fractional AGN contribution to the total infrared luminosity, $f_{\rm AGN}$, integrated over 8--1000~$\upmu$m.  Figure~\ref{fig:fracagn_hist} shows the distribution of \mbox{$f_{\rm AGN}(8\text{--}1000~\upmu\mathrm{m})$} for the full sample and for the two redshift subsamples. The AGN contribution to the total infrared luminosity is non-negligible, especially for the $z<0.5$ quasars. 
Therefore, if the total infrared luminosity is used to infer the SFR without decomposing the AGN component, the SFR may be overestimated.

Since MIR/FIR flux-density ratios trace the relative importance of AGN-heated warm dust and host galaxy cold dust, we also examine whether infrared colors can be used as empirical indicators of $f_{\rm AGN}(8$--$1000~\upmu\mathrm{m})$. Among the 24 WISE-to-Herschel colors examined in this work, W2/70~$\upmu\mathrm{m}$, W3/70~$\upmu\mathrm{m}$, and W3/100~$\upmu\mathrm{m}$ show the strongest correlations with $f_{\rm AGN}$. These three representative relations are shown in Figure~\ref{fig:strongestcolorfagn}, while the full set of WISE-to-Herschel color comparisons is provided in Appendix~\ref{app:wisehscolor}.

The observed correlations suggest that these colors can be used as practical empirical estimators of $f_{\rm AGN}$ for quasars with both WISE and Herschel detections. After estimating $f_{\rm AGN}$, the AGN-heated dust contribution to the total infrared luminosity can be separated from the cold dust emission associated with star formation, leading to a more reliable estimate of the host galaxy FIR SFR.

However, the reliability of these color--$f_{\rm AGN}$ relations depends on the quality of the FIR constraints. Sources with three or more Herschel detections generally follow the fitted relations more closely, whereas sources with two or fewer Herschel detections show larger deviations and tend to have a  higher inferred $f_{\rm AGN}$ at a given color. 
This offset may have two related origins. First, these objects may be intrinsically fainter in cold dust emission. A lower cold dust luminosity would naturally increase the relative AGN contribution to the total infrared luminosity, leading to a higher inferred $f_{\rm AGN}$. Second, with only a few Herschel detections, the FIR peak and long-wavelength side of the cold dust SED are poorly constrained, increasing the degeneracy between the AGN and cold dust components in the SED fitting.

Therefore, the color-based $f_{\rm AGN}$ indicators should be regarded as most reliable for sources with good FIR coverage. They should be used with caution for objects with two~or fewer Herschel detections. In addition, the observed colors can be affected by redshift, K-correction, photometric coverage, host galaxy contamination in the shorter WISE bands, and selection effects. Since the fitted relations are calibrated using radio-quiet sources with detections in at least three Herschel bands, they may not be directly applicable to radio-loud quasars or sources with poor FIR constraints.

\begin{figure}[H]
\begin{minipage}{0.32\textwidth}
    \centering
    \includegraphics[width=\textwidth]{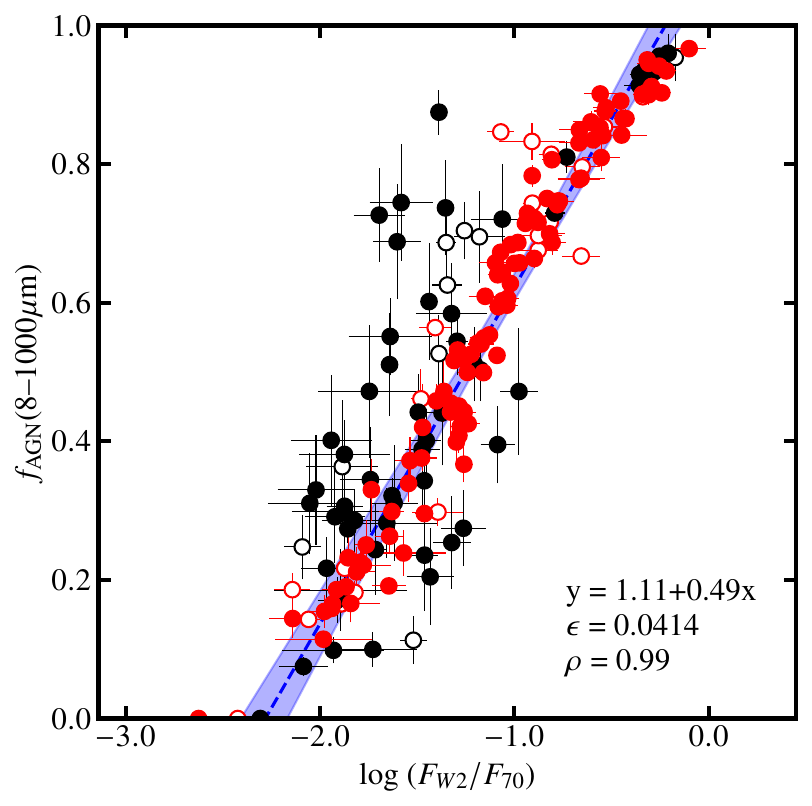}
    (\textbf{a}) W2/70~$\upmu\mathrm{m}$
\end{minipage}
\hfill
\begin{minipage}{0.32\textwidth}
    \centering
    \includegraphics[width=\textwidth]{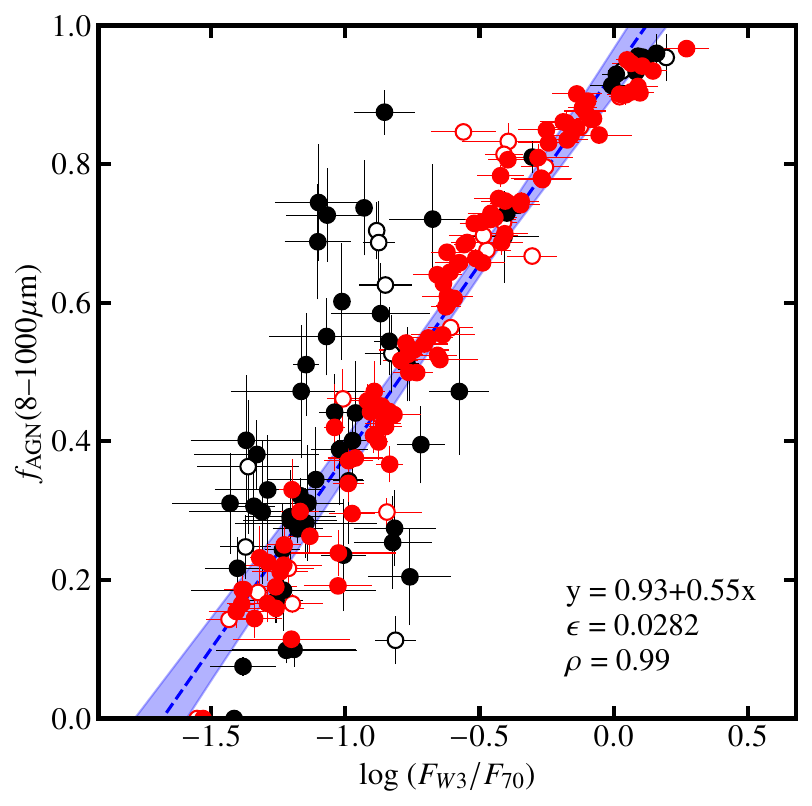}
    (\textbf{b}) W3/70~$\upmu\mathrm{m}$
\end{minipage}
\hfill
\begin{minipage}{0.32\textwidth}
    \centering
    \includegraphics[width=\textwidth]{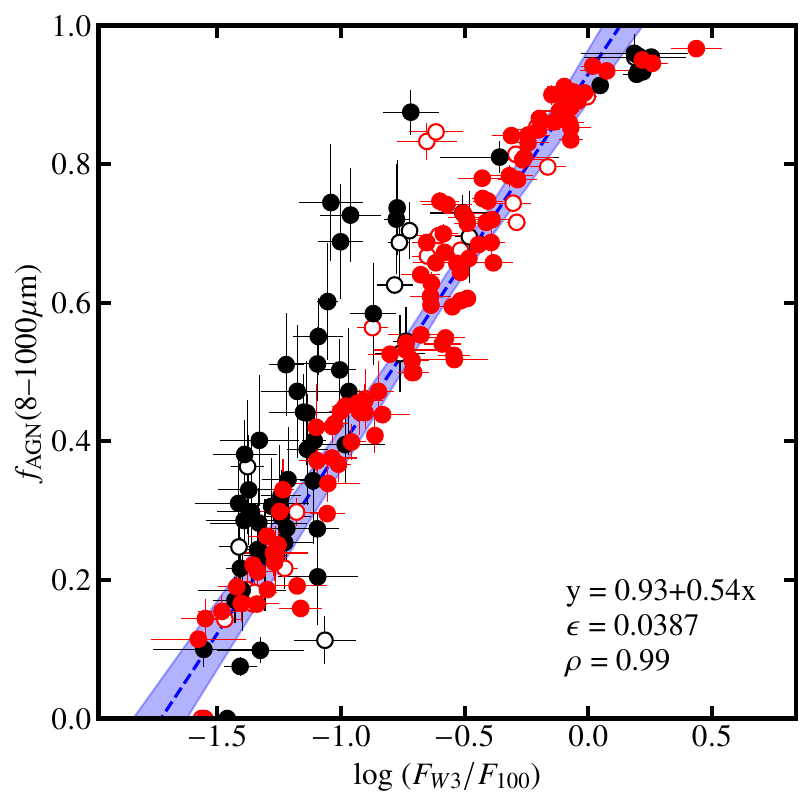}
    (\textbf{c}) W3/100~$\upmu\mathrm{m}$
\end{minipage}

\caption{
$f_{\rm AGN}(8$--$1000~\upmu\mathrm{m})$ {as} 
 a function of the three strongest WISE-to-Herschel color diagnostics: (\textbf{a}) W2/70~$\upmu\mathrm{m}$, (\textbf{b}) W3/70~$\upmu\mathrm{m}$, and (\textbf{c}) W3/100~$\upmu\mathrm{m}$. Red and black symbols denote sources detected in at least three and in no more than two Herschel bands, respectively. Filled and open symbols represent radio-quiet and radio-loud quasars, respectively. The dashed line and blue shaded region show the Bayesian linear regression fit and its uncertainty for the radio-quiet sources with detections in at least three Herschel bands. The fitted relation, intrinsic scatter $\epsilon$, and correlation coefficient $\rho$ are shown in each panel.
}
\label{fig:strongestcolorfagn}
\end{figure}
\section{Summary and Conclusions}
\label{sec:summaryconclusion}

We study the FIR SFRs and host galaxy properties of 202 SDSS and PG quasars at $0.02<z\lesssim0.8$ using multiwavelength SED decomposition. 
The photometric data cover the optical-to-FIR bands and include JCMT/SCUBA-2 observations at 450 and 850~$\upmu\mathrm{m}$. 
We fit the SEDs with CIGALE and AGNfitter and use different cold dust templates to estimate systematic uncertainties. 
Our main results are summarized as follows.

\begin{enumerate}

\item For each quasar, we derive five FIR SFR estimates using four CIGALE cold dust templates and AGNfitter. The median scatter among the four CIGALE template results is 0.10 dex, corresponding to approximately 26\%. AGNfitter gives SFRs lower than the mean CIGALE estimate by a median of 0.09 dex, or approximately 19\%. Considering all five estimates together, the median model-dependent scatter is 0.14~dex, corresponding to approximately 38\%. Thus, a systematic uncertainty of about 0.14~dex should be considered for sources with photometric coverage and modeling assumptions similar to those adopted here.

\item For the 58 quasars with SCUBA-2 coverage, removing the SCUBA-2 data changes the adopted FIR SFR by only $\sim$0.01 dex on average. The mean change is close to zero for both the SCUBA-2-detected and upper-limit-only subsets. However, the change for individual sources ranges from $-0.11$ to $+0.13$ dex and can be larger for quasars with limited Herschel coverage or radio-loud emission.

\item Within our FIR-constrained sample, many quasar hosts lie on or above the star-forming MS. However, for the SDSS subsample, the effective FIR selection threshold increases strongly with redshift, from $\Delta_{\rm MS,lim}\simeq-0.42$ dex at $z<0.1$ to \mbox{$\simeq0.86$--$0.87$ dex} at $z>0.3$. The higher-redshift bins therefore mainly represent the FIR-bright, high-SFR part of the quasar-host population. The full-sample $\Delta_{\rm MS}$--BHAR trend is not seen in all redshift bins, and no clear correlation is found between $\Delta_{\rm MS}$ or sSFR and the direct Eddington ratio.

\item As an additional test, $\Delta_{\rm MS}$ is positively correlated with the infrared-based $L_{\rm tor}/L_{\rm Edd}$ ratio in all four redshift bins, while the relation with $L_{\rm tor}$ alone is weaker. The $L_{\rm tor}/L_{\rm Edd}$ ratio is not a direct Eddington-ratio measurement and can also depend on torus properties and SED decomposition. The redshift-dependent FIR selection should also be considered when interpreting this relation.

\item The radiation field intensity, $U_{\rm min}$, tends to increase with quasar bolometric luminosity and cold dust temperature. 
This suggests that more luminous quasars may be related to stronger radiation fields. 
However, redshift effects, enhanced star formation, and SED-fitting degeneracies may also contribute to this trend.

\item We test 24 WISE-to-Herschel colors as indicators of $f_{\rm AGN}(8$--$1000~\upmu\mathrm{m})$. 
W2/70~$\upmu\mathrm{m}$, W3/70~$\upmu\mathrm{m}$, and W3/100~$\upmu\mathrm{m}$ show the strongest correlations. 
These colors may provide simple estimates of the AGN infrared contribution. But they should be used carefully for radio-loud sources or objects with weak FIR constraints.

\end{enumerate}

Overall, multi-component SED decomposition is important for obtaining FIR-based SFRs in quasar host galaxies. 
SCUBA-2 data do not strongly change the average SFRs of the SCUBA-2-observed subset but can improve the constraints for individual sources. 
The relations between host galaxy star formation and AGN properties should be interpreted with care because of redshift evolution, FIR selection effects, model dependence, and the different timescales traced by FIR SFRs and AGN accretion indicators. 

\vspace{6pt} 

\supplementary{The following supporting information can be downloaded at \linksupplementary{s1}, Table S1: Machine-readable photometric data used for the SED fitting of the 202 quasars. The accompanying README file describes the table columns, units, upper-limit convention, and the meaning of the Reference column.}

\authorcontributions{Conceptualization, X.F. and X.-B.W.; methodology, X.F., X.-B.W., Y.F. and Y.P.; software, X.F.; validation, X.F., Y.F. and Y.P.; formal analysis, X.F.; investigation, X.F., X.-B.W., Y.F., Y.P., R.Z. and H.W.; resources, X.F. and X.-B.W.; data curation, X.F. and Y.F.; writing---original draft preparation, X.F.; writing---review and editing, X.F., X.-B.W., Y.F., Y.P., R.Z. and H.W.; visualization, X.F.; supervision, X.-B.W.; project administration, X.-B.W.; funding acquisition, X.-B.W. All authors have read and agreed to the published version of the manuscript.}

\funding{
This research was funded by the National Key R\&D Program of China (2025YFA1614100, 2025YFA1614101) and the National Science Foundation of China (grant No. 12133001). 
}

\dataavailability{The complete photometric data used as input for the SED fitting are provided in Supplementary Table~S1 in machine-readable format. The public photometric, spectroscopic, and archival data used in this study are available from SDSS, IRSA, NED, CADC, and the relevant mission archives. Additional derived data supporting the findings of this study are available from the corresponding author upon reasonable request.}

\acknowledgments{{This} 
 work was supported by the High-performance Computing Platform of Peking University. 
We thank the staff of the James Clerk Maxwell Telescope for their support of the SCUBA-2 observations. The James Clerk Maxwell Telescope is operated by the East Asian Observatory on behalf of the National Astronomical Observatory of Japan, Academia Sinica Institute of Astronomy and Astrophysics, the Korea Astronomy and Space Science Institute, the National Astronomical Research Institute of Thailand, and organizations in the United Kingdom and Canada.
This work is based in part on observations obtained with SCUBA-2 on the JCMT under programs M18BP050, M19AP020, M19BP041, and M20AP044. This work also makes use of archival data from the Canadian Astronomy Data Centre (CADC), operated by the National Research Council of Canada with the support of the Canadian Space Agency.
This research has made use of the NASA/IPAC Infrared Science Archive (IRSA) and the NASA/IPAC Extragalactic Database (NED), which are funded by the National Aeronautics and Space Administration and operated by the California Institute of Technology. 
This work is based in part on observations made with the Herschel Space Observatory, an ESA space observatory with science instruments provided by European-led Principal Investigator consortia and with important participation from NASA. This publication also makes use of data products from the Wide-field Infrared Survey Explorer (WISE), a joint project of the University of California, Los Angeles, and the Jet Propulsion Laboratory/California Institute of Technology, funded by NASA. 
This work also makes use of data products from the Sloan Digital Sky Survey (SDSS), 2MASS, and the \textit{Spitzer} Space Telescope.}

\conflictsofinterest{The authors declare no conflicts of interest.} 


\abbreviations{Abbreviations}{
The following abbreviations are used in this manuscript:
\\

\noindent
\begin{tabular}{@{}ll}
AGN & Active galactic nucleus \\
BHAR & Black hole accretion rate \\
CADC & Canadian Astronomy Data Centre \\
CIGALE & Code Investigating GALaxy Emission \\
FIR & Far-infrared \\
FWHM & Full width at half maximum \\
IMF & Initial mass function \\
IRAC & Infrared Array Camera \\
IRSA & NASA/IPAC Infrared Science Archive \\
JCMT & James Clerk Maxwell Telescope \\
MCMC & Markov chain Monte Carlo \\
MIPS & Multiband Imaging Photometer for Spitzer \\
MIR & Mid-infrared \\
MS & Main sequence \\
NED & NASA/IPAC Extragalactic Database \\
NIR & Near-infrared \\
PACS & Photoconductor Array Camera and Spectrometer \\
PDF & Probability distribution function \\
PG & Palomar--Green \\
SCUBA-2 & Submillimetre Common-User Bolometer Array 2 \\
SDSS & Sloan Digital Sky Survey \\
SED & Spectral energy distribution \\
SFH & Star formation history \\
SFR & Star formation rate \\
sSFR & Specific star formation rate \\
SMBH & Supermassive black hole \\
SPIRE & Spectral and Photometric Imaging Receiver \\
SSP & Simple stellar population \\
WISE & Wide-field Infrared Survey Explorer \\
\end{tabular}
}
\newpage
\appendixtitles{yes} 
\appendixstart
\appendix

\section{SED-Fitting Results}
\label{app:sedresults}
\vspace{-6pt}
\begin{table}[H]
\caption{Selected SED-fitting results and adopted black hole properties for the 202 quasars}
\label{tab:sedresultpart_app}
\small
\begin{adjustwidth}{-\extralength}{0cm}
\begin{tabularx}{\fulllength}{@{}p{3.2cm}cccccccc@{}}
\toprule
\textbf{Name} &
\boldmath{$z$} &
\boldmath{{$\log M_\star$}} &
\boldmath{$\log {\rm SFR}_{\rm FIR}$} &
\boldmath{$T_{\rm dust}$} &
\boldmath{$U_{\rm min}$} &
\boldmath{$f_{\rm AGN}$} &
\boldmath{$\log M_{\rm BH}$} &
\boldmath{$\log\lambda_{\rm Edd}$} \\
&
&
\boldmath{($M_{\odot}$)} &
\boldmath{($M_{\odot}~{\rm yr}^{-1}$)} &
\textbf{(K)} &
&
&
\boldmath{($M_{\odot}$)} &
\\
\textbf{(1)} &
\textbf{(2)} &
\textbf{(3)} &
\textbf{(4)} &
\textbf{(5)} &
\textbf{(6)} &
\textbf{(7)} &
\textbf{(8)} &
\textbf{(9)} \\
\midrule

J001115.69+011459.23&0.580&10.86 $\pm$ 0.53&2.34 $\pm$ 0.15&43.52 $\pm$ 2.99&14.53 $\pm$ 8.68&0.03 $\pm$ 0.01&8.37&$-$0.95\\
J002421.84+002508.49&0.599&10.22 $\pm$ 0.12&2.12 $\pm$ 0.15&30.89 $\pm$ 3.56&2.73 $\pm$ 2.37&0.29 $\pm$ 0.05&7.36&$-$0.16\\
J005905.51+000651.67&0.719&11.56 $\pm$ 0.31&2.00 $\pm$ 0.38&45.86 $\pm$ 9.04&11.14 $\pm$ 8.83&0.70 $\pm$ 0.07&8.94&$-$0.64\\
J011430.24+000420.94&0.455&11.70 $\pm$ 0.08&1.43 $\pm$ 0.14&21.02 $\pm$ 3.12&1.18 $\pm$ 1.02&0.20 $\pm$ 0.07&8.42&$-$1.32\\
J011536.92{-}
000011.01&0.532&11.05 $\pm$ 0.18&1.72 $\pm$ 0.14&21.92 $\pm$ 2.72&1.06 $\pm$ 0.40&0.34 $\pm$ 0.09&8.24&$-$1.15\\
J013354.30+011034.00&0.704&11.03 $\pm$ 0.51&1.92 $\pm$ 0.24&29.08 $\pm$ 5.97&2.56 $\pm$ 2.62&0.25 $\pm$ 0.07&8.48&$-$1.09\\
J014003.99{-}002541.77&0.770&10.59 $\pm$ 0.17&2.40 $\pm$ 0.19&41.33 $\pm$ 4.08&2.79 $\pm$ 3.24&0.29 $\pm$ 0.07&8.76&$-$1.11\\
J015243.15+002039.65&0.578&11.74 $\pm$ 0.16&1.48 $\pm$ 0.26&17.30 $\pm$ 0.61&1.59 $\pm$ 2.43&0.11 $\pm$ 0.03&8.22&$-$0.96\\
J015950.25+002340.82&0.163&10.46 $\pm$ 0.08&2.40 $\pm$ 0.12&55.98 $\pm$ 0.22&30.00 $\pm$ 0.03&0.19 $\pm$ 0.00&8.06&$-$0.42\\
J021859.87+002855.80&0.351&10.19 $\pm$ 0.06&1.96 $\pm$ 0.18&32.74 $\pm$ 3.17&13.79 $\pm$ 7.91&0.51 $\pm$ 0.07&7.85&$-$0.37\\
J022430.61{-}000038.91&0.431&9.83 $\pm$ 0.26&1.41 $\pm$ 0.13&30.57 $\pm$ 4.08&2.73 $\pm$ 1.35&0.31 $\pm$ 0.08&7.40&$-$0.25\\
J090026.51+204158.70&0.706&11.20 $\pm$ 0.04&2.35 $\pm$ 0.06&42.50 $\pm$ 1.93&28.81 $\pm$ 3.69&0.63 $\pm$ 0.02&8.79&$-$0.60\\
J090158.88+002313.87&0.196&9.88 $\pm$ 0.24&1.52 $\pm$ 0.09&45.36 $\pm$ 3.68&18.82 $\pm$ 4.24&0.25 $\pm$ 0.04&8.38&$-$1.36\\
J090933.49+425346.51&0.670&11.96 $\pm$ 0.24&2.46 $\pm$ 0.14&17.19 $\pm$ 0.44&1.00 $\pm$ 0.02&0.56 $\pm$ 0.01&9.73&$-$2.02\\
J092159.40+450912.38&0.235&9.93 $\pm$ 0.07&1.88 $\pm$ 0.07&45.86 $\pm$ 0.71&30.00 $\pm$ 0.01&0.01 $\pm$ 0.00&8.64&$-$1.53\\
J092554.72+195405.13&0.192&10.98 $\pm$ 0.44&0.85 $\pm$ 0.14&57.07 $\pm$ 6.51&19.01 $\pm$ 9.30&0.93 $\pm$ 0.01&9.10&$-$1.19\\
J092635.12+072446.43&0.189&11.22 $\pm$ 0.08&1.25 $\pm$ 0.11&30.81 $\pm$ 3.91&2.85 $\pm$ 0.78&0.17 $\pm$ 0.01&9.26&$-$2.39\\
J094652.58+131953.83&0.133&10.29 $\pm$ 0.37&1.20 $\pm$ 0.19&61.04 $\pm$ 3.20&29.31 $\pm$ 2.86&0.78 $\pm$ 0.02&8.39&$-$0.80\\
J094745.14+072520.58&0.086&10.85 $\pm$ 0.01&0.04 $\pm$ 0.12&46.33 $\pm$ 3.60&3.49 $\pm$ 4.16&0.91 $\pm$ 0.01&8.23&$-$1.51\\
J095017.06+215022.41&0.455&10.28 $\pm$ 0.28&1.67 $\pm$ 0.07&33.01 $\pm$ 6.36&7.10 $\pm$ 6.23&0.50 $\pm$ 0.05&8.31&$-$0.54\\
J095240.17+515249.91&0.554&10.83 $\pm$ 0.59&1.63 $\pm$ 0.19&43.06 $\pm$ 8.28&6.64 $\pm$ 6.44&0.58 $\pm$ 0.07&8.50&$-$0.80\\
J095819.87+022903.51&0.345&10.06 $\pm$ 0.26&1.18 $\pm$ 0.04&30.39 $\pm$ 2.63&4.92 $\pm$ 0.75&0.37 $\pm$ 0.03&8.63&$-$1.65\\
J100043.14+020637.25&0.360&10.72 $\pm$ 0.11&1.14 $\pm$ 0.11&40.40 $\pm$ 5.48&9.38 $\pm$ 6.56&0.21 $\pm$ 0.04&7.89&$-$0.75\\
J100232.13+023537.33&0.658&10.98 $\pm$ 0.34&1.75 $\pm$ 0.11&40.45 $\pm$ 7.09&16.09 $\pm$ 8.13&0.32 $\pm$ 0.03&8.14&$-$0.42\\
J100420.13+051300.48&0.160&10.31 $\pm$ 0.44&0.93 $\pm$ 0.22&61.11 $\pm$ 2.96&22.30 $\pm$ 8.48&0.86 $\pm$ 0.02&7.74&$-$0.30\\
J103651.94+575950.96&0.500&10.46 $\pm$ 0.25&1.65 $\pm$ 0.13&34.05 $\pm$ 3.58&4.53 $\pm$ 3.08&0.30 $\pm$ 0.03&8.45&$-$1.03\\
J104009.33+560343.28&0.393&11.29 $\pm$ 0.05&0.98 $\pm$ 0.68&44.22 $\pm$ 11.79&11.11 $\pm$ 8.88&0.70 $\pm$ 0.04&7.89&$-$0.25\\
J104505.39+561118.34&0.428&11.04 $\pm$ 0.19&1.88 $\pm$ 0.10&47.59 $\pm$ 2.90&21.46 $\pm$ 6.41&0.01 $\pm$ 0.00&9.80&$-$2.48\\
J104739.49+563507.19&0.303&10.24 $\pm$ 0.40&1.71 $\pm$ 0.15&50.84 $\pm$ 1.91&6.32 $\pm$ 6.98&0.08 $\pm$ 0.01&8.07&$-$1.03\\
J105106.12+591625.24&0.767&10.82 $\pm$ 0.69&2.03 $\pm$ 0.16&29.89 $\pm$ 5.10&2.96 $\pm$ 2.63&0.40 $\pm$ 0.05&9.49&$-$1.42\\
J105143.89+335926.71&0.167&10.51 $\pm$ 0.20&0.86 $\pm$ 0.11&57.61 $\pm$ 6.17&10.12 $\pm$ 8.14&0.73 $\pm$ 0.03&8.27&$-$0.91\\
J105151.44{-}005117.66&0.359&10.83 $\pm$ 0.22&2.34 $\pm$ 0.12&65.93 $\pm$ 0.36&30.00 $\pm$ 0.22&0.60 $\pm$ 0.01&9.17&$-$0.72\\
J105705.41+580437.46&0.140&11.15 $\pm$ 0.06&1.17 $\pm$ 0.12&34.37 $\pm$ 4.61&4.72 $\pm$ 0.69&0.22 $\pm$ 0.02&7.86&$-$0.90\\
J105959.93+574848.17&0.453&10.01 $\pm$ 0.21&1.58 $\pm$ 0.18&27.65 $\pm$ 6.53&4.37 $\pm$ 4.75&0.60 $\pm$ 0.08&8.27&$-$0.68\\
J110036.64+564134.89&0.834&10.61 $\pm$ 0.18&2.42 $\pm$ 0.19&37.48 $\pm$ 6.26&9.56 $\pm$ 7.76&0.22 $\pm$ 0.05&8.38&$-$0.25\\
J111706.40+441333.30&0.144&10.79 $\pm$ 0.16&0.81 $\pm$ 0.13&65.25 $\pm$ 1.01&28.10 $\pm$ 4.82&0.93 $\pm$ 0.01&8.77&$-$1.13\\
J111830.29+402554.02&0.154&10.07 $\pm$ 0.26&1.41 $\pm$ 0.05&35.09 $\pm$ 1.54&6.31 $\pm$ 3.83&0.44 $\pm$ 0.01&8.45&$-$0.94\\
J112019.62+130320.10&0.314&10.29 $\pm$ 0.35&1.46 $\pm$ 0.12&59.43 $\pm$ 8.44&13.62 $\pm$ 8.88&0.46 $\pm$ 0.04&7.80&$-$0.57\\
J112048.99+133821.91&0.513&10.00 $\pm$ 0.19&1.60 $\pm$ 0.05&50.19 $\pm$ 10.36&4.99 $\pm$ 5.23&0.34 $\pm$ 0.03&8.63&$-$1.23\\
J112759.26+360207.00&0.667&10.61 $\pm$ 0.24&2.19 $\pm$ 0.26&29.24 $\pm$ 6.55&3.04 $\pm$ 4.31&0.38 $\pm$ 0.05&8.80&$-$0.98\\
J114121.76+014803.58&0.382&10.27 $\pm$ 0.05&1.82 $\pm$ 0.14&27.85 $\pm$ 3.06&2.75 $\pm$ 1.48&0.44 $\pm$ 0.05&8.59&$-$0.88\\
J114933.88+222227.07&0.554&10.29 $\pm$ 0.34&1.66 $\pm$ 0.07&36.79 $\pm$ 5.03&5.43 $\pm$ 6.54&0.40 $\pm$ 0.03&8.62&$-$0.99\\
J115253.68+000131.90&0.823&11.14 $\pm$ 0.28&2.23 $\pm$ 0.25&36.33 $\pm$ 7.13&7.52 $\pm$ 7.54&0.30 $\pm$ 0.10&8.50&$-$0.81\\
J120226.76{-}012915.28&0.150&10.21 $\pm$ 0.05&2.24 $\pm$ 0.10&37.58 $\pm$ 1.69&27.70 $\pm$ 4.84&0.14 $\pm$ 0.00&7.28&$-$0.26\\
J120312.14+015321.30&0.296&10.68 $\pm$ 0.37&1.45 $\pm$ 0.21&30.29 $\pm$ 3.74&1.99 $\pm$ 1.23&0.28 $\pm$ 0.05&8.86&$-$1.82\\
J120442.11+275411.80&0.165&10.06 $\pm$ 0.27&1.39 $\pm$ 0.14&65.72 $\pm$ 0.50&29.92 $\pm$ 1.01&0.68 $\pm$ 0.01&8.36&$-$1.34\\
J120734.63+150643.69&0.750&10.58 $\pm$ 0.32&2.07 $\pm$ 0.24&41.94 $\pm$ 12.34&11.19 $\pm$ 8.91&0.69 $\pm$ 0.06&8.63&$-$0.44\\
J121037.81+053805.88&0.436&10.07 $\pm$ 0.73&1.63 $\pm$ 0.09&46.50 $\pm$ 6.27&11.95 $\pm$ 8.56&0.33 $\pm$ 0.04&8.24&$-$0.95\\
J121728.37+065110.00&0.589&10.66 $\pm$ 0.13&1.94 $\pm$ 0.42&60.29 $\pm$ 7.12&9.53 $\pm$ 8.35&0.54 $\pm$ 0.05&8.57&$-$0.77\\
J121836.71+155908.43&0.766&10.33 $\pm$ 0.14&2.15 $\pm$ 0.11&39.50 $\pm$ 6.27&8.48 $\pm$ 6.85&0.42 $\pm$ 0.06&8.25&$-$0.39\\
J121906.57+160243.06&0.761&10.52 $\pm$ 0.17&2.28 $\pm$ 0.21&52.99 $\pm$ 10.16&22.71 $\pm$ 8.31&0.53 $\pm$ 0.06&7.93&$-$0.80\\
J121945.03+082117.95&0.228&10.41 $\pm$ 0.18&1.78 $\pm$ 0.13&57.41 $\pm$ 5.81&4.96 $\pm$ 2.33&0.22 $\pm$ 0.02&8.33&$-$1.18\\
J121946.54+145259.37&0.401&9.77 $\pm$ 0.11&1.17 $\pm$ 0.31&31.51 $\pm$ 6.09&4.81 $\pm$ 5.42&0.41 $\pm$ 0.03&8.79&$-$1.35\\
J122011.88+020342.21&0.240&11.05 $\pm$ 0.22&1.43 $\pm$ 0.32&27.19 $\pm$ 6.67&2.74 $\pm$ 3.78&1.00 $\pm$ 0.00&8.87&$-$0.87\\

\bottomrule
\end{tabularx}
\end{adjustwidth}
\end{table}

\begin{table}[H]\ContinuedFloat
\caption{\textit{Cont.}}
\label{tab:sedresultpart_app}
\small
\begin{adjustwidth}{-\extralength}{0cm}
\begin{tabularx}{\fulllength}{@{}p{3.2cm}cccccccc@{}}
\toprule
\textbf{Name} &
\boldmath{$z$} &
\boldmath{$\log M_\star$} &
\boldmath{{$\log {\rm SFR}_{\rm FIR}$}} &
\boldmath{$T_{\rm dust}$} &
\boldmath{$U_{\rm min}$} &
\boldmath{$f_{\rm AGN}$} &
\boldmath{$\log M_{\rm BH}$} &
\boldmath{$\log\lambda_{\rm Edd}$} \\
&
&
\boldmath{($M_{\odot}$)} &
\boldmath{($M_{\odot}~{\rm yr}^{-1}$)} &
\textbf{(K)} &
&
&
\boldmath{($M_{\odot}$)} &
\\
\textbf{(1)} &
\textbf{(2)} &
\textbf{(3)} &
\textbf{(4)} &
\textbf{(5)} &
\textbf{(6)} &
\textbf{(7)} &
\textbf{(8)} &
\textbf{(9)} \\
\midrule

J122026.72+062748.20&0.349&10.78 $\pm$ 0.34&1.64 $\pm$ 0.09&33.74 $\pm$ 3.47&4.83 $\pm$ 2.10&0.19 $\pm$ 0.02&8.63&$-$1.52\\
J122102.50+155447.04&0.229&10.10 $\pm$ 0.63&1.18 $\pm$ 0.21&35.46 $\pm$ 7.19&4.67 $\pm$ 5.59&0.55 $\pm$ 0.06&8.00&$-$0.83\\
J122102.95-000733.74&0.366&10.71 $\pm$ 0.17&2.21 $\pm$ 0.10&50.05 $\pm$ 3.60&21.78 $\pm$ 8.18&0.19 $\pm$ 0.02&7.75&$-$0.33\\
J122106.50+114625.46&0.340&11.04 $\pm$ 0.07&1.52 $\pm$ 0.07&39.79 $\pm$ 6.39&3.72 $\pm$ 4.29&0.17 $\pm$ 0.03&8.37&$-$1.19\\
J122307.37-002124.73&0.804&11.42 $\pm$ 0.22&2.23 $\pm$ 0.23&36.72 $\pm$ 5.31&4.05 $\pm$ 5.06&0.19 $\pm$ 0.06&--&--\\
J122312.17+095017.72&0.277&10.21 $\pm$ 0.31&1.55 $\pm$ 0.28&46.70 $\pm$ 4.90&4.42 $\pm$ 5.58&0.47 $\pm$ 0.10&8.05&$-$0.63\\
J122317.80+092306.94&0.682&10.96 $\pm$ 0.06&2.76 $\pm$ 0.10&48.72 $\pm$ 2.06&29.99 $\pm$ 0.26&0.17 $\pm$ 0.01&8.92&$-$0.62\\
J122404.62+045637.94&0.358&10.22 $\pm$ 0.25&1.76 $\pm$ 0.24&57.12 $\pm$ 9.15&5.11 $\pm$ 5.23&0.47 $\pm$ 0.04&8.25&$-$0.69\\
J122520.13+084450.76&0.535&10.52 $\pm$ 0.22&2.06 $\pm$ 0.04&34.01 $\pm$ 1.02&5.13 $\pm$ 1.55&0.42 $\pm$ 0.02&8.66&$-$0.60\\
J122526.21+141332.24&0.760&10.61 $\pm$ 0.22&2.35 $\pm$ 0.12&42.56 $\pm$ 4.87&21.34 $\pm$ 8.11&0.18 $\pm$ 0.02&8.03&$-$0.48\\
J122641.50+055906.81&0.290&11.36 $\pm$ 0.15&1.31 $\pm$ 0.18&31.93 $\pm$ 5.44&2.69 $\pm$ 2.09&0.24 $\pm$ 0.08&8.85&$-$1.56\\
J122822.10+114606.83&0.365&11.09 $\pm$ 0.26&1.86 $\pm$ 0.07&28.26 $\pm$ 2.65&4.83 $\pm$ 0.56&0.14 $\pm$ 0.03&8.84&$-$1.52\\
J122839.20+035749.29&0.608&10.57 $\pm$ 0.14&2.20 $\pm$ 0.10&39.38 $\pm$ 4.61&5.79 $\pm$ 5.14&0.37 $\pm$ 0.03&9.03&$-$1.32\\
J123436.54+123918.66&0.777&10.59 $\pm$ 0.13&2.44 $\pm$ 0.16&52.04 $\pm$ 3.17&26.37 $\pm$ 6.30&0.23 $\pm$ 0.05&8.73&$-$0.99\\
J123800.92+621336.09&0.440&10.33 $\pm$ 0.41&1.84 $\pm$ 0.16&39.80 $\pm$ 1.74&4.94 $\pm$ 0.77&0.17 $\pm$ 0.03&7.94&$-$0.59\\
J124511.26+335610.12&0.711&10.69 $\pm$ 0.24&2.19 $\pm$ 0.23&35.96 $\pm$ 8.06&5.32 $\pm$ 5.67&0.72 $\pm$ 0.08&9.01&$-$0.46\\
J125257.36+331555.81&0.367&11.36 $\pm$ 0.07&1.32 $\pm$ 0.16&21.81 $\pm$ 4.75&1.66 $\pm$ 1.07&0.27 $\pm$ 0.05&8.16&$-$1.03\\
J125317.57+310550.64&0.782&11.61 $\pm$ 0.06&2.39 $\pm$ 0.21&54.60 $\pm$ 5.54&23.04 $\pm$ 8.04&0.74 $\pm$ 0.07&9.02&$-$0.41\\
J125553.05+272405.23&0.316&11.48 $\pm$ 0.14&1.03 $\pm$ 0.14&28.67 $\pm$ 5.56&6.98 $\pm$ 6.59&0.40 $\pm$ 0.06&8.86&$-$1.45\\
J125703.80+250457.56&0.821&10.89 $\pm$ 0.21&2.63 $\pm$ 0.15&42.65 $\pm$ 4.85&6.94 $\pm$ 6.80&0.25 $\pm$ 0.05&8.19&$-$0.31\\
J125711.97+274216.45&0.793&10.55 $\pm$ 0.31&2.09 $\pm$ 0.14&40.90 $\pm$ 6.18&9.33 $\pm$ 8.17&0.44 $\pm$ 0.03&8.74&$-$0.88\\
J125757.23+322929.29&0.806&12.20 $\pm$ 0.16&2.60 $\pm$ 0.26&17.75 $\pm$ 2.03&1.00 $\pm$ 0.04&0.83 $\pm$ 0.03&8.74&$-$0.85\\
J130622.96+225752.95&0.758&10.65 $\pm$ 0.43&2.06 $\pm$ 0.31&31.47 $\pm$ 8.35&5.02 $\pm$ 5.48&0.36 $\pm$ 0.10&8.44&$-$1.15\\
J130947.00+081948.23&0.154&10.72 $\pm$ 0.31&1.00 $\pm$ 0.22&58.64 $\pm$ 8.67&29.72 $\pm$ 1.87&0.83 $\pm$ 0.01&8.59&$-$0.93\\
J131247.97+250756.71&0.425&10.99 $\pm$ 0.36&2.21 $\pm$ 0.21&40.35 $\pm$ 3.07&19.48 $\pm$ 9.57&0.31 $\pm$ 0.07&8.63&$-$1.09\\
J131312.13+284730.01&0.259&11.21 $\pm$ 0.18&0.93 $\pm$ 0.17&20.49 $\pm$ 3.52&1.54 $\pm$ 1.28&0.47 $\pm$ 0.09&8.21&$-$1.21\\
J131531.66+265414.73&0.620&10.24 $\pm$ 0.23&1.92 $\pm$ 0.21&37.98 $\pm$ 5.94&4.13 $\pm$ 5.32&0.33 $\pm$ 0.08&7.88&$-$0.65\\
J132919.84+250626.45&0.781&10.44 $\pm$ 0.19&2.22 $\pm$ 0.20&40.34 $\pm$ 7.64&4.14 $\pm$ 4.26&0.34 $\pm$ 0.08&8.81&$-$0.85\\
J133005.71+254243.77&0.598&10.85 $\pm$ 0.23&2.48 $\pm$ 0.34&54.85 $\pm$ 3.20&16.61 $\pm$ 9.25&0.73 $\pm$ 0.07&9.11&$-$1.34\\
J133442.29+320939.60&0.649&10.71 $\pm$ 0.10&2.39 $\pm$ 0.30&48.41 $\pm$ 6.56&27.62 $\pm$ 5.11&0.88 $\pm$ 0.03&8.52&$-$0.55\\
J134356.74+253847.69&0.086&10.24 $\pm$ 0.17&0.45 $\pm$ 0.08&59.64 $\pm$ 3.01&4.33 $\pm$ 4.46&0.69 $\pm$ 0.02&7.92&$-$1.09\\
J135632.80+210352.35&0.301&9.70 $\pm$ 0.32&1.28 $\pm$ 0.21&64.77 $\pm$ 2.82&22.78 $\pm$ 8.30&0.81 $\pm$ 0.02&8.82&$-$1.08\\
J140621.89+222346.54&0.098&9.96 $\pm$ 0.23&0.49 $\pm$ 0.10&30.74 $\pm$ 3.30&2.67 $\pm$ 1.10&0.70 $\pm$ 0.01&7.36&$-$0.32\\
J140655.66+015712.88&0.427&10.32 $\pm$ 0.12&2.10 $\pm$ 0.19&46.44 $\pm$ 5.30&26.06 $\pm$ 6.68&0.31 $\pm$ 0.07&7.96&$-$0.43\\
J140700.40+282714.65&0.077&9.53 $\pm$ 0.33&1.21 $\pm$ 0.16&56.24 $\pm$ 4.23&4.13 $\pm$ 3.16&0.63 $\pm$ 0.01&8.79&$-$1.63\\
J141637.45+003352.28&0.434&10.47 $\pm$ 0.15&2.32 $\pm$ 0.19&65.79 $\pm$ 0.52&25.26 $\pm$ 6.59&0.27 $\pm$ 0.02&8.67&$-$1.31\\
J141644.62+190541.92&0.365&9.85 $\pm$ 0.13&1.67 $\pm$ 0.11&48.70 $\pm$ 5.31&20.35 $\pm$ 7.44&0.23 $\pm$ 0.03&7.51&$-$0.16\\
J141700.83+445606.39&0.113&11.34 $\pm$ 0.09&0.73 $\pm$ 0.27&28.77 $\pm$ 5.10&2.00 $\pm$ 1.62&0.61 $\pm$ 0.01&7.93&$-$0.89\\
J142052.44+525622.42&0.677&10.69 $\pm$ 0.19&2.17 $\pm$ 0.29&49.66 $\pm$ 4.64&10.98 $\pm$ 8.75&0.69 $\pm$ 0.08&8.57&$-$0.63\\
J142648.78+005323.24&0.220&11.00 $\pm$ 0.11&1.40 $\pm$ 0.11&33.16 $\pm$ 3.41&4.11 $\pm$ 1.23&0.24 $\pm$ 0.07&8.40&$-$1.22\\
J142710.94+013023.22&0.704&10.80 $\pm$ 0.31&2.45 $\pm$ 0.27&44.87 $\pm$ 4.24&11.43 $\pm$ 8.79&0.40 $\pm$ 0.09&8.24&$-$0.37\\
J142753.79+345248.35&0.514&10.01 $\pm$ 0.47&1.58 $\pm$ 0.11&29.40 $\pm$ 4.75&3.24 $\pm$ 3.84&0.39 $\pm$ 0.08&8.00&$-$0.48\\
J142918.15+592106.65&0.739&11.08 $\pm$ 0.43&2.70 $\pm$ 0.14&17.78 $\pm$ 1.02&1.00 $\pm$ 0.00&0.11 $\pm$ 0.01&8.94&$-$1.50\\
J142943.07+474726.23&0.221&10.50 $\pm$ 0.58&1.27 $\pm$ 0.30&65.31 $\pm$ 3.38&23.42 $\pm$ 8.27&0.59 $\pm$ 0.01&7.95&$-$0.47\\
J143624.81-002905.36&0.325&9.98 $\pm$ 0.24&1.66 $\pm$ 0.31&43.24 $\pm$ 6.56&6.18 $\pm$ 6.58&0.74 $\pm$ 0.08&7.73&$-$0.30\\
J144231.82+014353.43&0.280&11.29 $\pm$ 0.20&0.99 $\pm$ 0.14&20.51 $\pm$ 3.16&1.33 $\pm$ 0.88&0.44 $\pm$ 0.08&8.62&$-$1.66\\
J145001.69+022006.75&0.520&10.45 $\pm$ 0.39&2.28 $\pm$ 0.14&36.81 $\pm$ 3.25&6.71 $\pm$ 5.15&0.15 $\pm$ 0.02&8.23&$-$0.63\\
J145108.76+270926.92&0.064&10.67 $\pm$ 0.18&0.63 $\pm$ 0.21&62.36 $\pm$ 1.33&13.99 $\pm$ 6.45&0.67 $\pm$ 0.01&7.29&$-$0.12\\
J145538.73+002238.06&0.434&11.43 $\pm$ 0.23&1.71 $\pm$ 0.21&26.64 $\pm$ 2.10&1.07 $\pm$ 0.36&0.10 $\pm$ 0.02&8.18&$-$0.91\\
J152114.26+222743.87&0.136&10.48 $\pm$ 0.32&1.02 $\pm$ 0.16&57.56 $\pm$ 2.47&6.54 $\pm$ 4.15&0.78 $\pm$ 0.01&7.90&$-$0.63\\
J154530.24+484608.98&0.399&10.84 $\pm$ 0.03&2.51 $\pm$ 0.08&64.33 $\pm$ 2.35&30.00 $\pm$ 0.16&0.55 $\pm$ 0.01&8.51&$-$0.26\\
J155444.58+082221.48&0.119&11.04 $\pm$ 0.10&0.22 $\pm$ 0.11&37.40 $\pm$ 4.42&16.60 $\pm$ 8.64&0.87 $\pm$ 0.01&7.73&$-$0.43\\
J161413.20+260416.21&0.131&10.05 $\pm$ 0.47&0.91 $\pm$ 0.26&34.42 $\pm$ 1.55&1.38 $\pm$ 0.86&0.54 $\pm$ 0.01&7.99&$-$0.62\\
J163352.34+402115.66&0.782&11.14 $\pm$ 0.54&2.14 $\pm$ 0.08&27.92 $\pm$ 1.51&3.49 $\pm$ 0.87&0.24 $\pm$ 0.03&8.99&$-$1.30\\
J163915.81+412833.70&0.690&11.12 $\pm$ 0.34&1.84 $\pm$ 0.05&17.17 $\pm$ 0.07&1.00 $\pm$ 0.03&0.30 $\pm$ 0.02&8.06&$-$0.80\\
J171033.22+584456.86&0.281&10.93 $\pm$ 0.39&1.53 $\pm$ 0.15&46.81 $\pm$ 2.70&4.89 $\pm$ 5.84&0.10 $\pm$ 0.02&7.79&$-$0.76\\
J171352.43+584201.25&0.521&10.94 $\pm$ 0.30&2.11 $\pm$ 0.20&53.23 $\pm$ 12.05&2.88 $\pm$ 2.00&0.46 $\pm$ 0.02&8.84&$-$0.92\\
J220759.39+001722.62&0.368&10.25 $\pm$ 0.18&1.40 $\pm$ 0.21&26.63 $\pm$ 4.29&4.26 $\pm$ 4.16&0.51 $\pm$ 0.05&7.75&$-$0.53\\
\bottomrule
\end{tabularx}
\end{adjustwidth}
\end{table}

\begin{table}[H]\ContinuedFloat
\caption{\textit{Cont.}}
\label{tab:sedresultpart_app}
\small
\begin{adjustwidth}{-\extralength}{0cm}
\begin{tabularx}{\fulllength}{@{}p{3.2cm}cccccccc@{}}
\toprule
\textbf{Name} &
\boldmath{$z$} &
\boldmath{{$\log M_\star$}} &
\boldmath{$\log {\rm SFR}_{\rm FIR}$} &
\boldmath{$T_{\rm dust}$} &
\boldmath{$U_{\rm min}$} &
\boldmath{$f_{\rm AGN}$} &
\boldmath{$\log M_{\rm BH}$} &
\boldmath{$\log\lambda_{\rm Edd}$} \\
&
&
\boldmath{($M_{\odot}$)} &
\boldmath{($M_{\odot}~{\rm yr}^{-1}$)} &
\textbf{(K)} &
&
&
\boldmath{($M_{\odot}$)} &
\\
\textbf{(1)} &
\textbf{(2)} &
\textbf{(3)} &
\textbf{(4)} &
\textbf{(5)} &
\textbf{(6)} &
\textbf{(7)} &
\textbf{(8)} &
\textbf{(9)} \\
\midrule

J223607.68+134355.32&0.326&10.15 $\pm$ 0.36&1.64 $\pm$ 0.13&64.76 $\pm$ 1.19&28.53 $\pm$ 4.15&0.81 $\pm$ 0.01&8.57&$-$0.52\\
J233741.32+001743.77&0.762&10.21 $\pm$ 0.07&2.08 $\pm$ 0.05&32.85 $\pm$ 1.88&4.83 $\pm$ 0.72&0.26 $\pm$ 0.02&8.53&$-$1.21\\
J235156.13{-}010913.34&0.174&11.06 $\pm$ 0.39&1.46 $\pm$ 0.15&50.62 $\pm$ 5.96&7.90 $\pm$ 5.74&0.68 $\pm$ 0.01&8.81&$-$1.18\\
PG0003+158&0.450&11.25 $\pm$ 0.25&1.45 $\pm$ 0.23&51.92 $\pm$ 10.72&11.18 $\pm$ 8.94&1.00 $\pm$ 0.00&9.45&$-$0.56\\
PG0003+199&0.025&10.22 $\pm$ 0.20&$-$0.09 $\pm$ 0.24&43.34 $\pm$ 5.65&28.06 $\pm$ 4.54&0.89 $\pm$ 0.00&7.52&$-$0.45\\
PG0007+106&0.089&11.19 $\pm$ 0.20&0.94 $\pm$ 0.17&24.38 $\pm$ 1.49&1.00 $\pm$ 0.07&0.70 $\pm$ 0.00&8.87&$-$1.18\\
PG0026+129&0.142&11.45 $\pm$ 0.16&0.49 $\pm$ 0.15&51.10 $\pm$ 5.96&16.84 $\pm$ 9.23&0.96 $\pm$ 0.01&8.12&$-$0.15\\
PG0043+039&0.384&10.68 $\pm$ 0.38&1.24 $\pm$ 0.04&50.24 $\pm$ 9.40&17.28 $\pm$ 9.36&1.00 $\pm$ 0.01&9.28&$-$0.87\\
PG0049+171&0.064&9.67 $\pm$ 0.19&$-$0.37 $\pm$ 0.29&33.89 $\pm$ 6.23&5.60 $\pm$ 5.77&0.90 $\pm$ 0.01&8.45&$-$1.58\\
PG0050+124&0.061&10.31 $\pm$ 0.49&1.39 $\pm$ 0.28&34.23 $\pm$ 2.80&4.71 $\pm$ 0.95&0.53 $\pm$ 0.00&7.57&0.09\\
PG0052+251&0.155&11.61 $\pm$ 0.36&1.13 $\pm$ 0.17&27.81 $\pm$ 2.06&2.67 $\pm$ 0.98&0.78 $\pm$ 0.00&8.99&$-$1.09\\
PG0157+001&0.164&10.64 $\pm$ 0.12&2.53 $\pm$ 0.10&61.62 $\pm$ 2.43&29.82 $\pm$ 1.47&0.16 $\pm$ 0.00&8.31&$-$0.46\\
PG0804+761&0.100&10.65 $\pm$ 0.69&0.73 $\pm$ 0.27&56.08 $\pm$ 4.18&28.77 $\pm$ 3.81&0.97 $\pm$ 0.00&8.55&$-$0.62\\
PG0838+770&0.131&10.85 $\pm$ 0.30&1.06 $\pm$ 0.30&29.78 $\pm$ 2.96&3.22 $\pm$ 1.29&0.53 $\pm$ 0.01&8.29&$-$0.69\\
PG0844+349&0.064&10.34 $\pm$ 0.22&0.30 $\pm$ 0.23&24.91 $\pm$ 1.10&1.00 $\pm$ 0.04&0.84 $\pm$ 0.00&8.03&$-$0.67\\
PG0921+525&0.035&10.23 $\pm$ 0.09&$-$0.32 $\pm$ 0.15&31.69 $\pm$ 2.40&4.73 $\pm$ 0.79&0.87 $\pm$ 0.00&7.45&$-$0.95\\
PG0923+201&0.190&11.05 $\pm$ 0.50&0.84 $\pm$ 0.16&55.80 $\pm$ 4.56&21.31 $\pm$ 8.80&0.95 $\pm$ 0.01&9.33&$-$1.42\\
PG0923+129&0.029&10.80 $\pm$ 0.08&0.56 $\pm$ 0.16&36.45 $\pm$ 3.15&4.94 $\pm$ 0.34&0.38 $\pm$ 0.00&7.52&$-$0.79\\
PG0934+013&0.050&10.15 $\pm$ 0.19&0.47 $\pm$ 0.16&39.13 $\pm$ 2.85&3.83 $\pm$ 1.04&0.30 $\pm$ 0.00&7.15&$-$0.40\\
PG0947+396&0.206&11.28 $\pm$ 0.29&1.23 $\pm$ 0.24&58.49 $\pm$ 6.77&19.71 $\pm$ 8.60&0.86 $\pm$ 0.01&8.81&$-$1.13\\
PG0953+414&0.239&10.35 $\pm$ 0.64&1.03 $\pm$ 0.13&40.22 $\pm$ 7.75&7.00 $\pm$ 6.54&0.94 $\pm$ 0.01&8.74&$-$0.49\\
PG1001+054&0.161&10.34 $\pm$ 0.49&0.84 $\pm$ 0.26&44.19 $\pm$ 10.84&14.43 $\pm$ 8.80&0.88 $\pm$ 0.01&7.87&$-$0.26\\
PG1004+130&0.240&10.60 $\pm$ 0.21&1.68 $\pm$ 0.18&64.44 $\pm$ 2.16&30.00 $\pm$ 0.02&0.74 $\pm$ 0.01&9.43&$-$1.02\\
PG1011{-}040&0.058&10.92 $\pm$ 0.23&0.50 $\pm$ 0.33&27.32 $\pm$ 2.74&3.59 $\pm$ 1.26&0.55 $\pm$ 0.00&7.43&$-$0.30\\
PG1012+008&0.185&11.15 $\pm$ 0.29&1.36 $\pm$ 0.17&41.43 $\pm$ 2.83&5.41 $\pm$ 2.37&0.73 $\pm$ 0.01&8.39&$-$0.51\\
PG1022+519&0.045&10.54 $\pm$ 0.19&0.33 $\pm$ 0.04&35.04 $\pm$ 1.31&5.16 $\pm$ 1.41&0.43 $\pm$ 0.00&7.25&$-$0.68\\
PG1048+342&0.167&11.29 $\pm$ 0.20&0.90 $\pm$ 0.17&30.95 $\pm$ 6.85&2.08 $\pm$ 1.72&0.75 $\pm$ 0.01&8.50&$-$0.92\\
PG1048{-}090&0.344&10.47 $\pm$ 0.70&1.04 $\pm$ 0.06&41.45 $\pm$ 10.56&14.47 $\pm$ 9.24&0.96 $\pm$ 0.05&9.37&$-$0.90\\
PG1049{-}005&0.357&10.83 $\pm$ 0.21&2.31 $\pm$ 0.15&61.63 $\pm$ 6.18&29.47 $\pm$ 2.61&0.64 $\pm$ 0.00&9.34&$-$0.84\\
PG1100+772&0.313&10.72 $\pm$ 0.44&1.65 $\pm$ 0.20&60.30 $\pm$ 5.17&24.21 $\pm$ 7.27&0.80 $\pm$ 0.01&9.44&$-$0.99\\
PG1103{-}006&0.425&10.24 $\pm$ 0.54&1.36 $\pm$ 0.25&54.52 $\pm$ 7.73&19.85 $\pm$ 9.13&0.93 $\pm$ 0.01&9.49&$-$0.95\\
PG1114+445&0.144&10.97 $\pm$ 0.11&0.79 $\pm$ 0.14&64.00 $\pm$ 2.48&26.85 $\pm$ 5.89&0.95 $\pm$ 0.00&8.72&$-$1.12\\
PG1115+407&0.154&10.33 $\pm$ 0.31&1.42 $\pm$ 0.04&34.67 $\pm$ 1.25&10.25 $\pm$ 6.17&0.45 $\pm$ 0.00&7.80&$-$0.31\\
PG1116+215&0.177&10.35 $\pm$ 0.70&1.16 $\pm$ 0.02&46.35 $\pm$ 5.05&24.49 $\pm$ 6.76&0.99 $\pm$ 0.00&8.69&$-$0.42\\
PG1119+120&0.049&10.62 $\pm$ 0.21&0.70 $\pm$ 0.31&49.80 $\pm$ 10.14&22.83 $\pm$ 7.04&0.50 $\pm$ 0.00&7.58&$-$0.58\\
PG1121+422&0.234&10.77 $\pm$ 0.28&0.72 $\pm$ 0.26&40.75 $\pm$ 11.68&12.26 $\pm$ 9.11&1.00 $\pm$ 0.00&8.17&$-$0.42\\
PG1126{-}041&0.060&10.90 $\pm$ 0.41&0.96 $\pm$ 0.22&41.61 $\pm$ 6.08&12.49 $\pm$ 6.37&0.60 $\pm$ 0.00&7.87&$-$0.61\\
PG1149{-}110&0.049&11.07 $\pm$ 0.18&0.43 $\pm$ 0.30&34.59 $\pm$ 3.90&3.06 $\pm$ 1.05&0.44 $\pm$ 0.01&8.04&$-$1.06\\
PG1151+117&0.176&10.89 $\pm$ 0.33&0.47 $\pm$ 0.03&47.17 $\pm$ 11.78&11.98 $\pm$ 8.99&0.96 $\pm$ 0.03&8.68&$-$1.05\\
PG1202+281&0.165&10.36 $\pm$ 0.26&1.33 $\pm$ 0.16&58.97 $\pm$ 8.51&26.02 $\pm$ 6.11&0.66 $\pm$ 0.01&8.74&$-$1.27\\
PG1211+143&0.085&9.94 $\pm$ 0.55&0.60 $\pm$ 0.14&58.37 $\pm$ 3.36&28.98 $\pm$ 3.48&0.95 $\pm$ 0.00&8.10&$-$0.16\\
PG1216+069&0.334&10.70 $\pm$ 0.40&0.95 $\pm$ 0.28&40.12 $\pm$ 12.53&9.79 $\pm$ 8.44&1.00 $\pm$ 0.00&9.36&$-$0.77\\
PG1226+023&0.158&11.83 $\pm$ 0.03&1.74 $\pm$ 0.02&57.58 $\pm$ 5.07&29.98 $\pm$ 0.53&1.00 $\pm$ 0.00&9.18&$-$0.29\\
PG1229+204&0.064&11.10 $\pm$ 0.16&0.67 $\pm$ 0.25&37.28 $\pm$ 3.41&4.19 $\pm$ 0.99&0.69 $\pm$ 0.00&8.26&$-$1.01\\
PG1244+026&0.048&9.53 $\pm$ 0.34&0.27 $\pm$ 0.29&51.83 $\pm$ 11.18&17.26 $\pm$ 8.64&0.52 $\pm$ 0.00&6.62&0.05\\
PG1259+593&0.472&11.45 $\pm$ 0.54&1.41 $\pm$ 0.16&39.74 $\pm$ 13.67&10.22 $\pm$ 8.54&1.00 $\pm$ 0.00&9.09&$-$0.31\\
PG1302{-}102&0.286&11.78 $\pm$ 0.39&1.74 $\pm$ 0.34&30.88 $\pm$ 0.14&1.00 $\pm$ 0.00&0.85 $\pm$ 0.01&9.05&$-$0.35\\
PG1307+085&0.155&10.55 $\pm$ 0.44&0.88 $\pm$ 0.29&40.43 $\pm$ 4.78&20.78 $\pm$ 8.52&0.88 $\pm$ 0.01&9.00&$-$1.12\\
PG1309+355&0.184&11.04 $\pm$ 0.14&1.22 $\pm$ 0.24&31.01 $\pm$ 3.23&2.18 $\pm$ 1.02&0.85 $\pm$ 0.01&8.48&$-$0.60\\
PG1310{-}108&0.035&9.72 $\pm$ 0.47&$-$0.14 $\pm$ 0.21&31.80 $\pm$ 9.32&2.08 $\pm$ 1.12&0.84 $\pm$ 0.00&7.99&$-$1.39\\
PG1322+659&0.168&10.65 $\pm$ 0.34&1.15 $\pm$ 0.10&54.11 $\pm$ 4.78&23.16 $\pm$ 7.38&0.72 $\pm$ 0.01&8.42&$-$0.57\\
PG1341+258&0.087&10.80 $\pm$ 0.14&0.48 $\pm$ 0.16&53.87 $\pm$ 5.92&4.93 $\pm$ 5.30&0.75 $\pm$ 0.01&8.15&$-$0.94\\
PG1351+236&0.055&10.91 $\pm$ 0.05&0.73 $\pm$ 0.05&35.07 $\pm$ 2.41&5.10 $\pm$ 1.09&0.19 $\pm$ 0.01&8.67&$-$1.75\\
PG1351+640&0.087&9.94 $\pm$ 0.21&1.51 $\pm$ 0.25&63.74 $\pm$ 5.61&29.83 $\pm$ 1.49&0.52 $\pm$ 0.00&8.97&$-$1.26\\
PG1352+183&0.158&10.68 $\pm$ 0.36&0.37 $\pm$ 0.15&52.06 $\pm$ 8.70&14.90 $\pm$ 9.50&0.96 $\pm$ 0.03&8.56&$-$0.87\\
PG1354+213&0.300&9.96 $\pm$ 0.51&1.20 $\pm$ 0.25&52.48 $\pm$ 10.55&20.80 $\pm$ 8.81&0.85 $\pm$ 0.01&8.77&$-$0.92\\
PG1402+261&0.164&10.32 $\pm$ 0.07&1.55 $\pm$ 0.09&49.04 $\pm$ 3.97&29.76 $\pm$ 1.71&0.71 $\pm$ 0.00&8.08&$-$0.23\\
PG1404+226&0.098&10.39 $\pm$ 0.25&0.51 $\pm$ 0.14&27.96 $\pm$ 2.63&1.93 $\pm$ 1.09&0.74 $\pm$ 0.01&7.01&0.24\\

\bottomrule
\end{tabularx}
\end{adjustwidth}
\end{table}

\begin{table}[H]\ContinuedFloat
\caption{\textit{Cont.}}
\label{tab:sedresultpart_app}
\small
\begin{adjustwidth}{-\extralength}{0cm}
\begin{tabularx}{\fulllength}{@{}p{3.2cm}cccccccc@{}}
\toprule
\textbf{Name} &
\boldmath{$z$} &
\boldmath{{$\log M_\star$}} &
\boldmath{$\log {\rm SFR}_{\rm FIR}$} &
\boldmath{$T_{\rm dust}$} &
\boldmath{$U_{\rm min}$} &
\boldmath{$f_{\rm AGN}$} &
\boldmath{$\log M_{\rm BH}$} &
\boldmath{$\log\lambda_{\rm Edd}$} \\
&
&
\boldmath{($M_{\odot}$)} &
\boldmath{($M_{\odot}~{\rm yr}^{-1}$)} &
\textbf{(K)} &
&
&
\boldmath{($M_{\odot}$)} &
\\
\textbf{(1)} &
\textbf{(2)} &
\textbf{(3)} &
\textbf{(4)} &
\textbf{(5)} &
\textbf{(6)} &
\textbf{(7)} &
\textbf{(8)} &
\textbf{(9)} \\
\midrule

PG1411+442&0.089&10.25 $\pm$ 0.52&0.57 $\pm$ 0.14&31.05 $\pm$ 1.80&4.12 $\pm$ 1.05&0.90 $\pm$ 0.00&8.20&$-$0.70\\
PG1415+451&0.114&11.48 $\pm$ 0.14&0.92 $\pm$ 0.33&33.72 $\pm$ 4.99&4.48 $\pm$ 1.96&0.66 $\pm$ 0.00&8.14&$-$0.71\\
PG1416{-}129&0.129&10.30 $\pm$ 0.48&0.70 $\pm$ 0.13&52.50 $\pm$ 6.96&21.43 $\pm$ 8.32&0.66 $\pm$ 0.01&9.19&$-$1.18\\
PG1425+267&0.366&10.87 $\pm$ 0.30&1.81 $\pm$ 0.09&64.89 $\pm$ 2.08&29.90 $\pm$ 1.13&0.72 $\pm$ 0.01&9.90&$-$1.27\\
PG1426+015&0.086&11.65 $\pm$ 0.20&1.17 $\pm$ 0.17&40.66 $\pm$ 1.69&5.03 $\pm$ 0.67&0.72 $\pm$ 0.00&9.15&$-$1.40\\
PG1427+480&0.221&10.39 $\pm$ 0.48&1.13 $\pm$ 0.34&54.60 $\pm$ 12.06&22.31 $\pm$ 8.62&0.61 $\pm$ 0.01&8.22&$-$0.59\\
PG1435{-}067&0.129&10.55 $\pm$ 0.46&0.31 $\pm$ 0.13&47.68 $\pm$ 6.88&2.13 $\pm$ 3.02&0.93 $\pm$ 0.01&8.50&$-$0.71\\
PG1440+356&0.077&11.31 $\pm$ 0.13&1.33 $\pm$ 0.08&43.37 $\pm$ 5.82&22.60 $\pm$ 6.20&0.44 $\pm$ 0.00&7.60&$-$0.18\\
PG1444+407&0.267&10.54 $\pm$ 0.28&1.96 $\pm$ 0.18&43.85 $\pm$ 12.16&12.45 $\pm$ 9.16&1.00 $\pm$ 0.00&8.44&$-$0.37\\
PG1448+273&0.065&10.81 $\pm$ 0.09&0.51 $\pm$ 0.28&47.17 $\pm$ 10.35&5.40 $\pm$ 4.66&0.64 $\pm$ 0.00&7.09&0.26\\
PG1501+106&0.036&10.50 $\pm$ 0.13&0.52 $\pm$ 0.23&64.71 $\pm$ 3.05&29.56 $\pm$ 2.32&0.52 $\pm$ 0.00&8.64&$-$1.48\\
PG1512+370&0.371&11.47 $\pm$ 0.23&1.47 $\pm$ 0.27&37.91 $\pm$ 8.96&3.21 $\pm$ 3.95&1.00 $\pm$ 0.06&9.53&$-$1.06\\
PG1519+226&0.137&10.68 $\pm$ 0.36&1.01 $\pm$ 0.22&50.59 $\pm$ 7.84&14.98 $\pm$ 7.36&0.83 $\pm$ 0.00&8.07&$-$0.49\\
PG1534+580&0.030&10.00 $\pm$ 0.15&0.09 $\pm$ 0.17&60.45 $\pm$ 5.01&17.60 $\pm$ 7.02&0.72 $\pm$ 0.00&8.30&$-$1.74\\
PG1535+547&0.038&10.62 $\pm$ 0.12&$-$0.31 $\pm$ 0.30&25.77 $\pm$ 1.96&1.22 $\pm$ 0.65&0.91 $\pm$ 0.00&7.30&$-$0.47\\
PG1543+489&0.400&10.84 $\pm$ 0.25&2.35 $\pm$ 0.18&57.39 $\pm$ 10.43&25.27 $\pm$ 7.03&0.54 $\pm$ 0.00&8.16&0.16\\
PG1545+210&0.266&10.45 $\pm$ 0.49&0.98 $\pm$ 0.03&50.82 $\pm$ 9.85&14.16 $\pm$ 9.40&0.95 $\pm$ 0.03&9.47&$-$1.17\\
PG1552+085&0.119&10.87 $\pm$ 0.22&0.24 $\pm$ 0.18&31.98 $\pm$ 7.65&5.81 $\pm$ 6.03&0.90 $\pm$ 0.01&7.67&$-$0.10\\
PG1612+261&0.131&10.42 $\pm$ 0.44&0.95 $\pm$ 0.31&40.16 $\pm$ 5.14&3.58 $\pm$ 2.30&0.50 $\pm$ 0.01&8.19&$-$0.60\\
PG1613+658&0.129&11.67 $\pm$ 0.37&1.76 $\pm$ 0.09&39.36 $\pm$ 3.63&15.08 $\pm$ 5.05&0.46 $\pm$ 0.00&9.32&$-$1.61\\
PG1617+175&0.114&9.70 $\pm$ 0.68&0.38 $\pm$ 0.15&51.27 $\pm$ 5.44&16.54 $\pm$ 9.24&0.95 $\pm$ 0.01&8.91&$-$1.20\\
PG1626+554&0.133&11.12 $\pm$ 0.16&0.16 $\pm$ 0.13&39.45 $\pm$ 11.53&12.01 $\pm$ 9.06&1.00 $\pm$ 0.00&8.63&$-$1.18\\
PG1700+518&0.282&11.37 $\pm$ 0.40&2.11 $\pm$ 0.26&41.60 $\pm$ 11.37&15.17 $\pm$ 9.65&0.66 $\pm$ 0.00&8.61&$-$0.02\\
PG1704+608&0.371&11.56 $\pm$ 0.17&2.17 $\pm$ 0.10&61.52 $\pm$ 2.25&30.00 $\pm$ 0.09&0.81 $\pm$ 0.00&9.55&$-$0.98\\
PG2112+059&0.466&10.88 $\pm$ 0.49&2.13 $\pm$ 0.28&47.45 $\pm$ 8.34&29.72 $\pm$ 1.87&0.90 $\pm$ 0.00&9.18&$-$0.12\\
PG2130+099&0.061&10.05 $\pm$ 0.43&1.01 $\pm$ 0.11&46.25 $\pm$ 2.22&5.53 $\pm$ 2.51&0.75 $\pm$ 0.00&8.04&$-$0.60\\
PG2209+184&0.070&11.35 $\pm$ 0.07&0.33 $\pm$ 0.08&24.54 $\pm$ 0.75&1.00 $\pm$ 0.01&0.67 $\pm$ 0.01&8.89&$-$1.55\\
PG2214+139&0.067&11.41 $\pm$ 0.09&0.20 $\pm$ 0.11&33.39 $\pm$ 4.33&3.20 $\pm$ 1.17&0.90 $\pm$ 0.00&8.68&$-$1.15\\
PG2233+134&0.325&10.01 $\pm$ 0.41&1.55 $\pm$ 0.21&60.58 $\pm$ 4.25&20.03 $\pm$ 8.38&0.85 $\pm$ 0.01&8.19&0.01\\
PG2251+113&0.323&11.87 $\pm$ 0.29&1.28 $\pm$ 0.10&45.53 $\pm$ 6.90&14.72 $\pm$ 8.90&0.90 $\pm$ 0.01&9.15&$-$0.59\\
PG2304+042&0.042&11.06 $\pm$ 0.07&$-$0.45 $\pm$ 0.18&57.51 $\pm$ 3.85&8.67 $\pm$ 7.92&0.81 $\pm$ 0.02&8.68&$-$1.74\\
PG2308+098&0.432&11.67 $\pm$ 0.35&1.29 $\pm$ 0.06&32.84 $\pm$ 9.93&6.52 $\pm$ 6.60&1.00 $\pm$ 0.00&9.76&$-$1.11\\
\bottomrule
\end{tabularx}
\end{adjustwidth}
\noindent\footnotesize{{Note}: Column (1): source name. Column (2): redshift. Column (3): stellar mass. Column (4): FIR SFR derived from the cold dust component integrated over $8$--$1000~\upmu\mathrm{m}$. Column (5): dust temperature. Column (6): minimum radiation field intensity. Column (7): fractional AGN contribution to the total infrared luminosity, measured over $8$--$1000~\upmu\mathrm{m}$; when a parameter is returned by multiple SED-fitting templates, we report the average value.
Columns (8) and (9) give the black hole mass and Eddington ratio, respectively. For the SDSS quasars, $M_{\rm BH}$ and $L_{\rm bol}$ are adopted from \citet{shen11}; for the PG quasars, they are adopted from \citet{shangguan2018}. The Eddington ratios are calculated as $\lambda_{\rm Edd}=L_{\rm bol}/L_{\rm Edd}$
}
\end{table}
\vspace{-6pt}

\section{Additional Comparisons of Dust-Related SED-Fitting~Parameters} \label{app:sedparcompare}

\begin{figure}[H]
\begin{minipage}{0.29\textwidth}
    \centering
    \includegraphics[width=\textwidth]{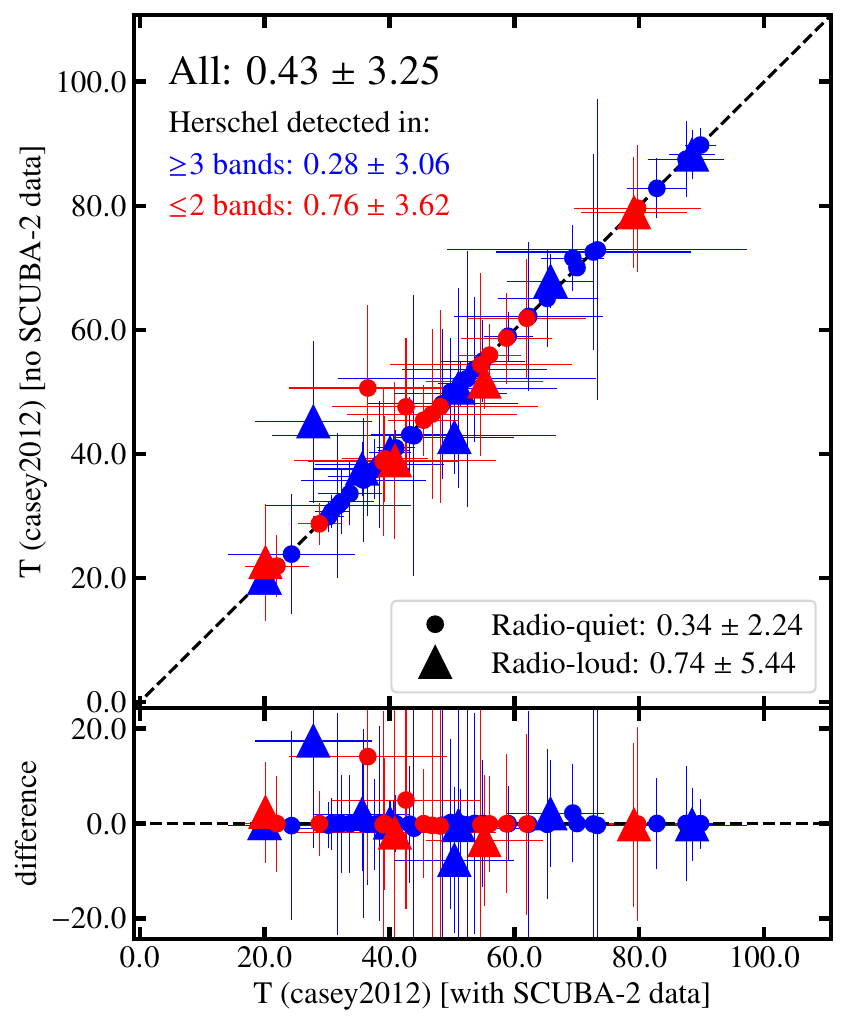}
    (\textbf{a})
\end{minipage}
\hfill
\begin{minipage}{0.29\textwidth}
    \centering
    \includegraphics[width=\textwidth]{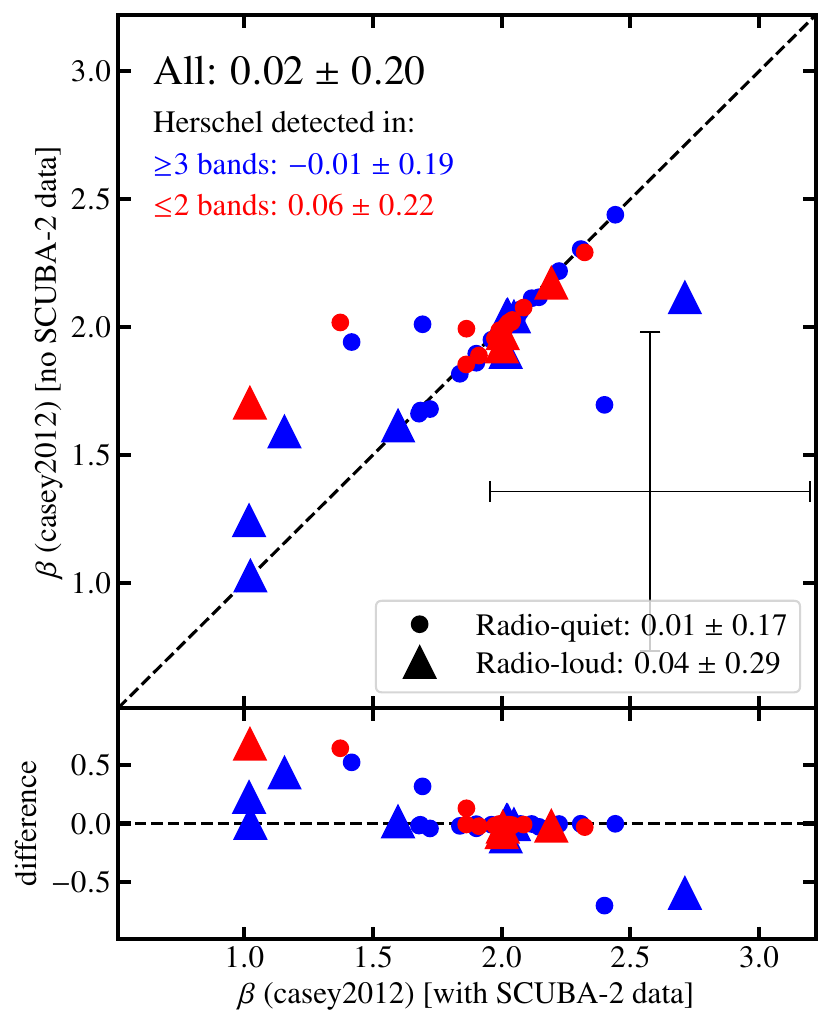}
    (\textbf{b})
\end{minipage}
\hfill
\begin{minipage}{0.29\textwidth}
    \centering
    \includegraphics[width=\textwidth]{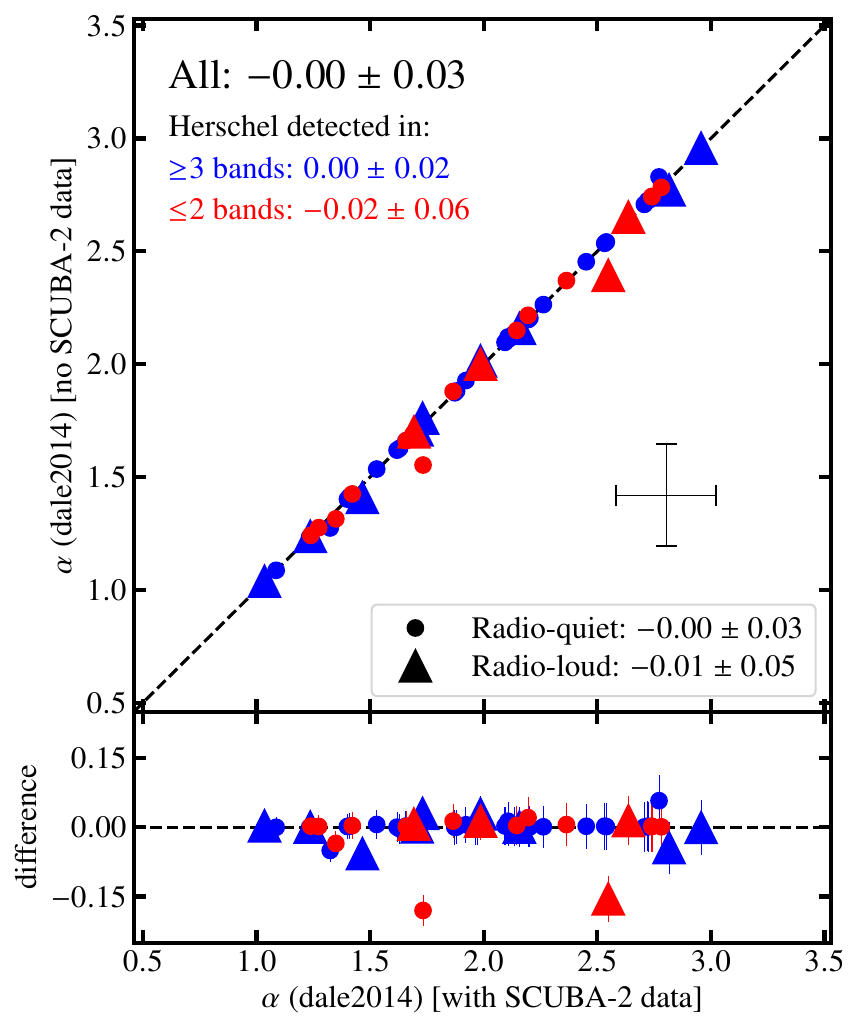}
    (\textbf{c})
\end{minipage}
\caption{
{Comparison} 
of the dust-related parameters derived with and without the SCUBA-2 data: (\textbf{a}) dust temperature, $T_{\rm dust}$, from the casey2012 template; (\textbf{b}) dust emissivity index, $\beta$, from the casey2012 template; and (\textbf{c}) dust-heating intensity parameter, $\alpha$, from the dale2014 template. The lower subpanel in each panel shows the difference between the two estimates. The symbols and colors are the same as in Figure~\ref{fig:firsfrwithornotscuba2}.
}
\label{fig:app_casey_dale_scuba2}
\end{figure}

\begin{figure}[H]
\begin{minipage}{0.29\textwidth}
    \centering
    \includegraphics[width=\textwidth]{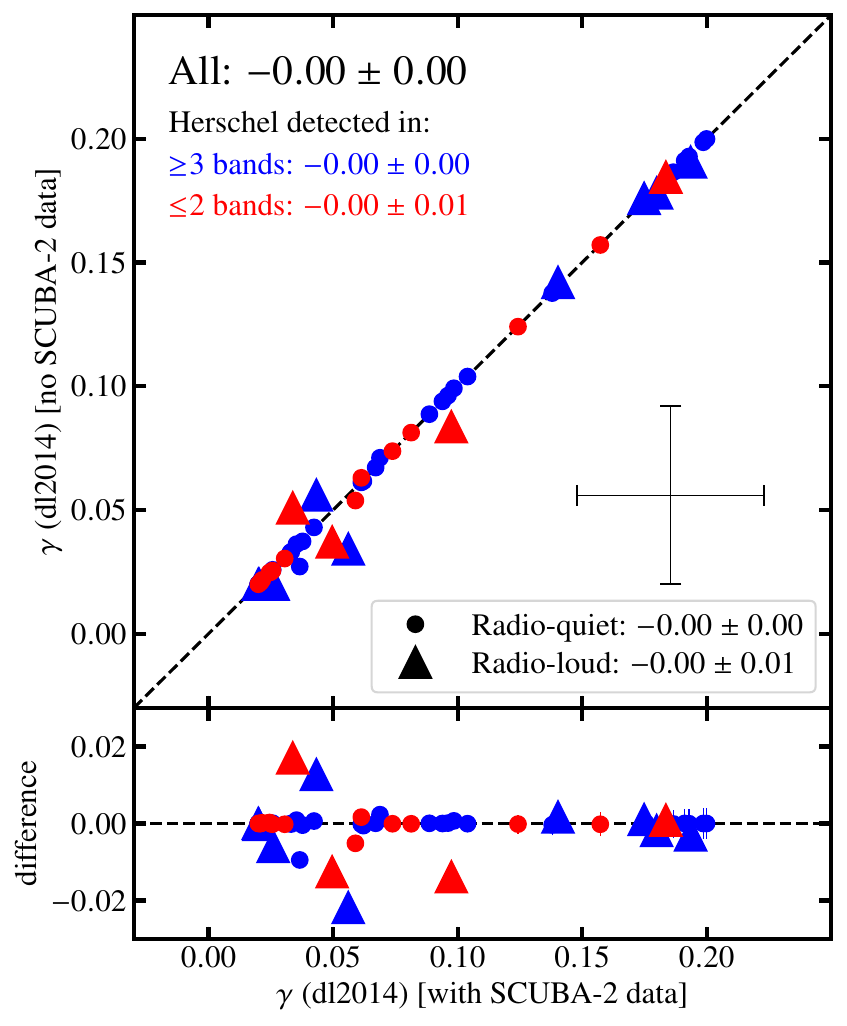}
    (\textbf{a})
\end{minipage}
\hfill
\begin{minipage}{0.29\textwidth}
    \centering
    \includegraphics[width=\textwidth]{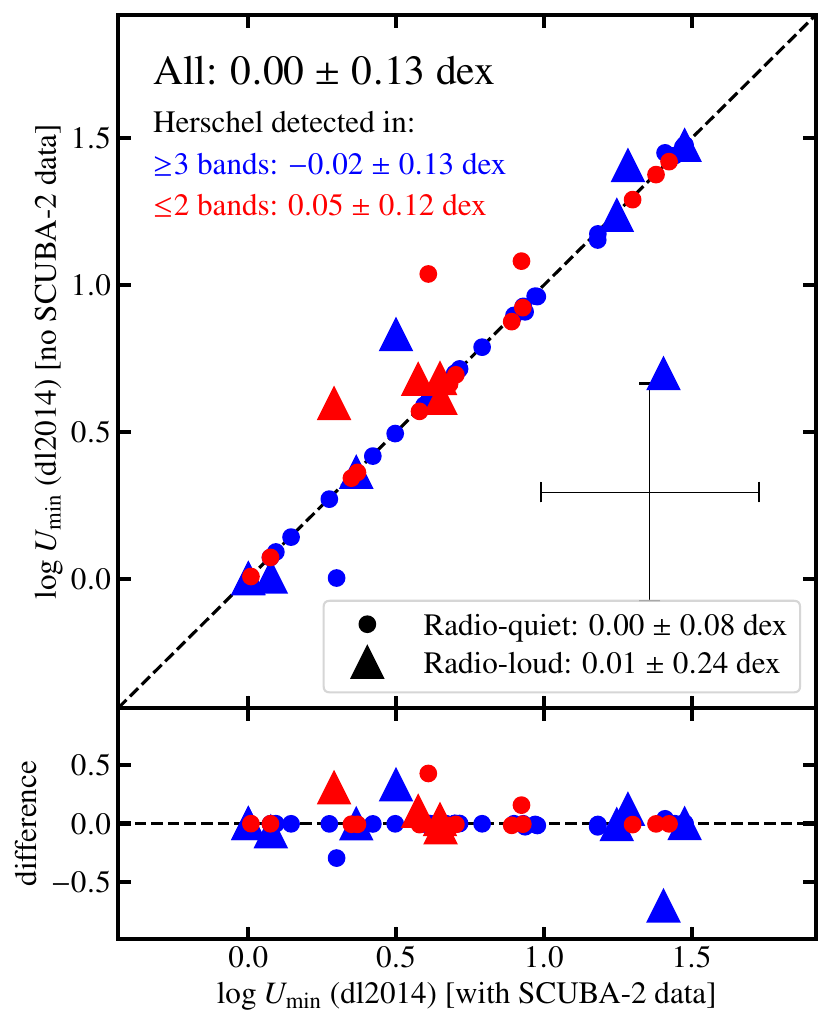}
    (\textbf{b})
\end{minipage}
\hfill
\begin{minipage}{0.29\textwidth}
    \centering
    \includegraphics[width=\textwidth]{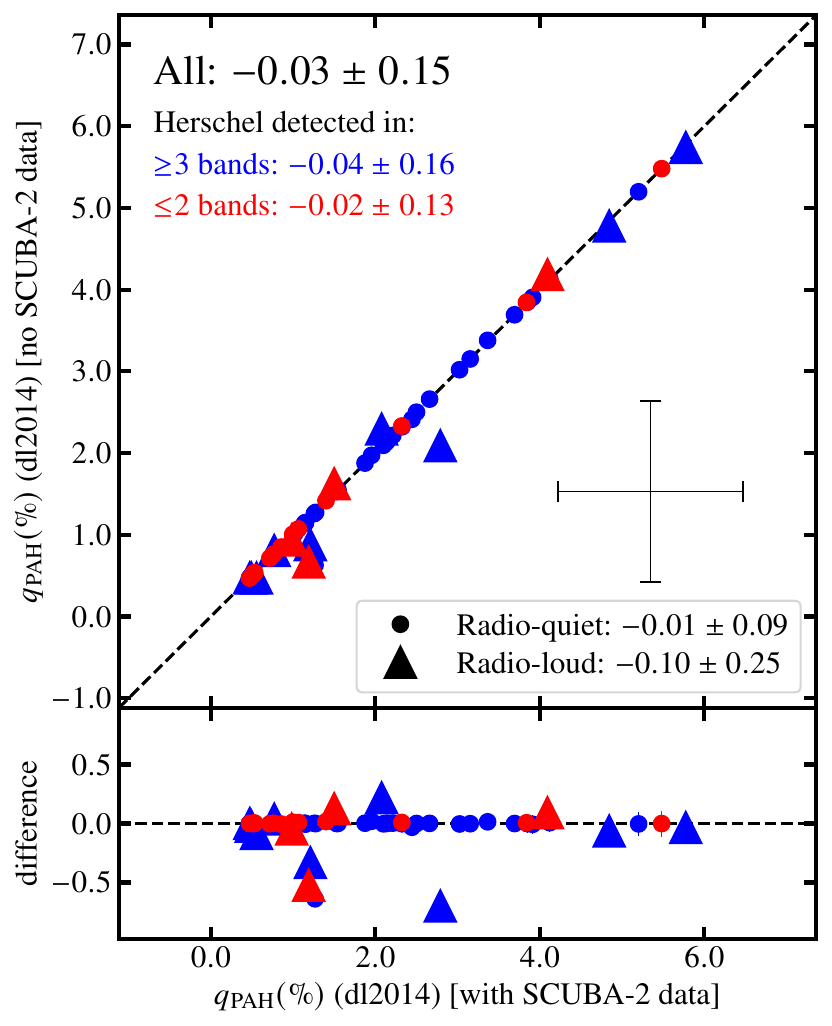}
    (\textbf{c})
\end{minipage}

\vspace{2mm}

\begin{minipage}{0.29\textwidth}
    \centering
    \includegraphics[width=\textwidth]{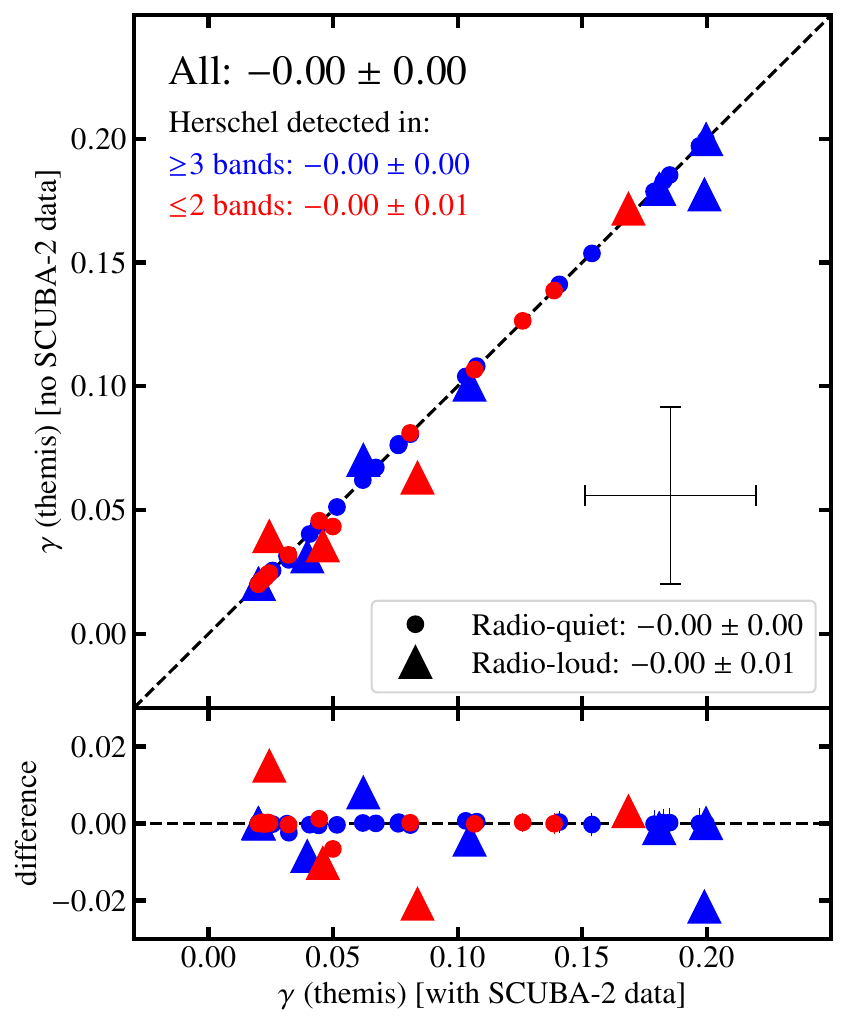}
    (\textbf{d})
\end{minipage}
\hfill
\begin{minipage}{0.29\textwidth}
    \centering
    \includegraphics[width=\textwidth]{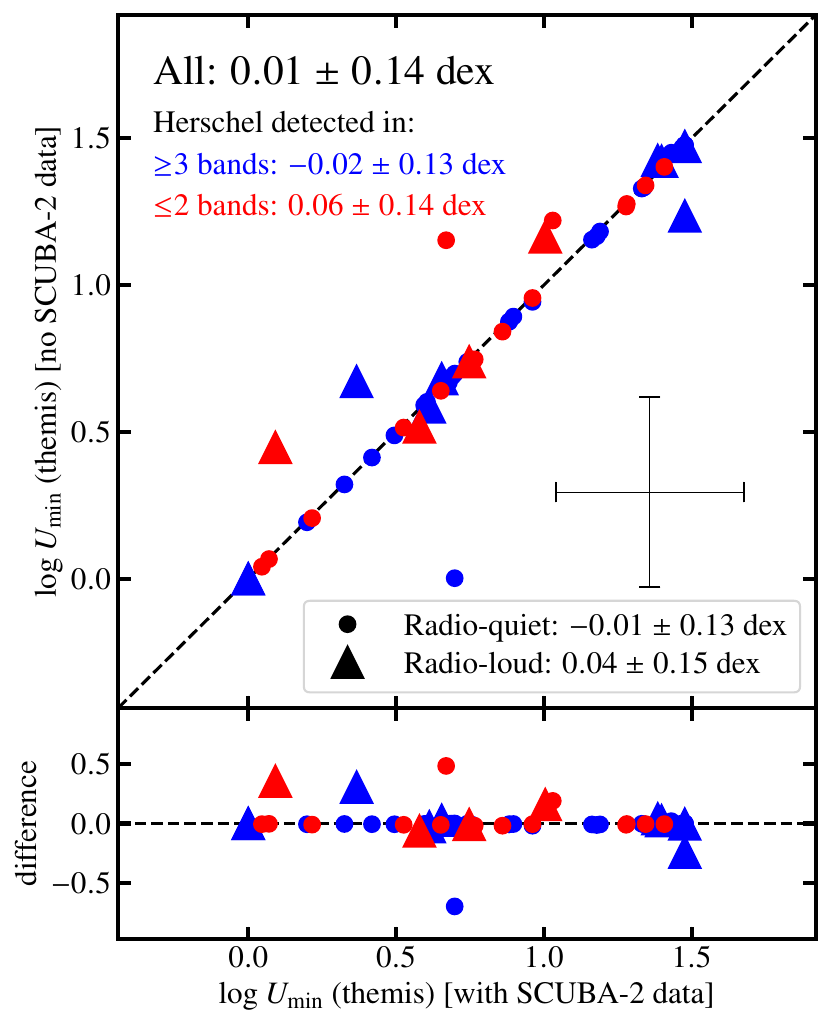}
    (\textbf{e})
\end{minipage}
\hfill
\begin{minipage}{0.29\textwidth}
    \centering
    \includegraphics[width=\textwidth]{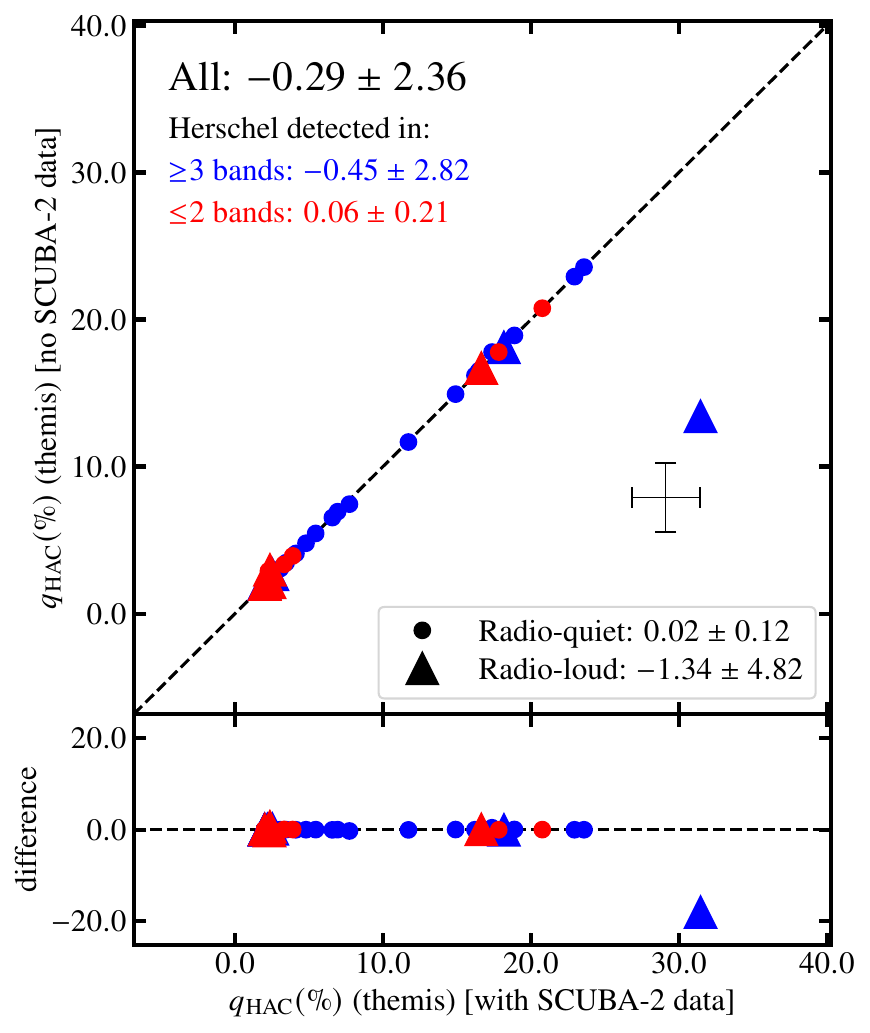}
    (\textbf{f})
\end{minipage}

\caption{
{Comparison} 
of the dust-related parameters derived with and without the SCUBA-2 data:
(\textbf{a}) $\gamma$ from the dl2014 template;
(\textbf{b}) $\log U_{\rm min}$ from the dl2014 template;
(\textbf{c}) $q_{\rm PAH}$ from the dl2014 template;
(\textbf{d}) $\gamma$ from the THEMIS template;
(\textbf{e}) $\log U_{\rm min}$ from the THEMIS template; and
(\textbf{f}) $q_{\rm HAC}$ from the THEMIS template. 
The lower subpanel in each panel shows the difference between the two estimates. The symbols and colors are the same as in Figure~\ref{fig:firsfrwithornotscuba2}.
}
\label{fig:app_dl2014_themis_scuba2}
\end{figure}

\begin{figure}[H]
\vspace{+4pt}
\begin{minipage}{0.45\textwidth}
    \centering
    \includegraphics[width=\textwidth]{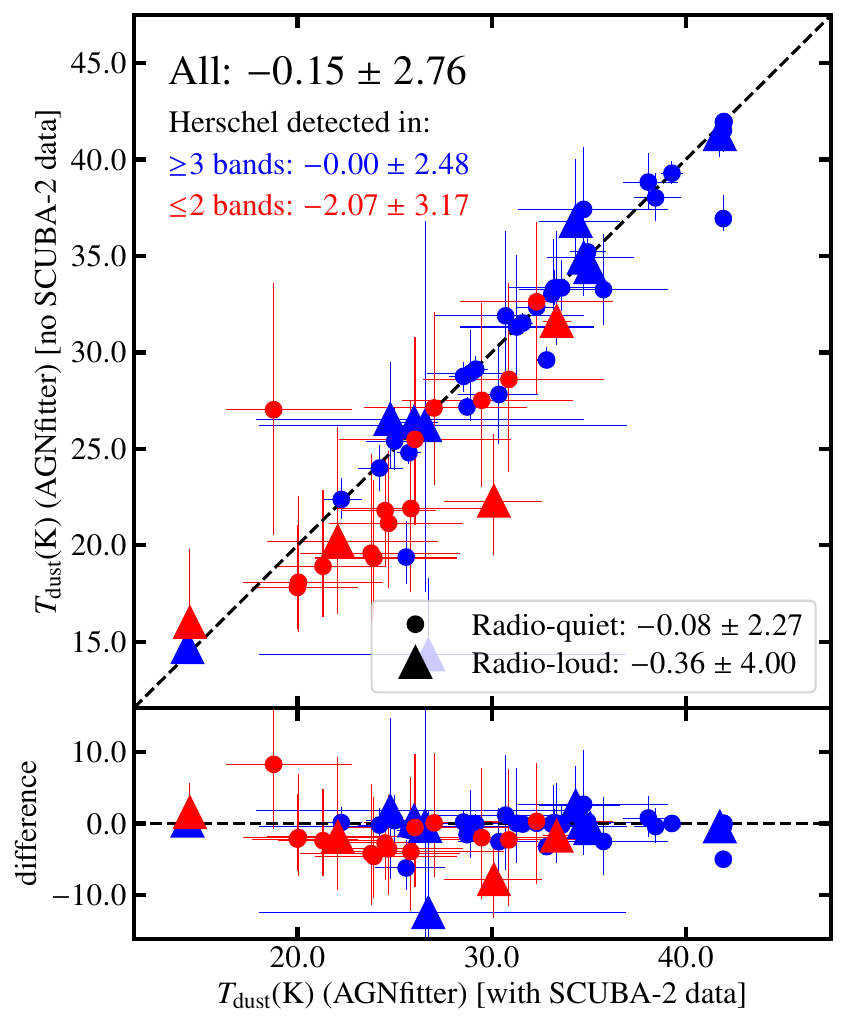}
    (\textbf{a})
\end{minipage}
\hfill
\begin{minipage}{0.45\textwidth}
    \centering
    \includegraphics[width=\textwidth]{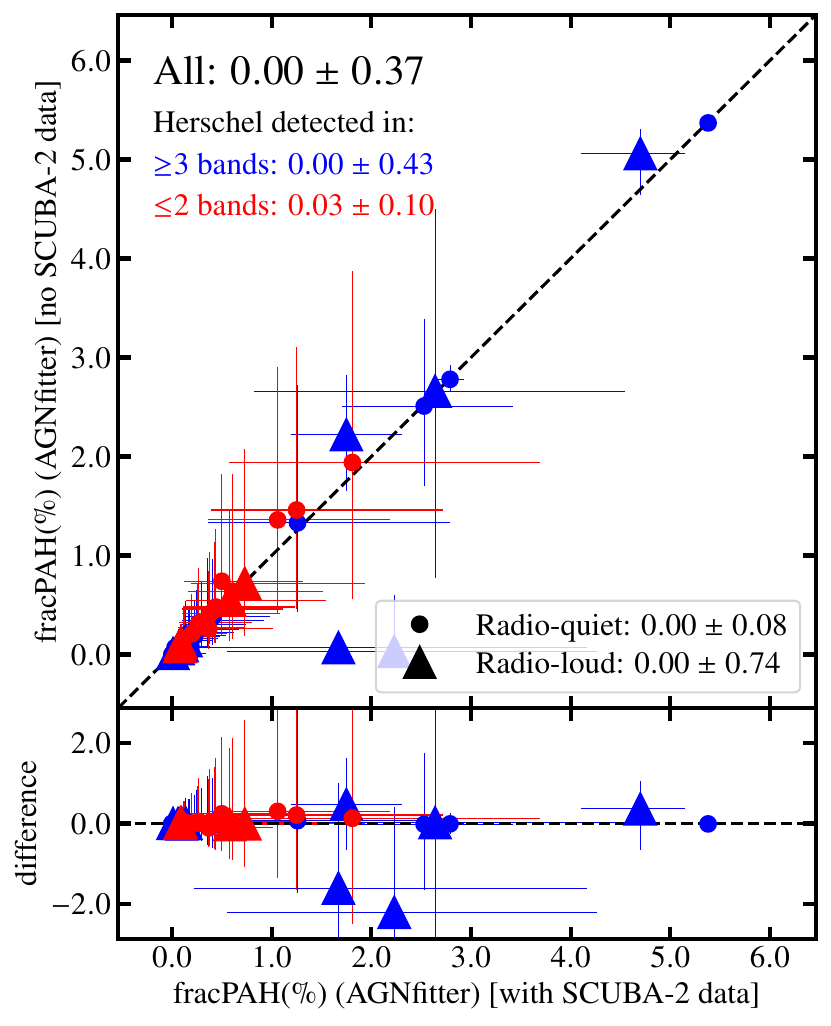}
    (\textbf{b})
\end{minipage}

\caption{
{Comparison} 
of the AGNfitter dust-related parameters derived with and without the SCUBA-2 data:
(\textbf{a}) dust temperature, $T_{\rm dust}$; and
(\textbf{b}) the PAH fraction, fracPAH. 
The lower subpanel in each panel shows the difference between the two estimates. The symbols and colors are the same as in Figure~\ref{fig:firsfrwithornotscuba2}.
}
\label{fig:app_agnfitter_scuba2}
\end{figure}

\textls[-25]{In this appendix, we present additional comparisons of the dust-related parameters derived with and without the SCUBA-2 data. The comparisons include the parameters from the CIGALE cold dust templates and AGNfitter. 
Figures \ref{fig:app_casey_dale_scuba2}--\ref{fig:app_agnfitter_scuba2} show the detailed comparisons.
In all panels, the symbols and colors are the same as in Figure \ref{fig:firsfrwithornotscuba2}.
Red and blue symbols denote sources detected in no more than two and in at least three of the six Herschel bands, respectively. Triangles and circles represent radio-loud and radio-quiet sources, respectively.} 

\textls[-25]{For the casey2012 template, excluding the SCUBA-2 data leads to
slightly higher dust temperatures and emissivity indices, with mean offsets of 0.43~K and 0.02 for the full sample, respectively. The offsets are larger for sources with limited Herschel detections and for radio-loud quasars,
with maximum differences reaching $\sim$18~K in $T_{\rm dust}$ and $\sim$0.7 in $\beta$.
For the dale2014 template, the mean offset in $\alpha$ is negligible for the full sample, but some sources with sparse Herschel coverage show larger deviations.
For the dl2014 and THEMIS templates, the mean offsets in $\gamma$ are close to zero for the full sample, although individual radio-loud sources can show noticeable differences. The effect on $\log U_{\rm min}$ is also small on average, but sources with fewer Herschel detections show larger scatter. The PAH- or HAC-related parameters are only weakly affected for the full sample, while radio-loud quasars exhibit larger object-to-object variations. 
For AGNfitter, excluding the SCUBA-2 data produces a small mean offset in $T_{\rm dust}$ for the full sample, but the offset becomes larger for sources with limited Herschel detections. The mean effect on fracPAH is small, although some radio-loud quasars show larger differences. }

\vspace{-12pt}

Overall, these comparisons confirm that the SCUBA-2 data have little effect on the average dust parameter estimates for the full sample, but they can provide important constraints for individual sources with poor far-infrared coverage or radio-loud emission.

\section{Kendall--$\tau$ Tests for Infrared-Based AGN Indicators}
\label{app:torus_kendall}

Table~\ref{tab:torus_kendall} lists the Kendall--$\tau$ test results for the relations between $\Delta_{\rm MS}$ and the AGNfitter torus indicators. For $L_{\rm tor}$, significant positive correlations are found in three redshift bins, while the $0.3<z<0.5$ bin shows no significant trend. In contrast, $\Delta_{\rm MS}$ is positively correlated with $L_{\rm tor}/L_{\rm Edd}$ in all four redshift bins.

We also tested the residual redshift dependence of $\log(L_{\rm tor}/L_{\rm Edd})$ within each redshift bin. No significant correlation with redshift is found, with $p$-values of 0.25, 0.31, 0.24, and 0.18 from the lowest- to highest-redshift bins. Therefore, the $\Delta_{\rm MS}$--$L_{\rm tor}/L_{\rm Edd}$ relation is not readily explained by a simple residual redshift trend within the bins.

\begin{table}[H]
\caption{{Kendall}
--$\tau$ test results for the relations between the main-sequence offset and AGNfitter torus indicators. The tests are based on individual sources within each redshift bin.}
\label{tab:torus_kendall}
\small
\begin{tabularx}{\textwidth}{LCCCCC}
\toprule
\textbf{Redshift Bin} & \boldmath{$N$} &
\multicolumn{2}{c}{\boldmath{$\Delta_{\rm MS}$}\textbf{--}\boldmath{$L_{\rm tor}$}} &
\multicolumn{2}{c}{\boldmath{$\Delta_{\rm MS}$}\textbf{--}\boldmath{$L_{\rm tor}/L_{\rm Edd}$}} \\
\cmidrule{3-6}
 & & \boldmath{$\tau$} & \boldmath{$p$} & \boldmath{$\tau$}& \boldmath{$p$} \\
\midrule
$z<0.1$       & 38 & 0.45 & $8.3\times10^{-5}$ & 0.30 & 0.009 \\
$0.1<z<0.3$   & 68 & 0.24 & 0.004              & 0.23 & 0.005 \\
$0.3<z<0.5$   & 49 & 0.16 & 0.110              & 0.39 & $1.1\times10^{-4}$ \\
$z>0.5$       & 47 & 0.30 & 0.003              & 0.21 & 0.044 \\
\bottomrule
\end{tabularx}
\end{table}

\section{Additional Stellar Mass Tests}
\label{app:mstar_tests}

\begin{figure}[H]
\begin{minipage}{0.48\textwidth}
    \centering
    \includegraphics[width=\textwidth]{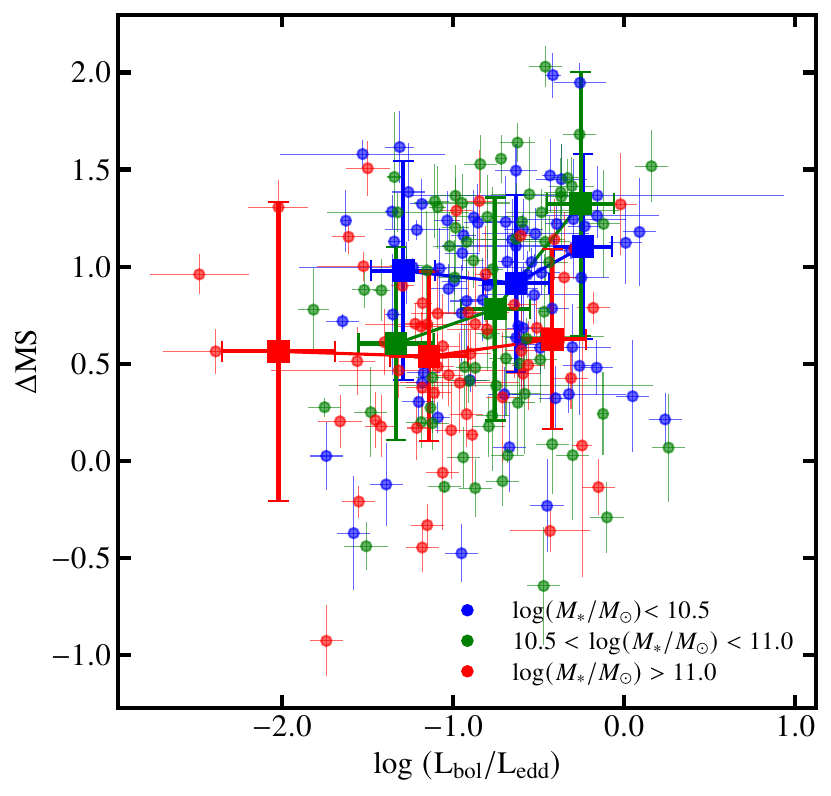}
    (\textbf{a})
\end{minipage}
\hfill
\begin{minipage}{0.48\textwidth}
    \centering
    \includegraphics[width=\textwidth]{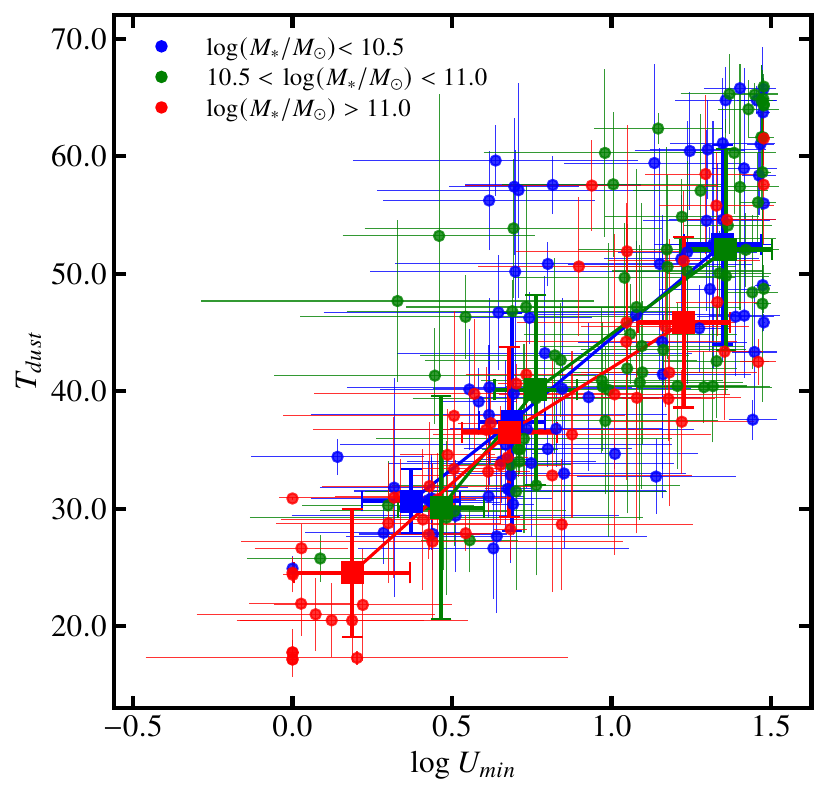}
    (\textbf{b})
\end{minipage}
\caption{\textls[-15]{{Additional} 
 stellar mass binned tests. 
(\textbf{a}) Relation between the main-sequence offset, $\Delta_{\rm MS}$, and Eddington ratio. 
(\textbf{b}) Relation between dust temperature, $T_{\rm dust}$, and the minimum radiation field intensity, $U_{\min}$. 
Blue, green, and red symbols represent sources with $\log(M_\ast/M_\odot)<10.5$, \mbox{$10.5<\log(M_\ast/M_\odot)<11.0$}, and $\log(M_\ast/M_\odot)>11.0$, respectively. 
Large squares and connecting lines show the median values in bins along the abscissa, and the error bars indicate the corresponding~scatter.}}
\label{fig:mstar_appendix}
\end{figure}

In Section~\ref{subsec:stellarmassdiscussion}, we discussed the possible effects of stellar mass on the relations involving star formation and dust heating. Here, we present two additional stellar mass binned tests. Figure~\ref{fig:mstar_appendix}a shows the relation between $\Delta_{\rm MS}$ and Eddington ratio, and Figure~\ref{fig:mstar_appendix}b shows the relation between $T_{\rm dust}$ and $U_{\min}$.

\section{Robustness Tests with SPIRE-Quality Subsamples}
\label{app:spiretests}

In this appendix, we compare the full SDSS FIR-constrained sample with two
SPIRE-quality subsamples defined in Section~\ref{subsec:spiretest}. The left panels show the subsample with $S/N_{250}\geq5$, while the right panels show the subsample with detections at $S/N\geq3$ in at least two SPIRE bands. Gray open circles represent the full SDSS sample. Blue-filled circles and orange-filled squares represent the two SPIRE-quality subsamples, respectively.

\begin{figure}[H]
\includegraphics[width=\textwidth]{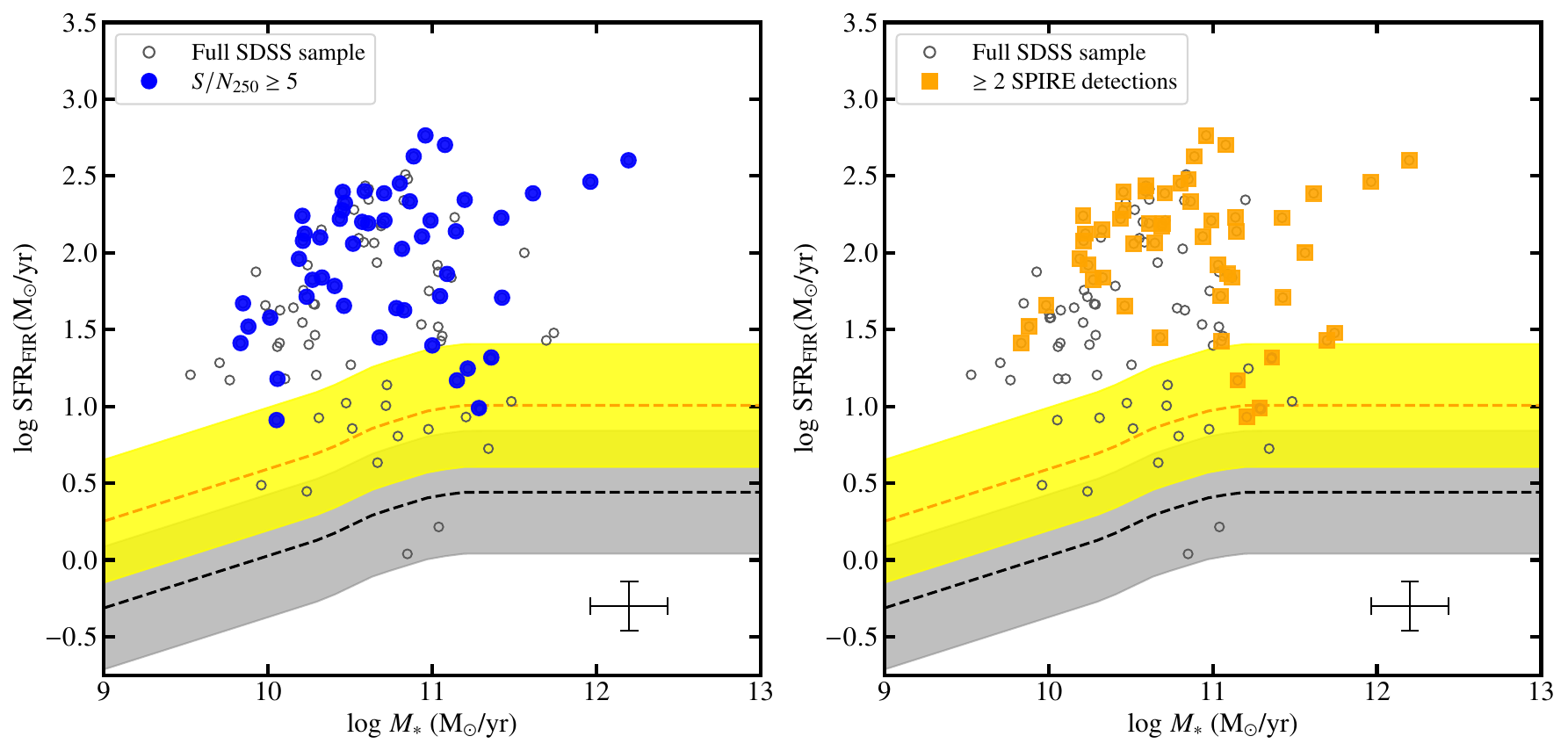}
\caption{Comparison of the FIR SFR--stellar mass relation between the full SDSS FIR-constrained sample and the two SPIRE-quality subsamples. The dashed curves and shaded regions show the adopted star-forming main sequence and its scatter. Gray open circles show the full SDSS sample. Blue-filled circles in the left panel represent sources with $S/N_{250}\geq5$, while orange-filled squares in the right panel represent sources detected at $S/N\geq3$ in at least two SPIRE bands.}
\label{fig:mainsequence_2subsample}
\end{figure}

\begin{figure}[H]
\vspace{+2pt}
\includegraphics[width=\textwidth]{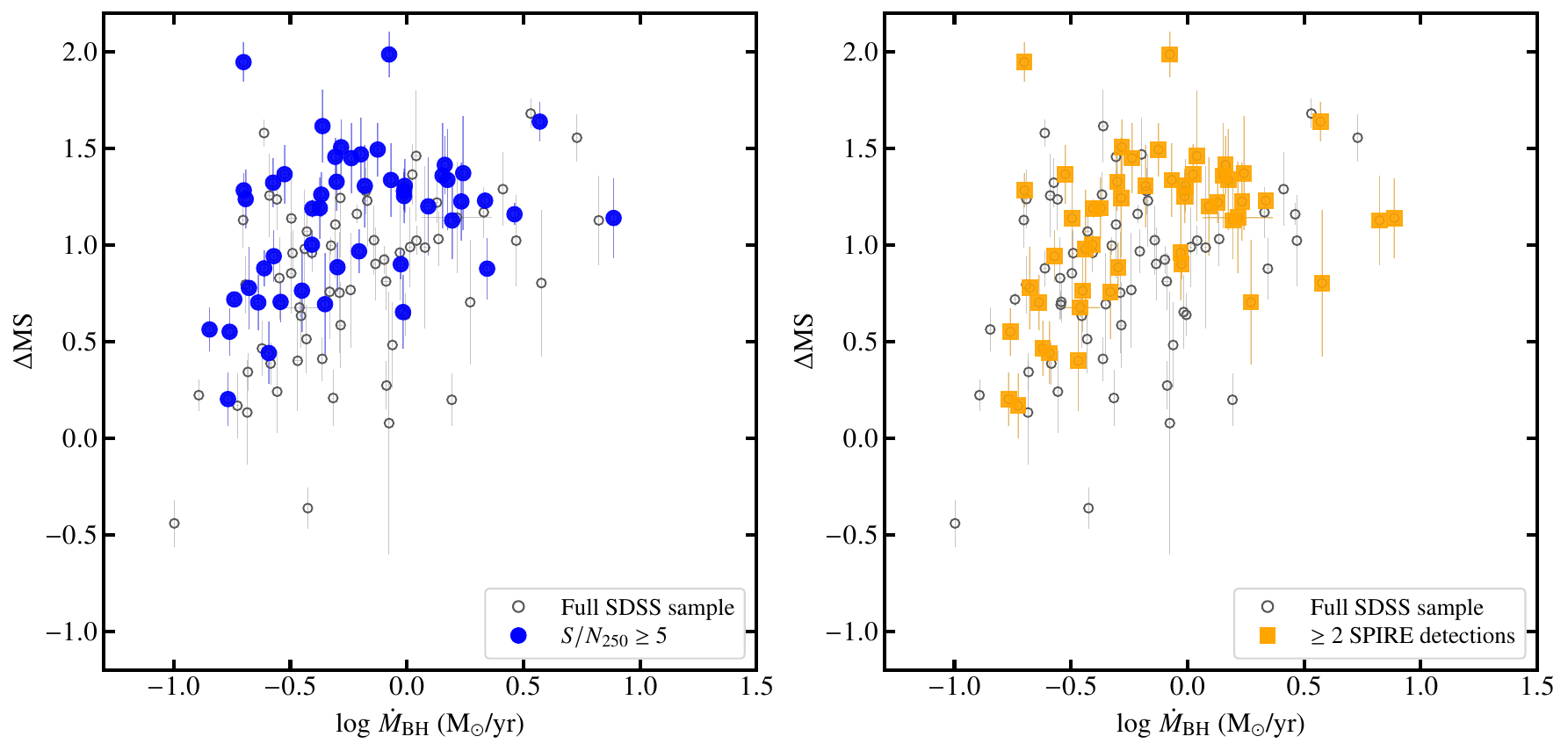}
\caption{Same as Figure~\ref{fig:mainsequence_2subsample} but for the relation between the main-sequence offset, $\Delta_{\rm MS}$, and black hole accretion rate.}
\label{fig:bhar_deltams_2subsample}
\end{figure}
\vspace{-6pt}
\begin{figure}[H]
\centering
\includegraphics[width=\textwidth]{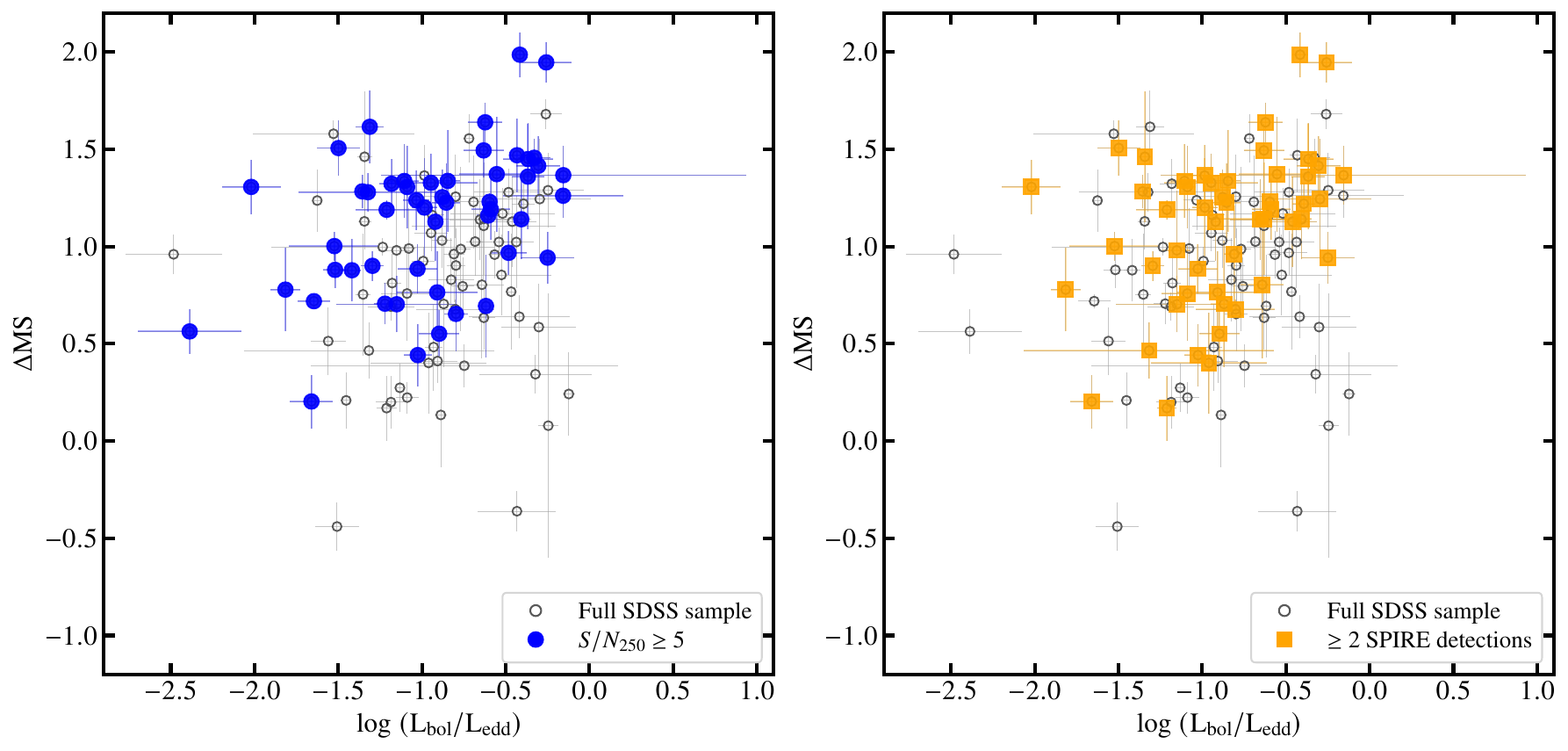}
\caption{Same as Figure~\ref{fig:mainsequence_2subsample} but for the relation between $\Delta_{\rm MS}$ and Eddington ratio.}
\label{fig:Leddratio_deltams_2subsample}
\end{figure}
\vspace{-6pt}
\begin{figure}[H]
\centering
\includegraphics[width=\textwidth]{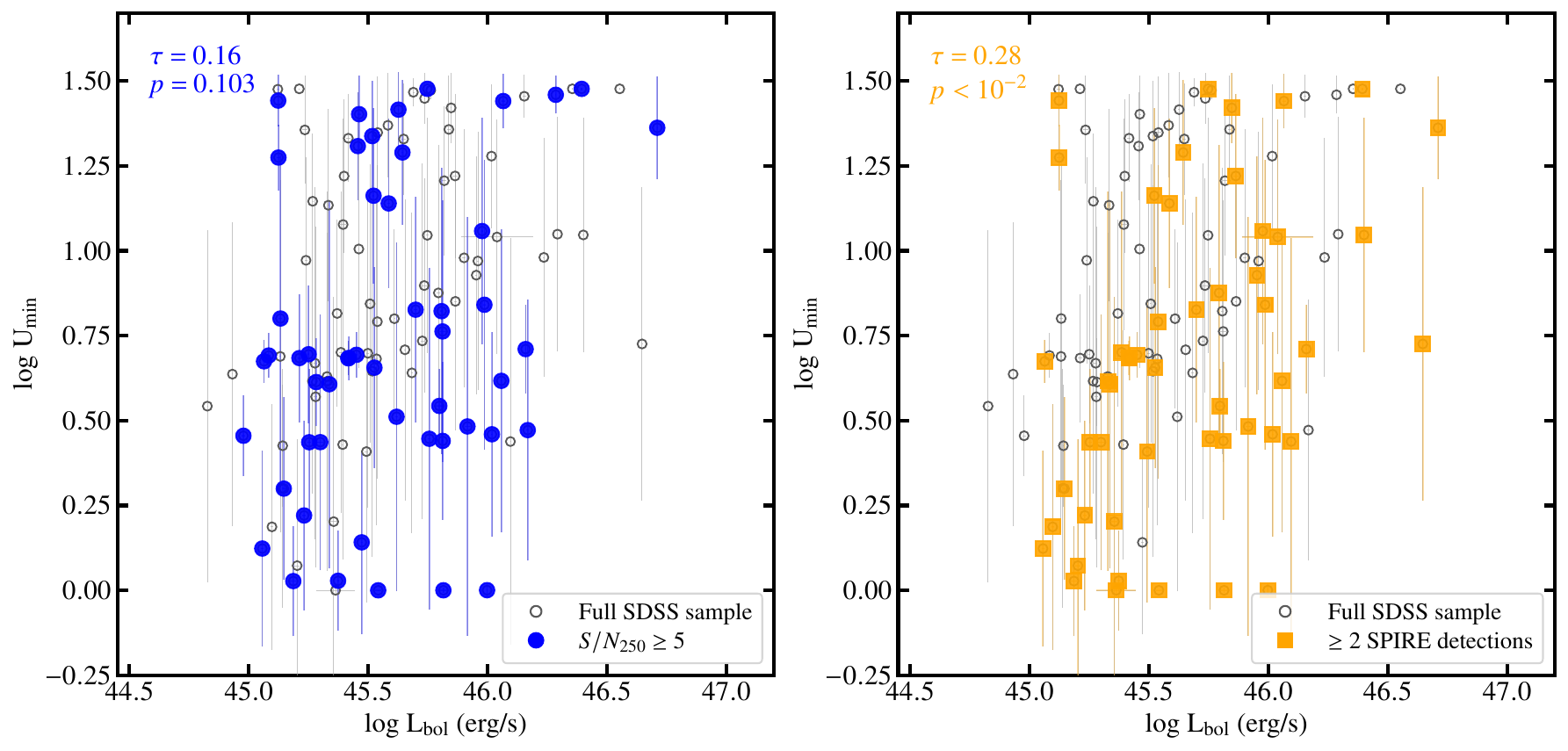}
\caption{Same as Figure~\ref{fig:mainsequence_2subsample} but for the relation between the minimum radiation field intensity, $U_{\min}$, and bolometric luminosity. The Kendall--$\tau$ coefficient and corresponding $p$ value are calculated for the highlighted SPIRE-quality subsample in each panel.} 
\label{fig:Lbol_Umin_2subsample}
\end{figure}

\section{WISE-to-Herschel Color Diagnostics}
\label{app:wisehscolor}

\begin{figure}[H]
\begin{minipage}{0.31\textwidth}
    \centering
    \includegraphics[width=\textwidth]{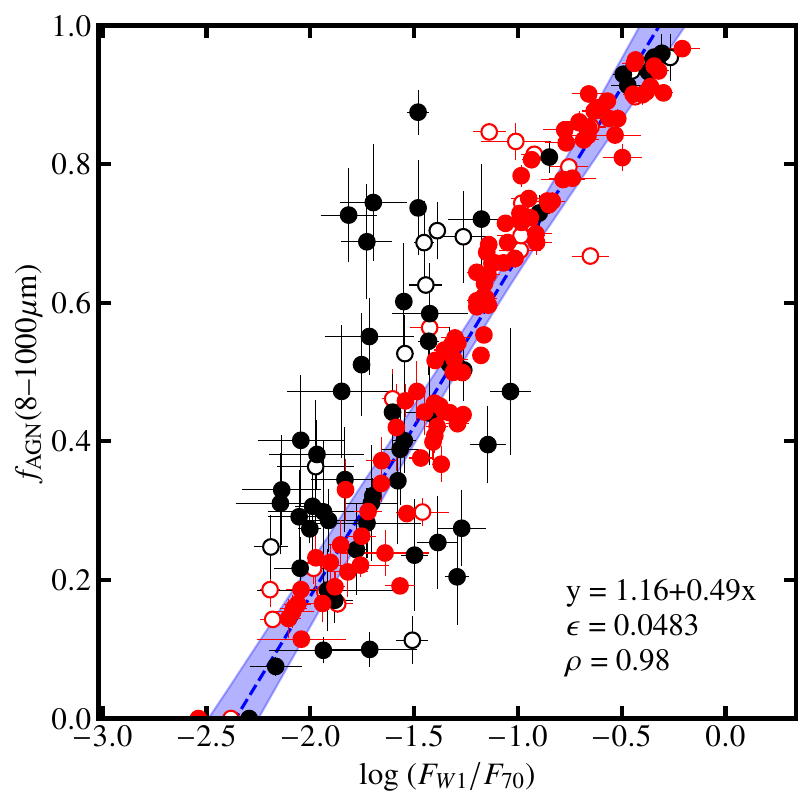}
    (\textbf{a})
\end{minipage}
\hfill
\begin{minipage}{0.31\textwidth}
    \centering
    \includegraphics[width=\textwidth]{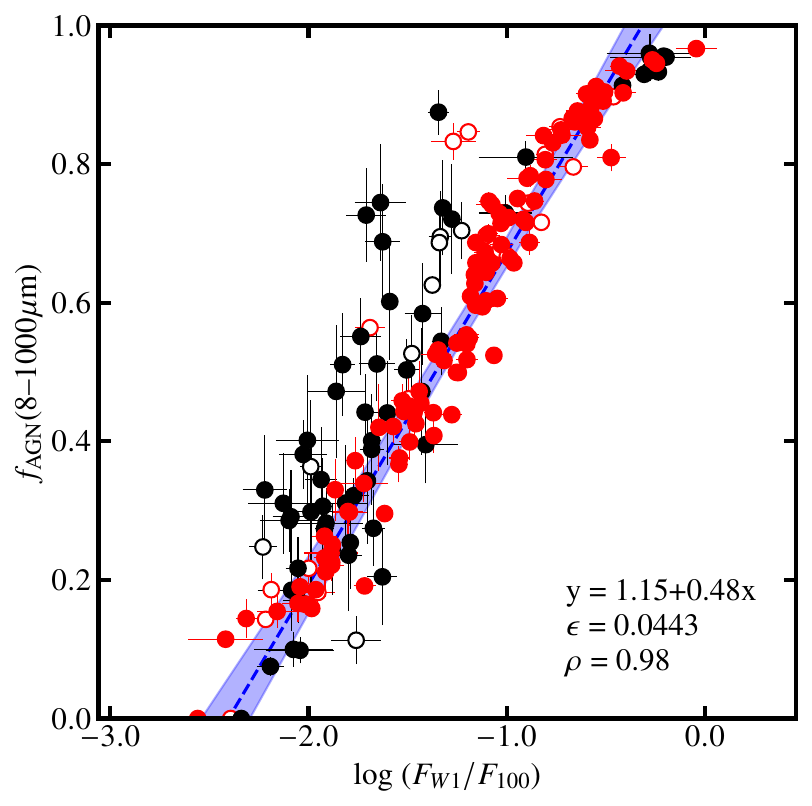}
    (\textbf{b})
\end{minipage}
\hfill
\begin{minipage}{0.31\textwidth}
    \centering
    \includegraphics[width=\textwidth]{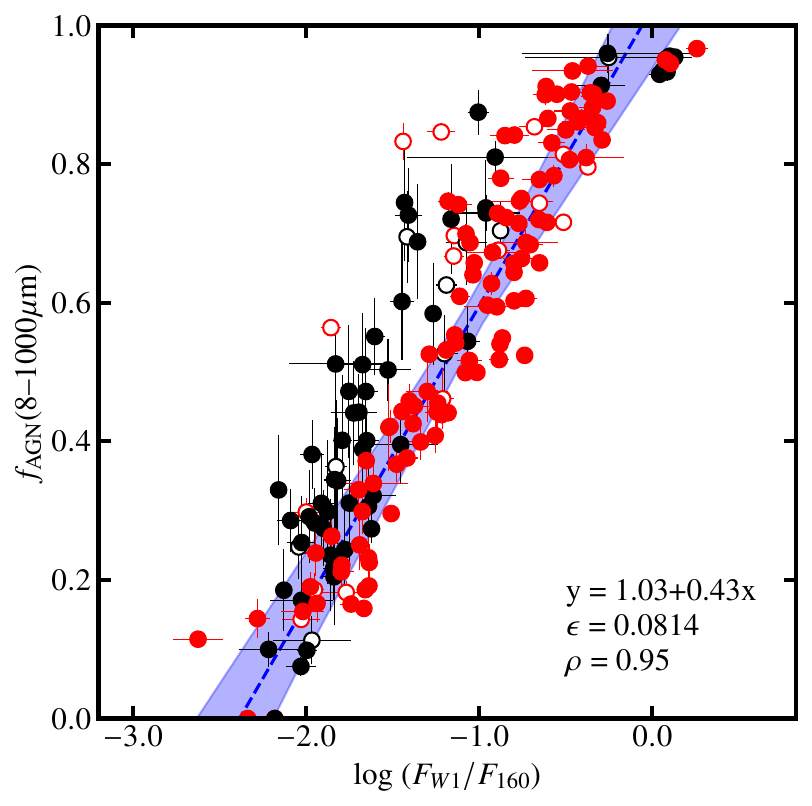}
    (\textbf{c})
\end{minipage}

\vspace{2mm}

\begin{minipage}{0.31\textwidth}
    \centering
    \includegraphics[width=\textwidth]{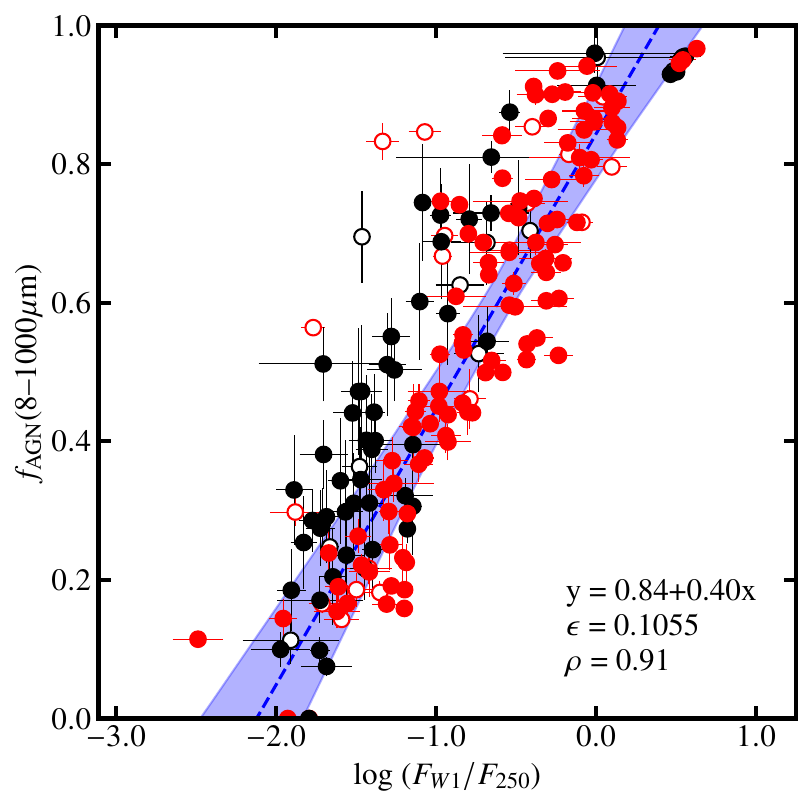}
    (\textbf{d})
\end{minipage}
\hfill
\begin{minipage}{0.31\textwidth}
    \centering
    \includegraphics[width=\textwidth]{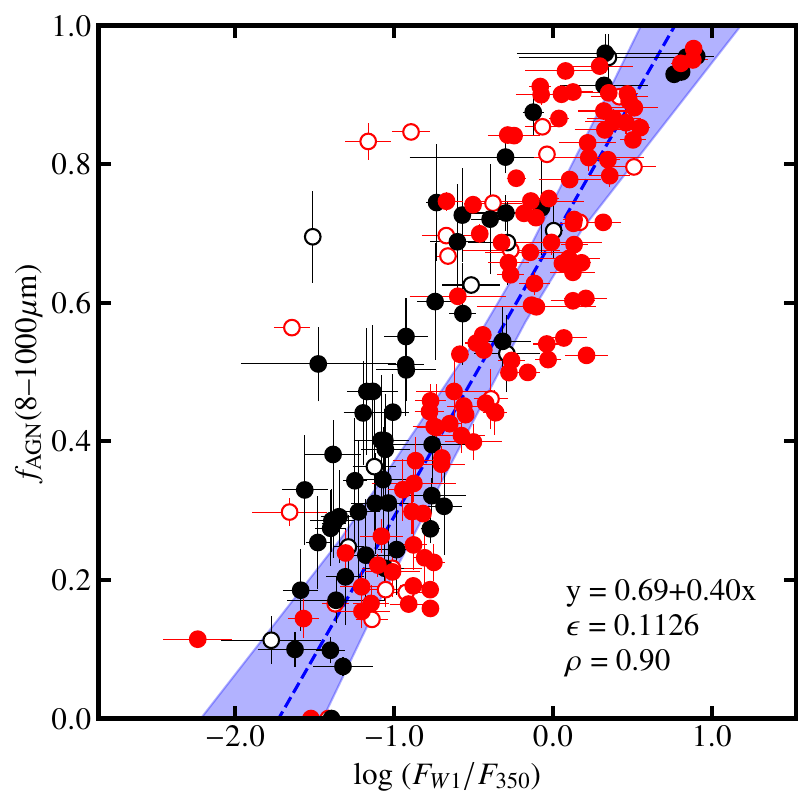}
    (\textbf{e})
\end{minipage}
\hfill
\begin{minipage}{0.31\textwidth}
    \centering
    \includegraphics[width=\textwidth]{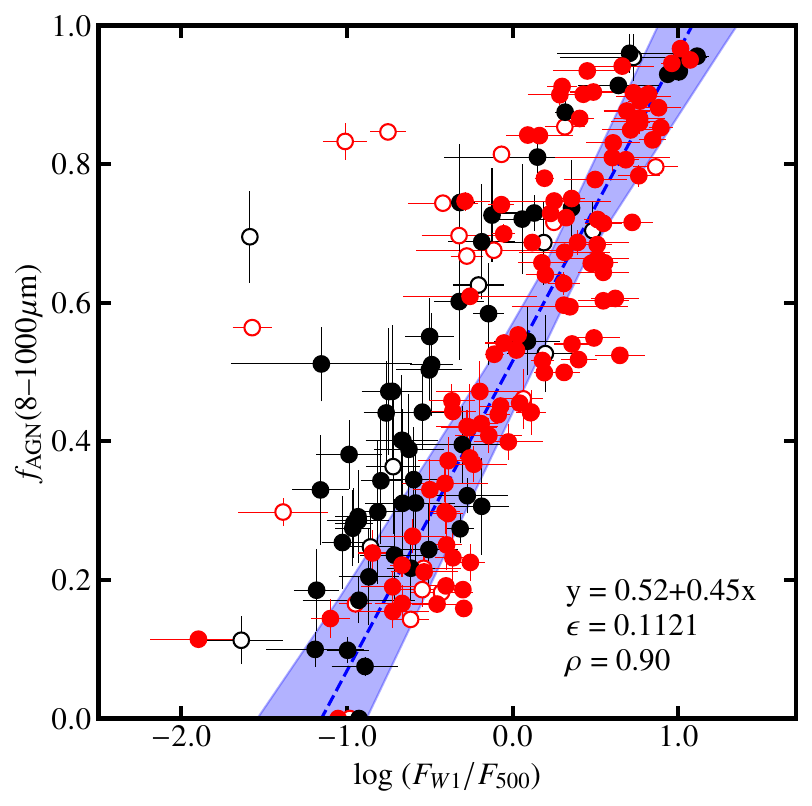}
    (\textbf{f})
\end{minipage}

\caption{{Same} 
as Figure~\ref{fig:strongestcolorfagn} but for WISE W1 and the six Herschel bands: 
(\textbf{a}) W1/70~$\upmu\mathrm{m}$,
(\textbf{b})~W1/100~$\upmu\mathrm{m}$,
(\textbf{c}) W1/160~$\upmu\mathrm{m}$,
(\textbf{d}) W1/250~$\upmu\mathrm{m}$,
(\textbf{e}) W1/350~$\upmu\mathrm{m}$, and
(\textbf{f}) W1/500~$\upmu\mathrm{m}$.
Red and black symbols denote sources detected in at least three and in no more than two Herschel bands, respectively. Filled and open symbols represent radio-quiet and radio-loud quasars, respectively. The dashed line and blue shaded region show the Bayesian linear regression fit and its uncertainty for the radio-quiet sources detected in at least three Herschel bands. The fitted relation, intrinsic scatter $\epsilon$, and correlation coefficient $\rho$ are shown in each panel.}
\label{fig:w1hscolorfagn}
\end{figure}

\begin{figure}[H]
\begin{minipage}{0.31\textwidth}
    \centering
    \includegraphics[width=\textwidth]{img/wisehscolor_fagn_W3_70_hyphentominus_correctendash.pdf}
    (\textbf{a})
\end{minipage}
\hfill
\begin{minipage}{0.31\textwidth}
    \centering
    \includegraphics[width=\textwidth]{img/wisehscolor_fagn_W3_100_hyphentominus_correctendash.pdf}
    (\textbf{b})
\end{minipage}
\hfill
\begin{minipage}{0.31\textwidth}
    \centering
    \includegraphics[width=\textwidth]{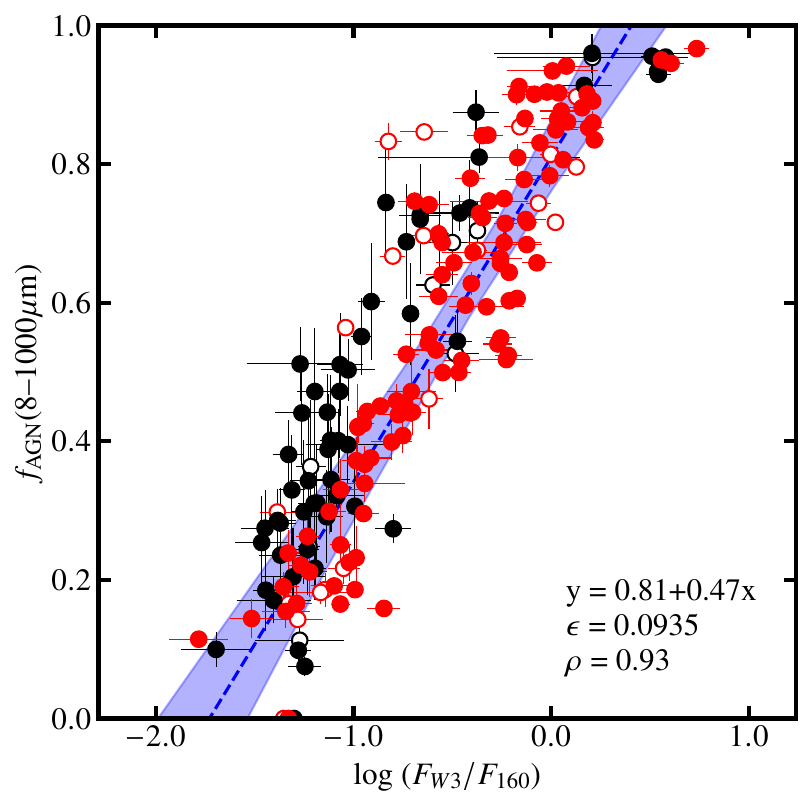}
    (\textbf{c})
\end{minipage}

\vspace{2mm}

\begin{minipage}{0.31\textwidth}
    \centering
    \includegraphics[width=\textwidth]{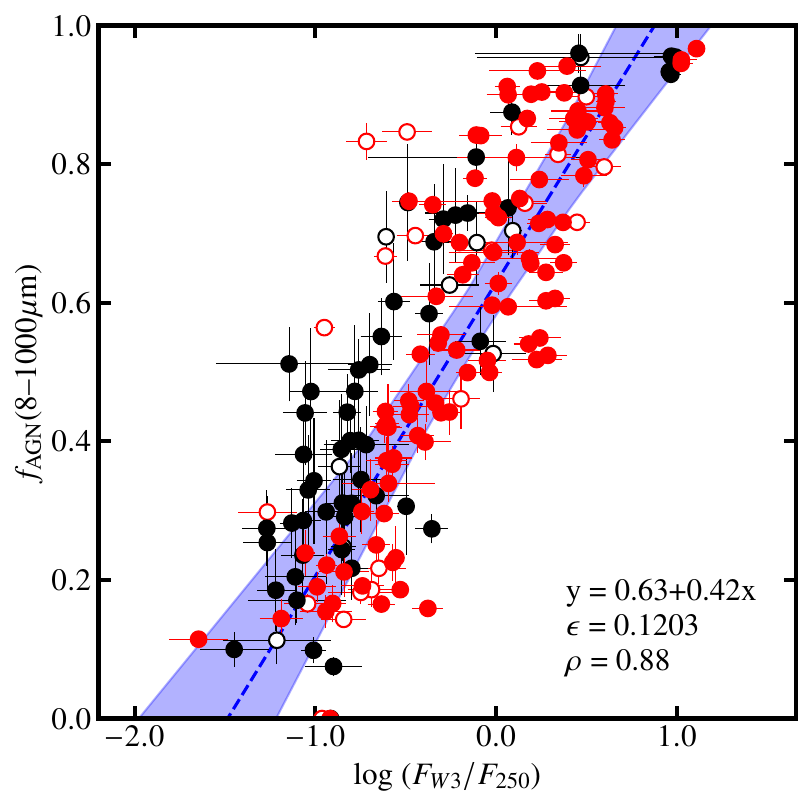}
    (\textbf{d})
\end{minipage}
\hfill
\begin{minipage}{0.31\textwidth}
    \centering
    \includegraphics[width=\textwidth]{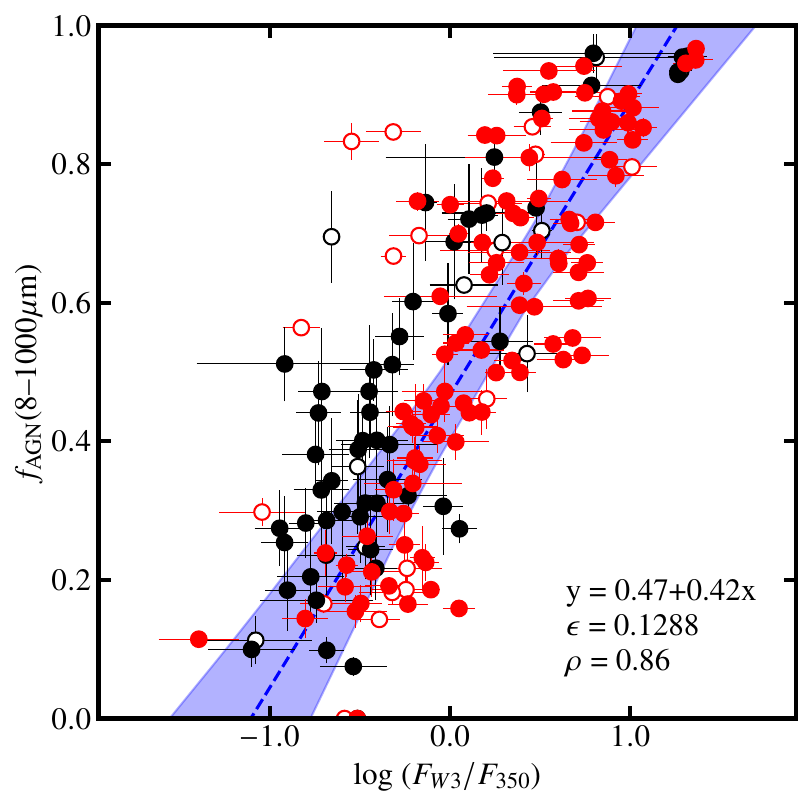}
    (\textbf{e})
\end{minipage}
\hfill
\begin{minipage}{0.31\textwidth}
    \centering
    \includegraphics[width=\textwidth]{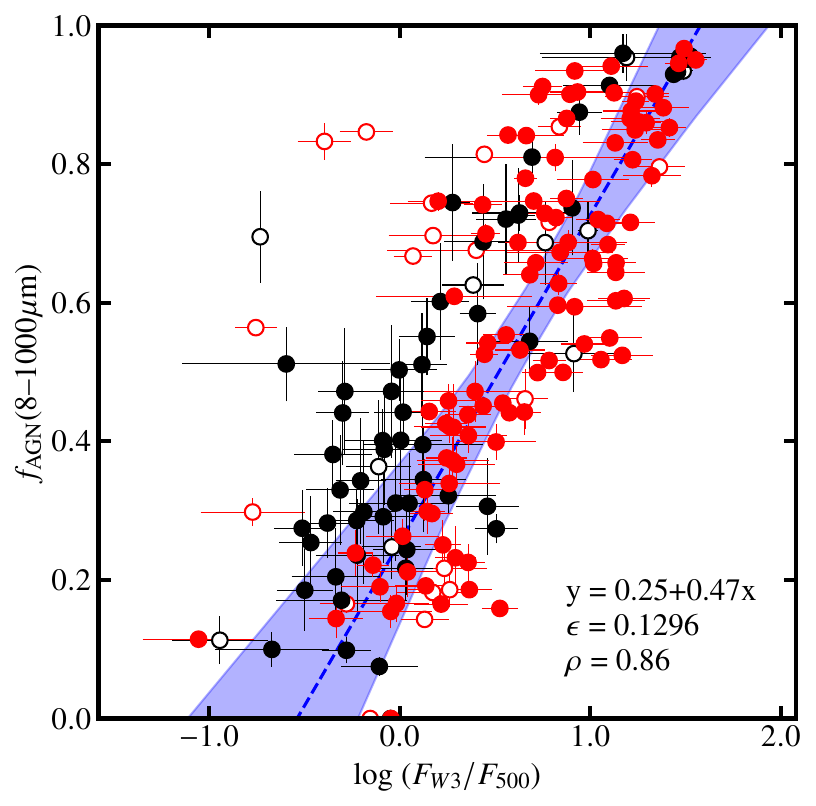}
    (\textbf{f})
\end{minipage}

\caption{{Same} 
as Figure~\ref{fig:strongestcolorfagn} but for WISE W3 and the six Herschel bands: 
(\textbf{a}) W3/70~$\upmu\mathrm{m}$,
(\textbf{b})~W3/100~$\upmu\mathrm{m}$,
(\textbf{c}) W3/160~$\upmu\mathrm{m}$,
(\textbf{d}) W3/250~$\upmu\mathrm{m}$,
(\textbf{e}) W3/350~$\upmu\mathrm{m}$, and
(\textbf{f}) W3/500~$\upmu\mathrm{m}$.
Red and black symbols denote sources detected in at least three and in no more than two Herschel bands, respectively. Filled and open symbols represent radio-quiet and radio-loud quasars, respectively. The dashed line and blue shaded region show the Bayesian linear regression fit and its uncertainty for the radio-quiet sources detected in at least three Herschel bands. The fitted relation, intrinsic scatter $\epsilon$, and correlation coefficient $\rho$ are shown in each panel.}
\label{fig:w3hscolorfagn}
\end{figure}

\begin{figure}[H]
\begin{minipage}{0.31\textwidth}
    \centering
    \includegraphics[width=\textwidth]{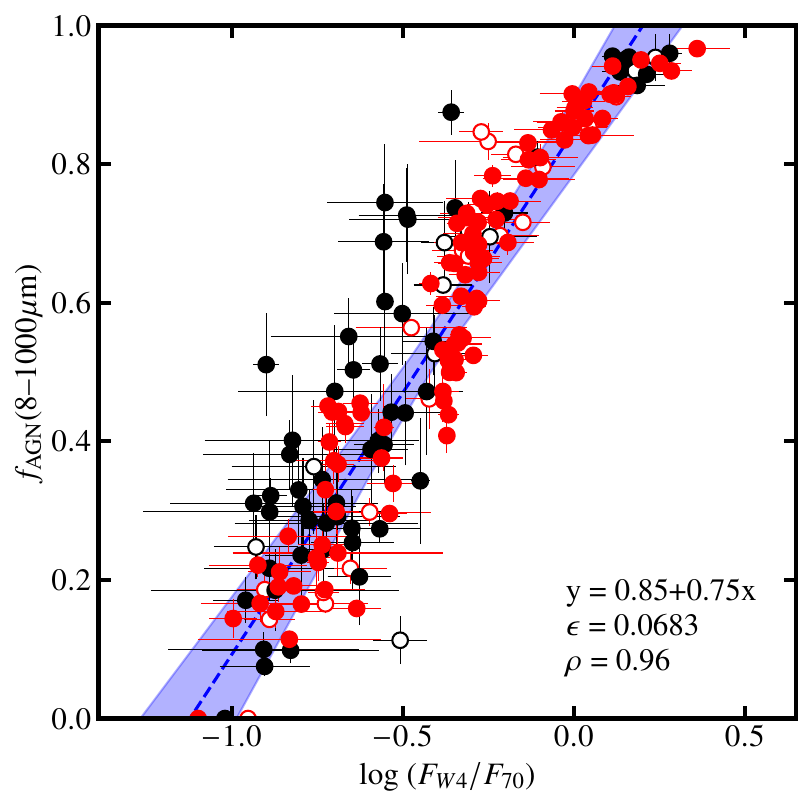}
    (\textbf{a})
\end{minipage}
\hfill
\begin{minipage}{0.31\textwidth}
    \centering
    \includegraphics[width=\textwidth]{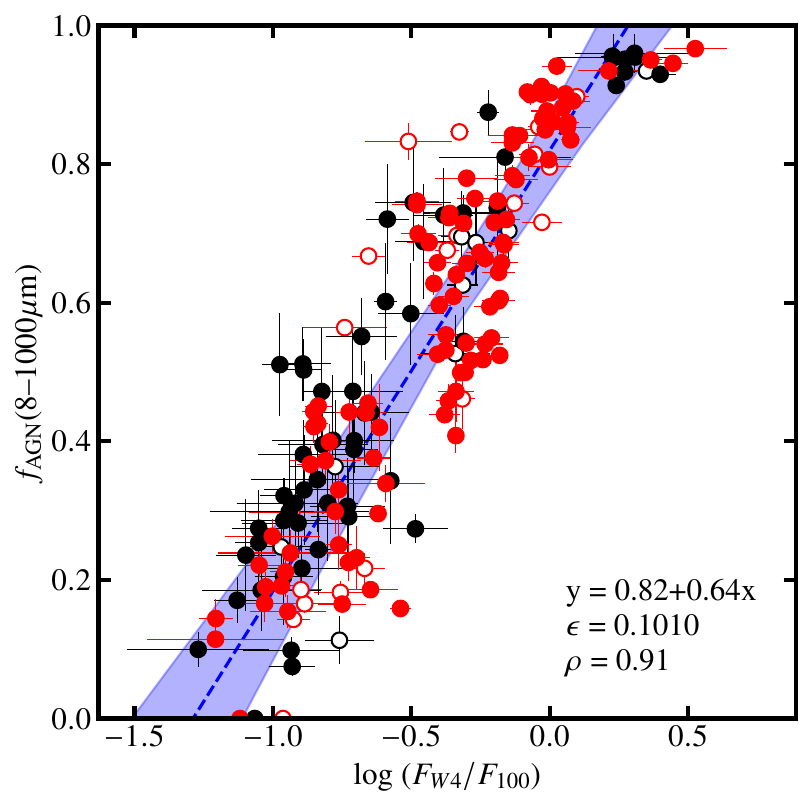}
    (\textbf{b})
\end{minipage}
\hfill
\begin{minipage}{0.31\textwidth}
    \centering
    \includegraphics[width=\textwidth]{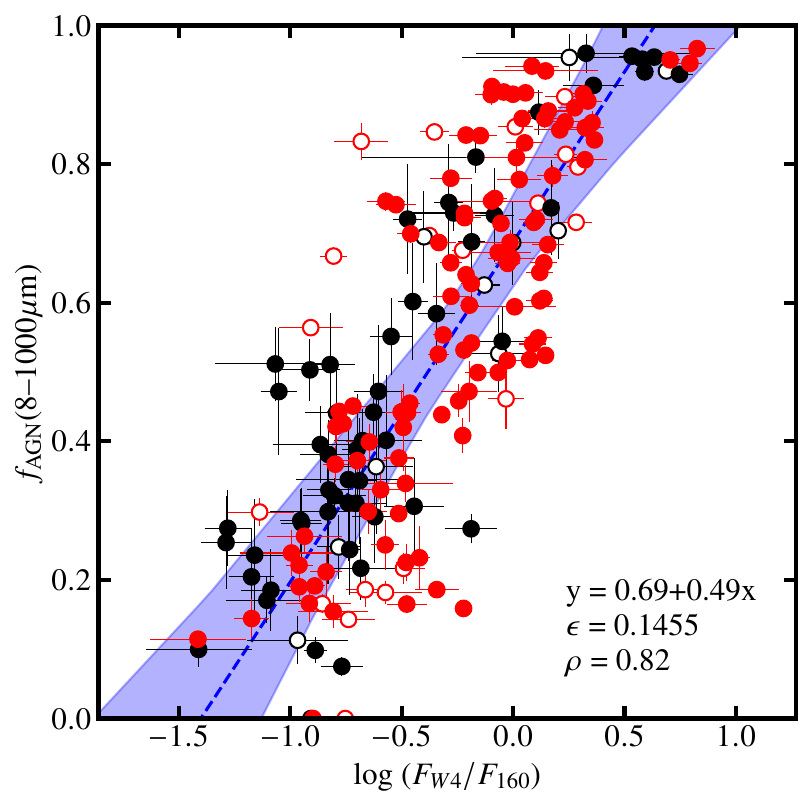}
    (\textbf{c})
\end{minipage}

\vspace{2mm}

\begin{minipage}{0.31\textwidth}
    \centering
    \includegraphics[width=\textwidth]{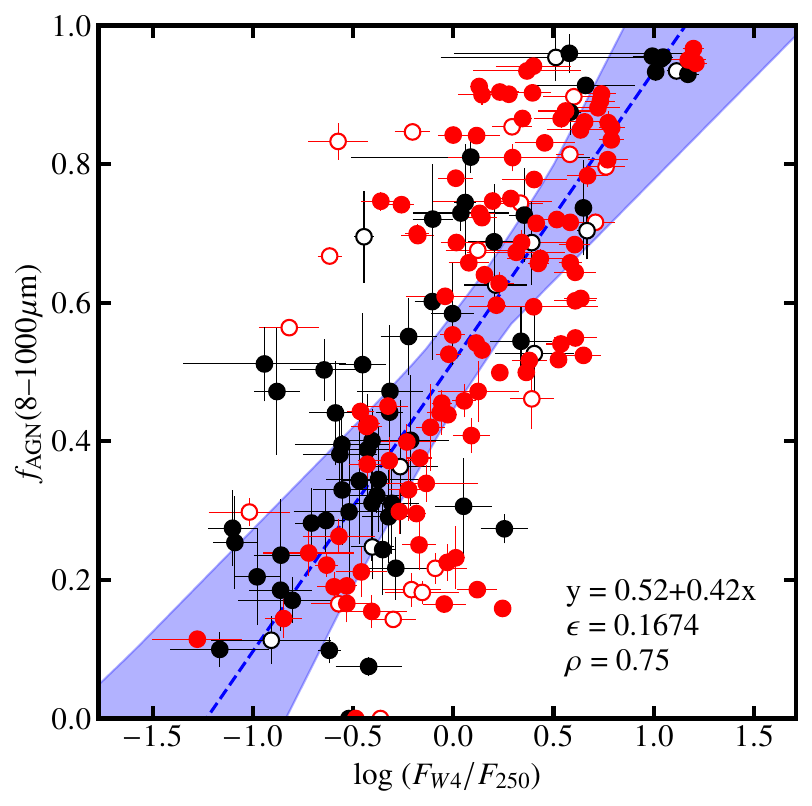}
    (\textbf{d})
\end{minipage}
\hfill
\begin{minipage}{0.31\textwidth}
    \centering
    \includegraphics[width=\textwidth]{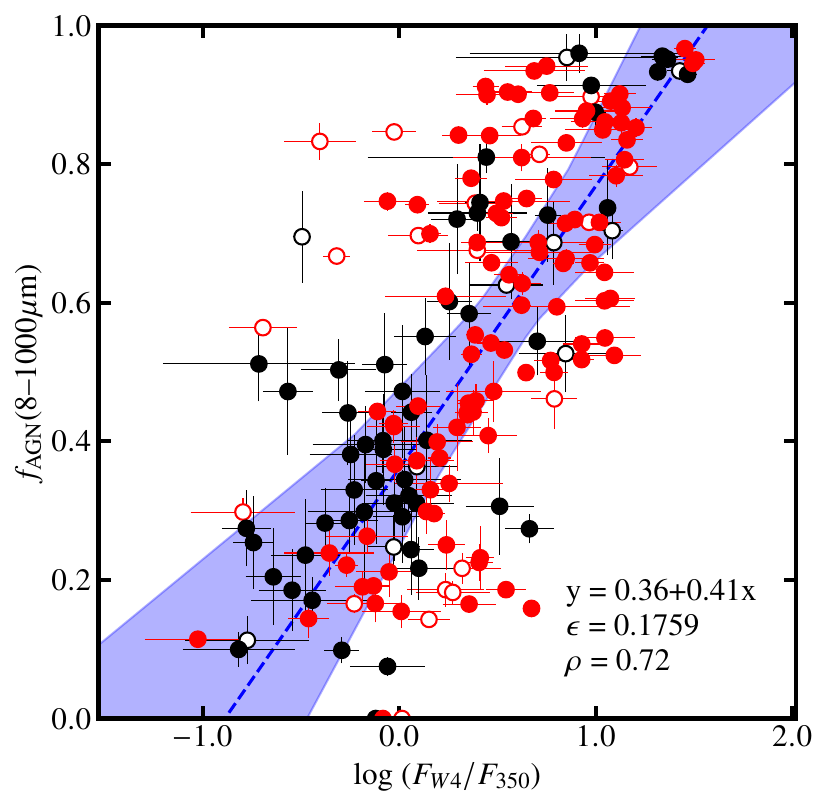}
    (\textbf{e})
\end{minipage}
\hfill
\begin{minipage}{0.31\textwidth}
    \centering
    \includegraphics[width=\textwidth]{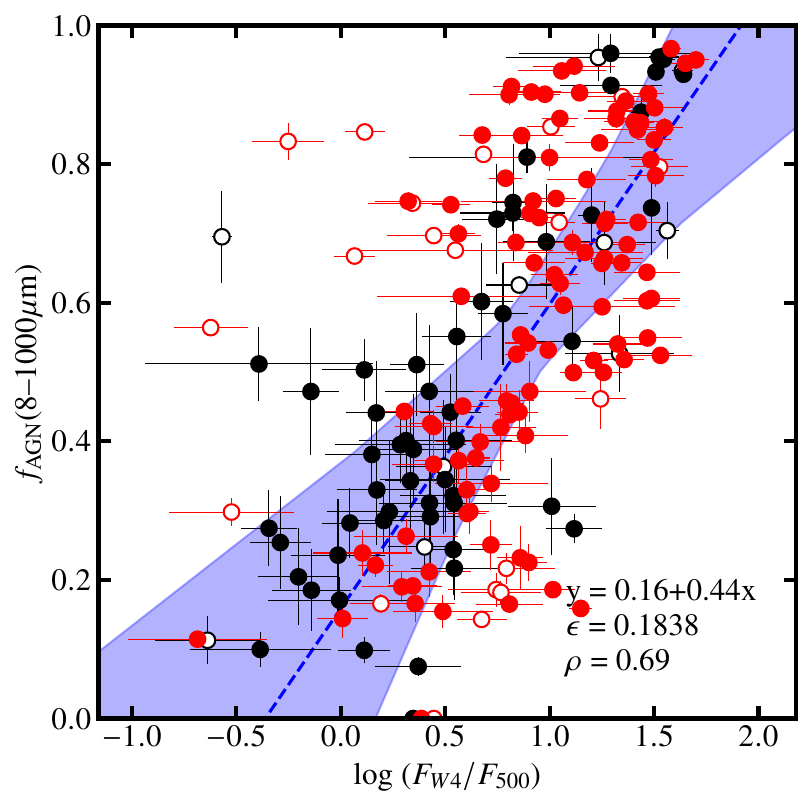}
    (\textbf{f})
\end{minipage}

\caption{{Same} 
as Figure~\ref{fig:strongestcolorfagn} but for WISE W4 and the six Herschel bands: 
(\textbf{a}) W4/70~$\upmu\mathrm{m}$,
(\textbf{b})~W4/100~$\upmu\mathrm{m}$,
(\textbf{c}) W4/160~$\upmu\mathrm{m}$,
(\textbf{d}) W4/250~$\upmu\mathrm{m}$,
(\textbf{e}) W4/350~$\upmu\mathrm{m}$, and
(\textbf{f}) W4/500~$\upmu\mathrm{m}$.
Red and black symbols denote sources detected in at least three and in no more than two Herschel bands, respectively. Filled and open symbols represent radio-quiet and radio-loud quasars, respectively. The dashed line and blue shaded region show the Bayesian linear regression fit and its uncertainty for the radio-quiet sources detected in at least three Herschel bands. The fitted relation, intrinsic scatter $\epsilon$, and correlation coefficient $\rho$ are shown in each panel.}
\label{fig:w4hscolorfagn}
\end{figure}

{Figures}
~\ref{fig:w1hscolorfagn}--\ref{fig:w4hscolorfagn} show the full set of relations between $f_{\rm AGN}(8$--$1000~\upmu\mathrm{m})$ and the flux-density ratios between the four WISE bands and the six Herschel bands.

\vspace{-6pt}
\begin{figure}[H]
\begin{minipage}{0.31\textwidth}
    \centering
    \includegraphics[width=\textwidth]{img/wisehscolor_fagn_W2_70_hyphentominus_correctendash.pdf}
    (\textbf{a})
\end{minipage}
\hfill
\begin{minipage}{0.31\textwidth}
    \centering
    \includegraphics[width=\textwidth]{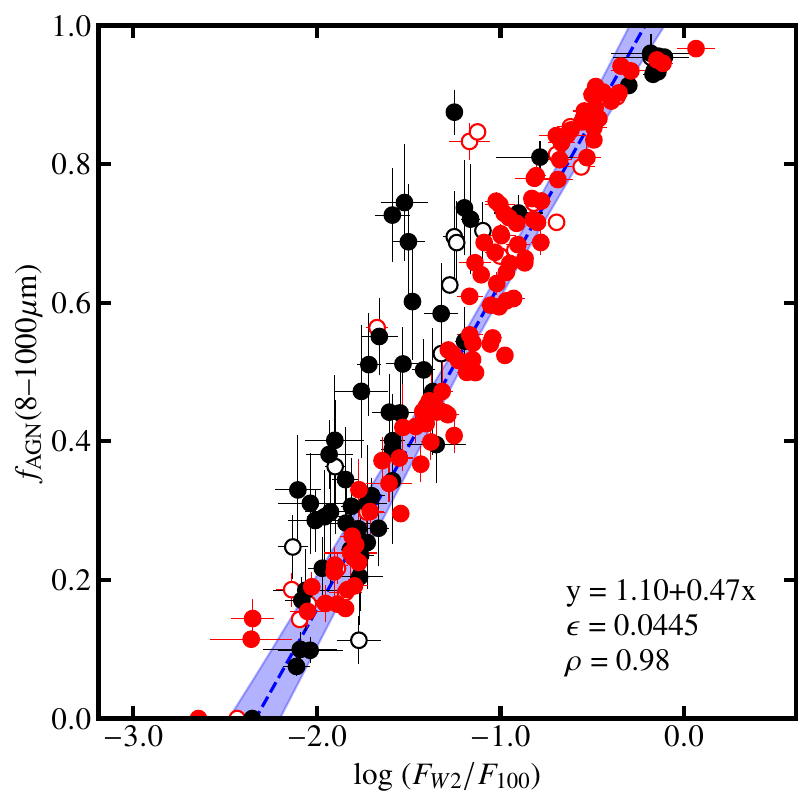}
    (\textbf{b})
\end{minipage}
\hfill
\begin{minipage}{0.31\textwidth}
    \centering
    \includegraphics[width=\textwidth]{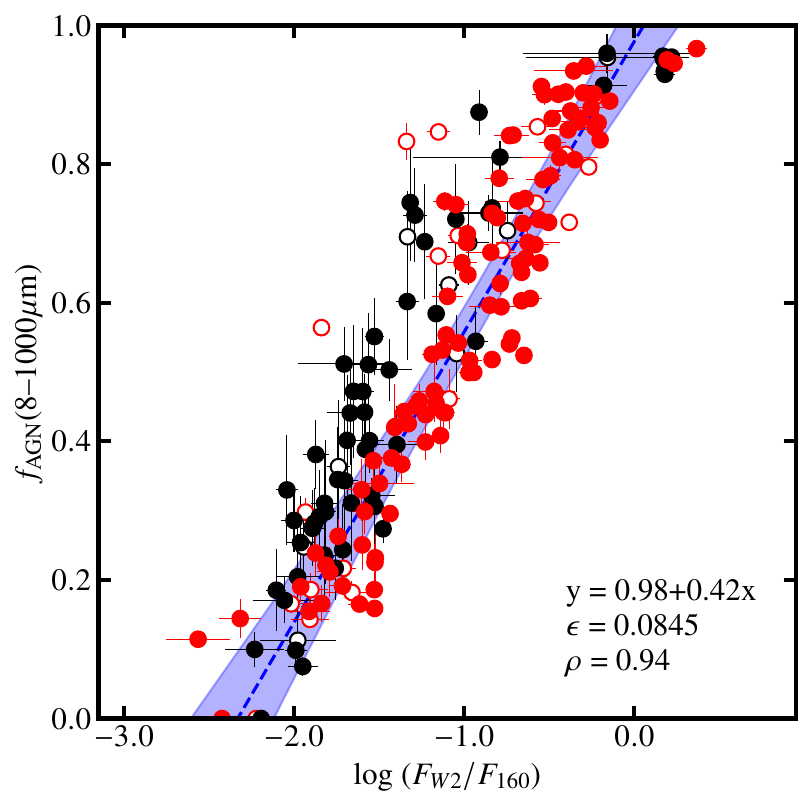}
    (\textbf{c})
\end{minipage}

\caption{\textit{Cont}.}
\end{figure}

\begin{figure}[H]\ContinuedFloat


\begin{minipage}{0.31\textwidth}
    \centering
    \includegraphics[width=\textwidth]{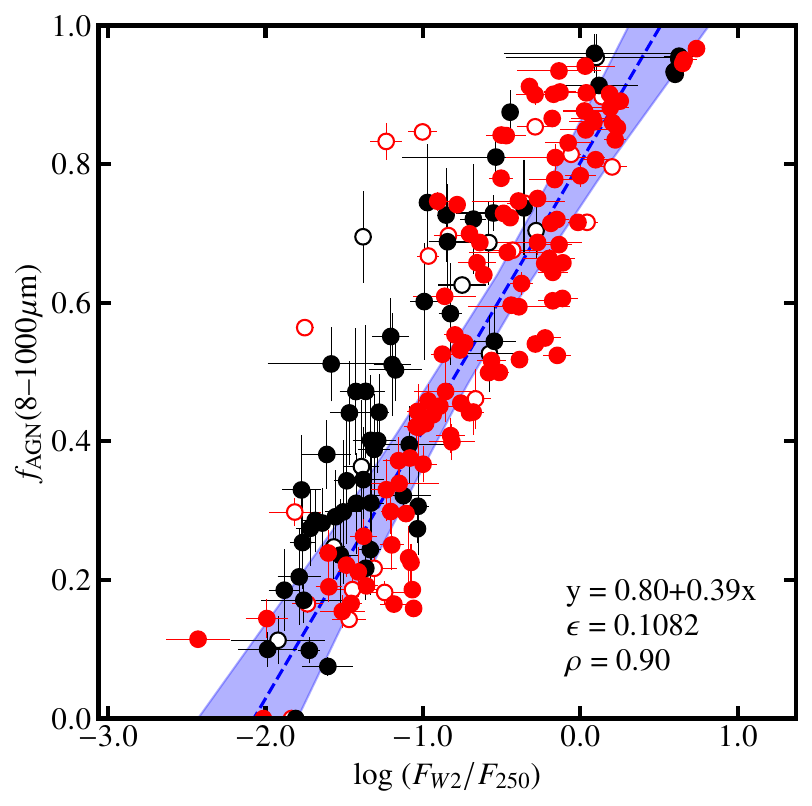}
    (\textbf{d})
\end{minipage}
\hfill
\begin{minipage}{0.31\textwidth}
    \centering
    \includegraphics[width=\textwidth]{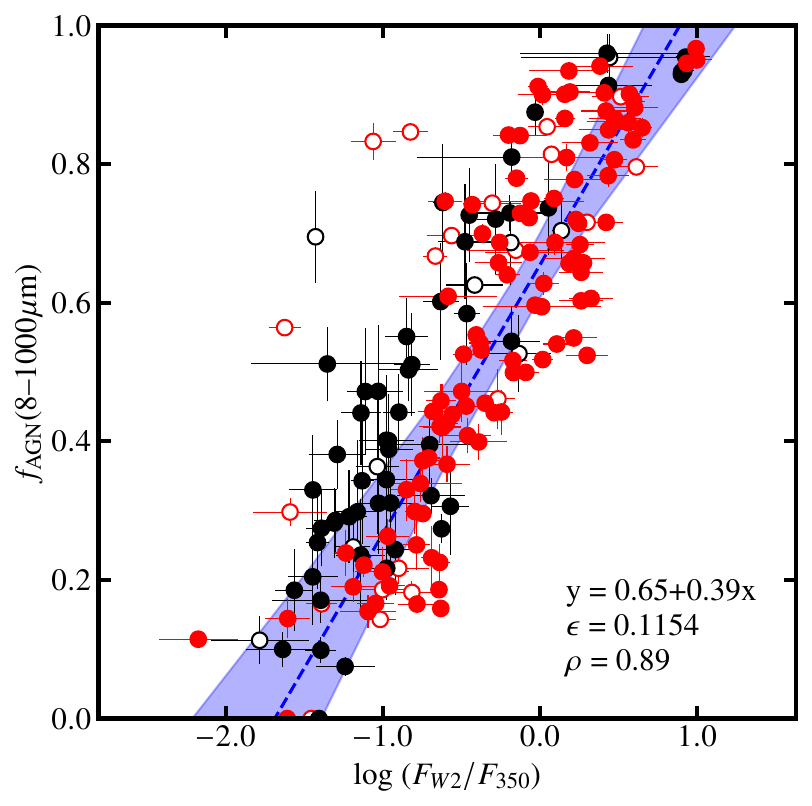}
    (\textbf{e})
\end{minipage}
\hfill
\begin{minipage}{0.31\textwidth}
    \centering
    \includegraphics[width=\textwidth]{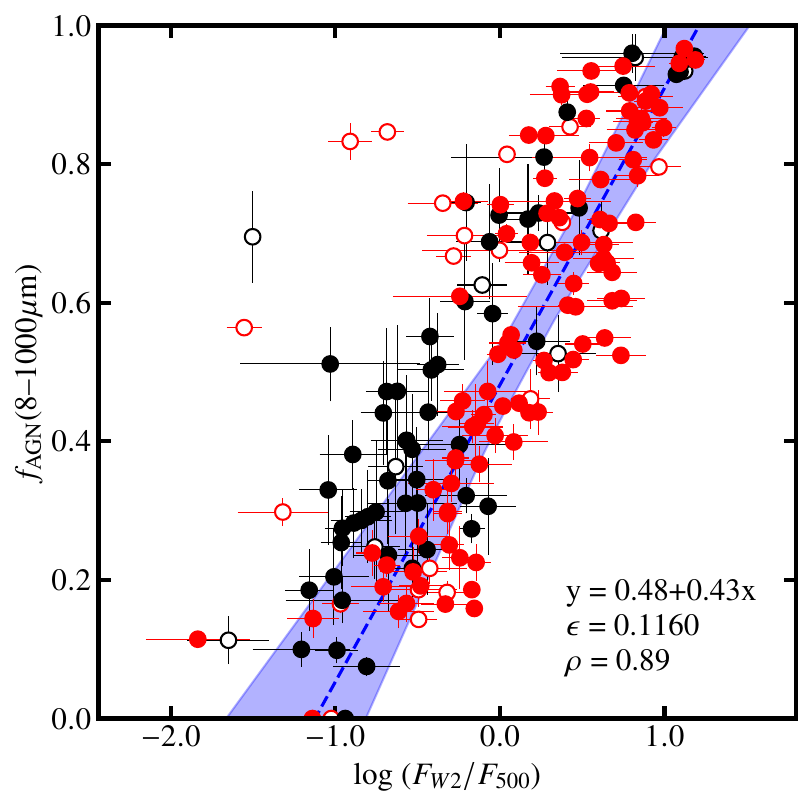}
    (\textbf{f})
\end{minipage}

\caption{{Same} 
as Figure~\ref{fig:strongestcolorfagn} but for WISE W2 and the six Herschel bands: 
(\textbf{a}) W2/70~$\upmu\mathrm{m}$,
(\textbf{b})~W2/100~$\upmu\mathrm{m}$,
(\textbf{c}) W2/160~$\upmu\mathrm{m}$,
(\textbf{d}) W2/250~$\upmu\mathrm{m}$,
(\textbf{e}) W2/350~$\upmu\mathrm{m}$, and
(\textbf{f}) W2/500~$\upmu\mathrm{m}$.
Red and black symbols denote sources detected in at least three and in no more than two Herschel bands, respectively. Filled and open symbols represent radio-quiet and radio-loud quasars, respectively. The dashed line and blue shaded region show the Bayesian linear regression fit and its uncertainty for the radio-quiet sources detected in at least three Herschel bands. The fitted relation, intrinsic scatter $\epsilon$, and correlation coefficient $\rho$ are shown in each panel.}
\label{fig:w2hscolorfagn}
\end{figure}
\vspace{-6pt}
\vspace{-6pt}

\begin{adjustwidth}{-\extralength}{0cm}
\printendnotes[custom]
\reftitle{References}

\PublishersNote{}
\end{adjustwidth}
\end{document}